\documentclass[12pt]{article}
\usepackage{jheppub}
\usepackage{graphicx}
\usepackage{bm}
\usepackage{bbold}
\usepackage{amssymb}
\usepackage{amsmath,latexsym}
\usepackage[utf8]{inputenc} 
\usepackage{bmpsize}
\usepackage{eufrak}
\newcommand{\besp}{\begin{split}}
\newcommand{\eesp}{\end{split}}

\newcommand{\g}{\hat{g}_{\mu\nu}}
\newcommand{\gup}{\hat{g}^{\mu\nu}}
\newcommand{\Du}{\hat{\Delta}^{\mu\nu}}

\newcommand{\ds}{dS_3\otimes R}

\newcommand{\gub}{SO(3)_q\otimes SO(1,1)\otimes Z_2}

\newcommand{\um}{\hat{u}^\mu}
\newcommand{\uum}{\hat{u}_\mu}
\newcommand{\un}{\hat{u}^\nu}
\newcommand{\unn}{\hat{u}_\nu}
\newcommand{\tmn}{\hat{T}^{\mu\nu}}

\newcommand{\pit}{\hat{\pi}}

\newcommand{\pr}{\partial_\rho}

\newcommand{\pp}{{\hat{p}}}
\newcommand{\po}{\hat{p}_\Omega^2}
\newcommand{\pv}{\hat{p}_\varsigma}
\newcommand{\pth}{\hat{p}_\theta}
\newcommand{\pph}{\hat{p}_\phi}
\newcommand{\ph}{\hat{p}^\rho}

\newcommand{\tem}{\hat{T}}
\newcommand{\lam}{\hat{\Lambda}}

\newcommand{\tdf}{\delta \tilde{f}}

\newcommand{\ene}{\hat{\epsilon}}
 
\newcommand{\R}{\hat{\mathcal{R}}}
\newcommand{\Ih}{\hat{I}}
\newcommand{\It}{\tilde{I}}

\newcommand{\J}{\hat{\mathcal{J}}}
\newcommand{\lu}{\hat{l}^\mu}

\newcommand{\xh}{\hat{x}}
\newcommand{\lh}{\hat{l}}
\newcommand{\xu}{\hat{\Xi}^{\mu\nu}}

\newcommand{\she}{\hat{\pi}}
\newcommand{\Pt}{\hat{\mathcal{P}}_\perp}
\newcommand{\Pl}{\hat{\mathcal{P}}_L}
\newcommand{\E}{\hat{E}_{\bold{\pp}\,\hat{u}}}
\newcommand{\El}{\hat{E}_{\bold{\pp}\,\lh}}

\newcommand{\feq}{f_\mathrm{eq}}
\newcommand{\be}{\begin{equation}}
\newcommand{\ee}{\end{equation}}
\newcommand{\bea}{\begin{eqnarray}}
\newcommand{\eea}{\end{eqnarray}}
\newcommand{\bs}{\begin{subequations}}
\newcommand{\es}{\end{subequations}}
\newcommand{\beal}{\begin{align}}
\DeclareMathAlphabet\mathbfcal{OMS}{cmsy}{b}{n}

\begin{document}
	
	\title{Far-from-equilibrium attractors and nonlinear dynamical systems approach to the Gubser flow }
	\date{\today }
	\author{Alireza Behtash$^{1,2}$}
	\author{C.~N.~Cruz-Camacho$^3$}
	\author{M.~Martinez$^1$}
	\affiliation{$^1$ Department of Physics, North Carolina State University, Raleigh, NC 27695, USA}
	\affiliation{$^2$Kavli Institute for Theoretical Physics University of California, Santa Barbara, CA 93106, USA}
	\affiliation{$^3$Universidad Nacional de Colombia, Sede Bogot\'a, Facultad de Ciencias,
		Departamento de F\'isica, Grupo de F\'isica Nuclear, Carrera 45 $N^o$ 26-85,
		Edificio Uriel Guti\'errez, Bogot\'a D.C. C.P. 1101, Colombia}
	
	\emailAdd{abehtas@ncsu.edu}
		\emailAdd{cncruzc@gmail.com}
	\emailAdd{mmarti11@ncsu.edu}

	\abstract {
		The non-equilibrium attractors of systems undergoing Gubser flow within relativistic kinetic theory are studied. In doing so we employ well-established methods of nonlinear dynamical systems which rely on finding the fixed points, investigating the structure of the flow diagrams of the evolution equations, and characterizing the basin of attraction using a Lyapunov function near the stable fixed points. We obtain the  attractors of anisotropic hydrodynamics, Israel-Stewart (IS) and transient fluid (DNMR) theories and show that they are indeed non-planar and the basin of attraction is essentially three dimensional. The attractors of each hydrodynamical model are compared with the one obtained from the exact Gubser solution of the Boltzmann equation within the relaxation time approximation. We observe that the anisotropic hydrodynamics is able to match up to high numerical accuracy the attractor of the exact solution while the second order hydrodynamical theories fail to describe it. We show that the IS and DNMR asymptotic series expansion diverge and use resurgence techniques to perform the resummation of these divergences. We also comment on a possible link between the manifold of steepest descent paths in path integrals and basin of attraction for the attractors via Lyapunov functions that opens a new horizon toward effective field theory description of hydrodynamics. Our findings indicate that anisotropic hydrodynamics is an effective theory for far-from-equilibrium fluid dynamics which resums the Knudsen and inverse Reynolds numbers to all orders.
}
	

	\maketitle
	
	%Alphabetical order ...
	%%%%%%%%%%%%%%%%%%%%%%%%%%%%%%%%%%%%
	\section{Introduction and summary}
	%%%%%%%%%%%%%%%%%%%%%%%%%%%%%%%%%%%%
	Hydrodynamics is an effective theory which describes the long wavelength phenomena and/or small frequency of physical systems. The fluid dynamical equations of motion are derived by assuming that the mean free path is smaller than the typical size of the system~\cite{landaufluid}. The existence of a large separation between the microscopic to macroscopic scales can be reformulated as a small gradient expansion around a background (usually the thermal equilibrium state) which varies slowly. Thus, hydrodynamics would be invalid in far-from-equilibrium situations where the gradients of the hydrodynamical fields are large. This might be the case in systems of small size. However, recent experimental results of $p-p$ collisions at high energies~\cite{Aad:2015gqa,Khachatryan:2015lva} have shown evidence of collective flow behavior similar to the one observed in ultra-relativistic heavy ion collisions. The experimental data measured in $p-p$ collisions can be described quantitatively by using hydrodynamical models~\cite{Weller:2017tsr,Bozek:2010pb,Werner:2010ss}. On the other hand, different theoretical toy models for both weak and strong coupling~\cite{Kurkela:2015qoa,Critelli:2017euk,Denicol:2014xca,Florkowski:2014sfa,Florkowski:2013lza,Florkowski:2013lya,Denicol:2014tha,Chesler:2009cy,Heller:2011ju,Wu:2011yd,vanderSchee:2012qj,Casalderrey-Solana:2013aba} have presented an overwhelming evidence that hydrodynamics becomes valid even in non-equilibrium situations where large gradients are present during the space-time evolution of the fluid. The success of hydrodynamical models to describe small systems as well as applying it to far-from-equilibrium situations calls for a better theoretical understanding of the foundations of hydrodynamics. 
	
	For different strongly coupled theories it has been shown that the hydrodynamical gradient series expansion has zero radius of convergence and thus, it diverges~\cite{Heller:2014wfa}. On the other hand, the divergent behavior of the hydrodynamical series expansion is also well known in weakly coupled systems based on the Boltzmann equation for relativistic and non-relativistic systems~\cite{Heller:2016rtz,Denicol:2016bjh,Bazow:2016oky,santosdufty,sant}. A more detailed mathematical analysis of the origin of this divergence has unveiled the existence of a unique universal solution, the so-called attractor~\cite{Heller:2015dha}. The novel attractor solution is intrinsically related with the mathematical theory of resurgence~\cite{Heller:2016rtz,Basar:2015ava,Aniceto:2015mto} and the details of this solution depends on the particular theory under consideration~\cite{Berges:2013eia,Micha:2004bv,Baier:2000sb,Mehtar-Tani:2016bay,Berges:2013fga,Spalinski:2017mel,Buchel:2016cbj,Romatschke:2017vte,Strickland:2017kux,Florkowski:2017jnz}. In simple terms, the attractor is a set of points in the phase space of the dynamical variables to which a family of solutions of an evolution equation merge after transients have died out. In relativistic hydrodynamics it has been found in recent years that for far-from-equilibrium initial conditions the trajectories in the phase space merge quickly towards a non-thermal attractor before the system reaches the full thermal equilibrium. This type of non-equilibrium attractor can be fully determined by very few terms of the gradient series of relatively large size which involves transient non-hydrodynamical degrees of freedom~\cite{Heller:2014wfa,Heller:2016rtz}.  This property of the attractor solution indicates that the system reaches its hydrodynamical behavior and thus, {\it hydrodynamizes} at scales of time shorter than the typical thermalization and isotropization scales. The fact that hydrodynamization happens in different size systems at short scales of time while exhibiting this degree of universality in far-from-equilibrium initial conditions might explain the unreasonable success of hydrodynamics in small systems such as $p-p$ and heavy-light ion collisions~\cite{Romatschke:2016hle}. 
	
	In this work we continue exploring the properties of attractors for rapidly expanding systems within relativistic kinetic theory. Previous works have focused on fluids undergoing Bjorken flow~\cite{Heller:2016rtz,Basar:2015ava,Aniceto:2015mto,Strickland:2017kux,Florkowski:2017jnz} and non-homogeneous expanding plasmas~\cite{Romatschke:2017acs}. We expand these studies by investigating the properties of the attractors in the plasmas undergoing Gubser flow~\cite{Gubser:2010ze,Gubser:2010ui}. 
	
The Gubser flow describes a conformal system which expands azimuthally symmetrically in the transverse plane together with boost-invariant longitudinal expansion. The symmetry of the Gubser flow becomes manifest in the de Sitter space times a line $\ds$~\cite{Gubser:2010ze,Gubser:2010ui} and thus, the dynamics of this system is studied in this curved spacetime. The search for attractors in the Gubser flow poses new challenges due to the geometry and the symmetries associated to this velocity profile. We determine the location of the attractors with the help of well-known methods of nonlinear dynamical systems~\cite{kloeden,tournier}. Our results bring new features and tools to the study of attractors which were not addressed previously in the context of relativistic hydrodynamics. For instance, in the $2d$ system of ordinary differential equations (ODE) derived from different hydrodynamical truncation schemes, we study the flow diagrams - streamline plots of the velocity vector fields in the space of state variables- and carefully examine the early- and late-time behavior of the flow lines near each fixed point and show that the system is exponentially asymptotically stable. We observe that the attractor is a $1d$ non-planar manifold only asymptotically because the $2d$ system cannot be dimensionally reduced to a $1d$ non-autonomous one by any reparametrization of the variables, which is a signature of Gubser flow geometry governed by a true $3d$ autonomous dynamical system as opposed to the Bjorken model. This is made more precise in the context of this $3d$ dynamical system, where the linearization problem is reviewed and a mathematically rigorous definition of attractor is given. We estimate the shape of the basin of attraction by giving an approximate Lyapunov function near the attracting fixed point. There we 
touch upon an important link between the basin of attraction and attractors with manifold of steepest descent paths in the path integral formalism of quantum field theories and briefly discuss the role of a Lyapunov function as an analogue of some effective action for the stable hydrodynamical and non-hydrodynamical modes. 

 We finally discuss the properties of attractors of second order hydrodynamical theories (Israel-Stewart (IS)~\cite{Israel:1979wp} and transient fluid (DNMR) theory~\cite{Denicol:2012cn}), anisotropic hydrodynamics (aHydro)~\cite{Martinez:2010sc,Ryblewski:2010ch,Martinez:2017ibh,Alqahtani:2017tnq,Alqahtani:2017jwl,Bluhm:2015raa,Bluhm:2015bzi,Alqahtani:2016rth,Molnar:2016vvu,Molnar:2016gwq,Bazow:2015cha,Nopoush:2014pfa,Tinti:2013vba,Bazow:2013ifa,Ryblewski:2012rr,Martinez:2012tu} and the exact kinetic theory solution of the Gubser flow~\cite{Denicol:2014xca,Denicol:2014tha}. The numerical comparisons lead us to conclude that aHydro reproduces with high numerical accuracy the universal asymptotic attractor obtained from the exact kinetic Gubser solution. Finally we show that the asymptotic series solution of DNMR and IS diverges asymptotically and we briefly comment how to cure this problem by using a resurgent transseries.
	
	The paper is structured as follows: In Sec.~\ref{sec:setup} we review the Gubser flow, its exact solution for the Boltzmann equation within the relaxation time approximation (RTA) as well as different hydrodynamical truncation schemes. In Sec.~\ref{sec:flow} we study extensively the flow lines of the IS theory in 2 and 3 dimensions. Our numerical studies of the attractors for different theories and their comparisons are discussed in Sec.~\ref{sec:attractors}. In this section we also address the issue of the divergence of the IS and DNMR hydrodynamical theories and how to fix it using resurgent asymptotics. Our findings are summarized in Sec.~\ref{sec:conclusion}. Some technical details of our calculations are presented in the Appendices.
	
	%%%%%%%%%%%%%%%%%%%%%%%%%%%%%%%%%%%%
	\section{Setup}
	\label{sec:setup}
	%%%%%%%%%%%%%%%%%%%%%%%%%%%%%%%%%%%%

	Before starting our discussion we first introduce the notation used in this paper. We work in natural units where $\hbar=c=k_B=1$. The metric signature is taken to be ``mostly plus'' $(-,+,+,+)$. In Minkowski space with Milne coordinates $x^{\mu }=(\tau, r, \phi, \varsigma)$ the line element is given by
	\begin{equation}
		ds^{2}=g_{\mu \nu }dx^{\mu }dx^{\nu }=-d\tau^{2}+dr^{2}+r^2d\phi^{2}+\tau^2 d\varsigma^{2}\,.
	\end{equation}
	where the longitudinal proper time $\tau$, spacetime rapidity $\varsigma$ and polar coordinates $r$ and $\phi$ are related with the usual Cartesian coordinates ($t,x,y,z$) through the following expressions
	\begin{equation}
		\begin{split}
			&\tau =\sqrt{t^{2}{-}z^{2}}\,,\hspace{2cm}\varsigma ={\rm arctanh}{\left(\frac{z}{t}\right)}\\
			&r=\sqrt{x^{2}{+}y^{2}}\,,\hspace{1.85cm}\phi=\arctan\left(\frac{y}{x}\right)\,.
		\end{split}
	\end{equation}
	It is better to study the Gubser flow in de Sitter space times a line ($dS_3\otimes R$) that is a curved spacetime in which the flow is static and the symmetries are manifest~\cite{Gubser:2010ze,Gubser:2010ui}. This is obtained by applying a conformal map between $dS_3\otimes R$ and Minkowski space, which consists of rescaling the metric $ds^2\to d\hat{s}^2=e^\Omega ds^2$ with $\Omega=\log\tau^{-2}$. Afterwards one performs the following coordinate transformation $x^\mu=(\tau,r,\phi,\eta)\mapsto \xh^\mu=(\rho,\theta,\phi,\eta)$~\footnote{We assign variables with a hat to all quantities defined in $dS_3\otimes R$.} with
	\be
	\label{eq:rhotheta}
	\rho(\tilde\tau,\tilde r) =-\mathrm{arcsinh}\left( \frac{1-\tilde\tau^2+\tilde r^2}
	{2\tilde\tau }\right)\,,\hspace{1cm}
	\theta (\tilde\tau,\tilde r) =\mathrm{arctan}\left(\frac{2\tilde r}{1+\tilde\tau^2-\tilde r^2}\right) \,.
	\ee
	Here, $\tilde\tau = q\tau$ and $\tilde r = q r$ with $q$ being an arbitrary energy scale that sets the transverse size of the system~\cite{Gubser:2010ze,Gubser:2010ui}. The time-like coordinate is the variable $\rho\in (-\infty,\infty)$ and the polar coordinate is $ \theta\in [0,2\pi)$. Therefore, the line element in $dS_3\otimes R$ reads as
	\begin{equation}
		d\hat{s}^{2}=-d\rho ^{2}+\cosh ^{2}\!\rho \left( d\theta ^{2}+\sin
		^{2}\theta\, d\phi ^{2}\right) +d\varsigma ^{2}\,, 
		\label{eq:linedS3R}
	\end{equation}
	so the metric is $\g=\mathrm{diag}\bigl(-1,\cosh^2\rho, \cosh^2\rho\sin^2\theta,1\bigr)$. In $dS_3\otimes R$ the flow velocity is the normalized time-like vector $\hat{u}^\mu=(1,0,0,0)$, which is invariant under the $q$-deformed symmetry group $\gub$ ~\cite{Gubser:2010ze,Gubser:2010ui}. We choose the fluid velocity to be defined in the Landau frame, i.e., $\tmn\,\hat{u}_\nu\equiv\ene\,\um$.

	%%%%%%%%%%%%%%%%%%%%%%%%%%%%%%%%%%%%
	\subsection{Relativistic kinetic theory for the Gubser flow}
	\label{subsec:relkin}
	%%%%%%%%%%%%%%%%%%%%%%%%%%%%%%%%%%%%
	
	In kinetic theory the macroscopic hydrodynamic variables are calculated as momentum moments of the distribution function. The symmetry of the Gubser flow restricts the phase-space distribution $f(\xh,\pp)=f(\rho,\po,\pp_\varsigma)$, i.e., the distribution function depends only on invariants of the $\gub$  symmetry group: the de Sitter time $\rho$, the combination of momentum components $\po=\pp_\theta^2+\pp_\phi^2/\sin^2\theta$ where $\pp_\theta$ and the longitudinal momentum component $\pp_\phi$ are conjugate to the coordinates $\theta,\phi$, respectively, as well as $\pp_\varsigma$, that is conjugate to the coordinate $\varsigma$~\cite{Denicol:2014xca,Denicol:2014tha}~\footnote{A more detailed derivation of this exact solution can be found in the original references~\cite{Denicol:2014xca,Denicol:2014tha}.}. Thus, in $\ds$ the RTA Boltzmann equation reduces to a one-dimensional relaxation type equation~\cite{Denicol:2014xca,Denicol:2014tha}
	\be
	\label{eq:boltzgubeq}
	\partial_\rho f(\rho,\po,\pp_\varsigma)=-\frac{1}{\hat{\tau}_r(\rho)}
	\biggl(f(\rho,\po,\pp_\varsigma)-\feq\biggl(\frac{-\hat{u}\,{\cdot}\,\pp}{\tem(\rho)}\biggr)\biggr)\,,
	\ee
	where $\tem$ is the temperature. In this work we take $\feq(z){\,=\,}e^{-z}$ as the local thermal equilibrium distribution. The conformal symmetry demands $\hat{\tau}_r(\rho)=c/\tem(\rho)$ with $c=5\,\eta/s$ where $\eta$ and $s$ are the shear viscosity and entropy density, respectively. Given a solution for the Boltzmann equation~\eqref{eq:boltzgubeq} one calculates the energy-momentum tensor as a second-rank tensor moment defined as~\footnote{Any phase-space observable $\hat{\mathcal{O}}(\hat{x},\hat{p}^\mu)$  obtained from a given phase-space distribution $f_X$ will be denoted as $\langle\mathcal{O}(\hat{x},\hat{p})\rangle_{X}$.}
	\be
	\label{eq:tmn}
	\tmn(\rho)=\langle\,\pp^\mu\pp^\nu\,\rangle\,.
	\ee
	The relaxation equations of $\tmn$ are obtained by either solving Eq.~\eqref{eq:boltzgubeq} exactly or by finding an approximate perturbative solution. In the next section we review different approximate methods to obtain the fluid dynamical equations within kinetic theory as well as the exact solution of the kinetic Eq.~\eqref{eq:boltzgubeq}.
	
	%%%%%%%%%%%%%%%%%%%%%%%%%%%%%%%%%%%%
	\subsection{Fluid dynamical theories}
	\label{subsec:fluidtheories}
	%%%%%%%%%%%%%%%%%%%%%%%%%%%%%%%%%%%%
	
	For weakly coupled systems the equations of the macroscopic fluid dynamical variables are derived from a microscopic underlying kinetic theory based on the Boltzmann equation. The derivation of these equations assumes that the Boltzmann equation admits a generic solution of the form
	\be
	\label{eq:generexp}
	f(x^\mu,p_i)=f_b(x^\mu,p_i)\,\sum_{\alpha,l}\,a_{\alpha}(x^\mu)\,M_\alpha^{(l)}(x^\mu,p_i)\,,
	\ee
	where $f_b$ is a distribution function that describes the evolution of an existing background, $M^{(l)}_\alpha$ are orthogonal polynomials of degree $l$, whose orthogonality properties depend on $f_0$, and $a_\alpha$ are moments of the full distribution function $f$. The leading order background distribution function is chosen based on the problem at hand~\cite{mintzer}. In the rest of the section we shall briefly describe two choices of the leading order background distribution function which leads to different relaxation equations for the components of the energy-momentum tensor.  
	%%%%%%%%%%%%%%%%%%%%%%%%%%%%%%%%%%%%
	\subsubsection{Expansion around an equilibrium background}
	\label{subsubsec:IS-DNMR}
	%%%%%%%%%%%%%%%%%%%%%%%%%%%%%%%%%%%%
	
	The canonical derivation of fluid dynamics assumes an expansion around a local equilibrium background
	\be
	\label{eq:isotropic}
	f(\hat{x},\hat{p})=f_\mathrm{eq}\left(-\frac{\hat{u}\cdot\,\pp}{\tem(\rho)}\right) + \delta f (\hat{x}\,,\hat{p})\,.
	\ee
	For relativistic systems $f_\mathrm{eq}$ is taken to be the equilibrium J\"uttner distribution function $f(x^\mu,p_i)=1/\left(e^{\beta\left[(u\cdot p)-\mu\right]}+a\right)$ where $\beta=1/T$ with $T$ being the temperature, $\mu$ the chemical potential and $a=+1,-1,0$ for particles following Fermi-Dirac, Bose-Einstein or Maxwell-Boltzmann statistics, respectively. The energy of the particle $-\hat{u}\cdot\,\pp$ is taken to be isotropic in the local rest frame (LRF) while $\delta f$ encodes the deviations from the thermal equilibrium state. It is implicitly assumed that $\delta f\ll f_\mathrm{eq}$. For systems close to equilibrium the four-momentum is decomposed as $\pp^\mu=\E\um+\pp^{\langle\mu\rangle}$ where $\E=-(\hat{u}\cdot \pp)$ (in LRF $\E\equiv\,\ph$) while $\pp^{\langle\mu\rangle}=\Du\pp_\nu$ projects over the spatial momentum component orthogonal to the flow velocity. For the Gubser flow this vectorial decomposition of the four-momentum allows to write the most general conformal isotropic energy-momentum tensor~\cite{Grad} 
	\be
	\label{eq:enemomvh}
	\tem^{\mu\nu}=\ene\,\hat{u}^\mu\hat{u}^\mu+\hat{\mathcal{P}}\hat{\Delta}^{\mu\nu}+\hat{\pi}^{\mu\nu}\,,
	\ee
	where $\ene$ is the energy density, $\hat{\mathcal{P}}$ is the isotropic pressure and $\she^{\mu\nu}$ is the shear stress tensor, which are the momentum moments of the distribution function given by
	\begin{subequations}
		\label{eq:eqtmn}
		\begin{align}
			&\ene=\langle (-\hat{u}\cdot\pp)^2\rangle\,,\\
			&\hat{\mathcal{P}}=\frac{1}{3}\langle\hat{\Delta}_{\mu\nu}\pp^\mu\pp^\nu\rangle\,,\\
			\label{eq:shearvisc}
			&\she^{\mu\nu}=\langle\pp^{\langle\mu}\pp^{\nu\rangle}\rangle\,.
		\end{align}
	\end{subequations}
	In the third line of Eqs.~\eqref{eq:eqtmn} we introduce the symmetric, orthogonal to $\um$ and traceless operator $A^{\langle\mu}B^{\nu\rangle}=\Du_{\alpha\beta}A^\alpha B^\beta$ with $\Du_{\alpha\beta}=(\hat{\Delta}^{\mu}_{\alpha}\hat{\Delta}^{\nu}_{\beta}+\hat{\Delta}^{\mu}_{\beta}\hat{\Delta}^{\nu}_{\alpha}-\frac{2}{3}\Du\hat{\Delta}_{\alpha\beta})/2$. For the Gubser flow there is only one independent shear stress component $\hat\pi{\,\equiv\,}\she^{\varsigma\varsigma}$~\cite{Gubser:2010ze,Gubser:2010ui}. For conformal systems the equation of state reads as $\hat{\mathcal{P}}=\ene/3$. 
	
	The local temperature of the system introduced in Eq.~\eqref{eq:isotropic} is found from the Landau matching condition $\ene = \ene_{eq.}(T)$~\cite{landaufluid}. This phenomenological constraint ensures that the parameter $\tem$ in $f_\mathrm{eq}$ is adjusted such that the dissipative corrections encoded in $\delta f$ do not shift the value of the energy density. As a consequence, deviations from the local thermal equilibrium are captured entirely by the shear stress tensor $\she^{\mu\nu}\equiv\langle\pp^{\langle\mu}\pp^{\nu\rangle}\rangle_{\delta}$ where $\langle\cdots\rangle_{\delta}$ indicates a momentum moment weighted by $\delta f$.
	
    The energy-momentum conservation law  $\hat{D}_\mu\tmn=0$ for the energy-momentum tensor~\eqref{eq:eqtmn} gives us the following evolution equation for the energy density $\ene$~\cite{Gubser:2010ui,Gubser:2010ze}
	\be
	\label{eq:conlawalt}
	\frac{\pr\ene}{\ene}\,+\,\frac{8}{3}\tanh\rho\,=\,\frac{\hat{\pi}}{\ene}\tanh\rho\, , 
	\ee
	which can be rewritten in terms of the temperature since for conformal systems $\ene\sim \tem^4$ and thus, $\partial_\rho \tem/\tem=4\partial_\rho\ene/\ene$. As a result one finds~\cite{Gubser:2010ui,Gubser:2010ze}
	\be
	\label{eq:conlaw}
	\frac{\pr\tem}{\tem}\,+\,\frac{2}{3}\tanh\rho\,=\,\frac{\bar{\pi}}{3}\tanh\rho\,,  
	\ee
	where $\bar{\pi}:=\hat\pi^{\varsigma\varsigma}/(\ene+\hat{\mathcal{P}})$. We point out that for the Gubser flow there is only one independent component of the shear viscous tensor~\cite{Gubser:2010ze,Gubser:2010ui}. It is needed an additional evolution equation for $\bar{\pi}$ obtained by expanding $\delta f$ within some approximation. The form of $\delta f$ is found by expanding it in terms of a set of orthogonal polynomials in energy $\E$, and a set of irreducible tensors in momentum invariant under the little group $SO(3)$ of the Lorentz transformations, i.e., 1, $p^{\langle\mu\rangle}$, $p^{\langle\mu}p^{\nu\rangle}$, etc. Afterwards, the relaxation equations of the dissipative macroscopic quantities are obtained by applying a systematic truncation method based on a power counting in the Knudsen $Kn$ and inverse Reynolds $Re^{-1}$ numbers~\cite{Denicol:2012cn}. The lowest truncation order provides the IS equations while the inclusion of all other second order terms, i.e. in $Kn^{2}$, $Re^{-2}$ and $Kn\cdot Re^{-1}$, gives the DNMR equations. A detailed discussion of this power counting schemes can be found in Ref.~\cite{Denicol:2012cn}.
	
	For the normalized shear stress $\bar{\pi}=\hat\pi/(\ene+\hat{\mathcal{P}})$ one finds the following evolution equation for the IS and DNMR theory respectively~\cite{Marrochio:2013wla,Denicol:2014tha}
	\bs
	\label{eq:secthPieq}
	\beal
	\label{eq:ISpi}
	\text{For IS theory:}&\hspace{.75cm}\hat{\tau}_{\hat{\pi}}\left(\frac{d\bar\pi}{d\rho}\,+\,\frac{4}{3}\left(\bar{\pi}\right)^2\tanh\rho\right)\,+\,\bar\pi=\frac{4}{3}\frac{\eta}{s\,\tem}\,\tanh\rho\,,\\
	\label{eq:DNMRpi}
	\text{For DNMR theory:}
	&\hspace{.75cm}\hat\tau_{\hat\pi}\left(\pr\bar{\pi}+\frac{4}{3}\left(\bar{\pi}\right)^2\,\tanh\rho\right)+\,\bar{\pi}=\frac{4}{3}\frac{\eta}{s\,\tem}\,\tanh\rho\,\,+\,\frac{10}{7}\hat\tau_{\hat\pi}\bar{\pi}\,\tanh\rho\,.
\end{align}
\es
The conformal symmetry implies $\hat{\tau}_{\hat\pi}=c/\tem$ with $c=5\eta/s$.

%%%%%%%%%%%%%%%%%%%%%%%%%%%%%%%%%%%%
\subsection{Anisotropic hydrodynamics}
\label{subsec:aniso}
%%%%%%%%%%%%%%%%%%%%%%%%%%%%%%%%%%%%
Anisotropic hydrodynamics~\cite{Martinez:2010sc,Ryblewski:2010ch,Martinez:2017ibh,Alqahtani:2017tnq,Alqahtani:2017jwl,Bluhm:2015raa,Bluhm:2015bzi,Alqahtani:2016rth,Molnar:2016vvu,Molnar:2016gwq,Bazow:2015cha,Nopoush:2014pfa,Tinti:2013vba,Bazow:2013ifa,Ryblewski:2012rr,Martinez:2012tu} attempts to describe systems with a large expansion rate along some particular direction $\lu$. In these situations there is a momentum anisotropy along $\lu$ driving the system far from equilibrium. Thus, instead of expanding around an equilibrium configuration, it is better to consider an anisotropic background as a leading order term for the generic expansion of the Boltzmann equation~\eqref{eq:generexp}, i.e., 
\be
\label{eq:anisexp}
f(\hat{x},\hat{p})=f_a\left(-\frac{\hat{u}\cdot\pp}{\hat\Lambda},\frac{\xi}{\hat\Lambda}(\hat{l}\cdot\pp)\right)+\tdf(\hat{x},\hat{p})\,.
\ee
The parameter $\xi$ measures the strength of momentum anisotropy along the $\lu$ direction, $\hat\Lambda$ is an arbitrary momentum scale and $\delta \tilde{f}$ takes into account residual dissipative corrections. In principle, the parameter $\xi$ measures the size of the microscopic momentum-space anisotropies. For anisotropic systems the four-momentum of particles is decomposed as $\pp^\mu=\E \hat u^\mu+\El\lu+\pp^{\{\mu\}}$ where $\El=(\lh\cdot\pp)$ is the spatial component of the particle momentum along the space-like vector $\lu$ ($\lu\lh_\mu$=1) while $\pp^{\{\mu\}}=\xu\pp_\nu$ are the spatial components of the momentum which are orthogonal to both $\um$ and $\lu$. Here, we introduce the projection tensor $\xu=\gup+\um\un-\lu\hat{l}^\nu=\Du-\lu\hat{l}^\nu$ which is orthogonal to $\um$ and $\lu$~\cite{Huang:2009ue,Huang:2011dc,Gedalin1,Gedalin2,Molnar:2016vvu,Molnar:2016gwq,Martinez:2017ibh}. We also define the symmetric traceless operator $A^{\{\mu}B^{\nu\}}=\xu_{\alpha\beta}A^\alpha\,B^\beta$ with $\xu_{\alpha\beta}=\left(\Xi^\mu_\alpha\Xi^\nu_\beta+\Xi^\nu_\beta\Xi^\mu_\alpha-\xu\Xi_{\alpha\beta}\right)/2$. By construction $\xu_{\alpha\beta}$ is also orthogonal to $\um$ and $\lu$. 

The anisotropic decomposition for $\pp^\mu$ allows us to write the most general anisotropic energy-momentum tensor as follows~\cite{Molnar:2016gwq,Martinez:2017ibh}
\be
\label{eq:MNRemt}
\tmn=\ene\,\um\,\un + \Pl\,\lu\lh^{\nu} + \Pt\,\xu\,,
\ee
where the energy density $\ene$, transverse and longitudinal pressures $\Pt$ and $\Pl$, respectively, are the following momentum moments
\bs
\label{eq:anisodecomp}
\beal
\ene&=\uum\unn\tmn\,\equiv \langle\,(-\,\hat{u}\cdot\pp)^2\,\rangle\,,\\
\Pt&=\frac{1}{2}\hat{\Xi}_{\mu\nu}\tem^{\mu\nu}\,\equiv\frac{1}{2}\langle\,\hat{\Xi}_{\mu\nu}\pp^\mu\pp^\nu\,\rangle\,,\\
\label{eq:planis}
\Pl&=\lh_\mu\lh_\nu\tmn\,\equiv\langle\,(\hat{l}\cdot\pp)^2\,\rangle\,.
\end{align}
\es
In this case the conformal symmetry implies that $\ene=2\Pt+\Pl$. For the Gubser flow the shear stress tensor can be related with the total pressure anisotropy through~\cite{Molnar:2016gwq,Martinez:2017ibh} 
\be
\label{eq:shearreduced}
\begin{split}
\hat{\pi}^{\mu\nu}&=\frac{2}{3}(\Pl-\Pt)\left(\lu\hat{l}^\nu{-}\textstyle{\frac{1}{2}}\xu\right)\,.
\end{split}
\ee  
%

%%%%%%%%%%%%%%%%%%%%%%%%%%%%%%%%%%%%
\subsubsection{$\mathcal{P}_L$ matching}
\label{subsubsec:PLmat}
%%%%%%%%%%%%%%%%%%%%%%%%%%%%%%%%%%%%
When introducing the leading order anisotropic distribution function defined in Eq.~\eqref{eq:anisexp} we did not comment on how to determine the parameters $\hat{\Lambda}$ and $\xi$. The different variants of aHydro have been proposed in recent years in order to perform this matching procedure~\cite{Martinez:2010sc,Ryblewski:2010ch,Martinez:2017ibh,Alqahtani:2017tnq,Alqahtani:2017jwl,Bluhm:2015raa,Bluhm:2015bzi,Alqahtani:2016rth,Molnar:2016vvu,Molnar:2016gwq,Bazow:2015cha,Nopoush:2014pfa,Tinti:2013vba,Bazow:2013ifa,Ryblewski:2012rr,Martinez:2012tu}. In a previous study of the Gubser flow, it was shown that the $\Pl$ matching scheme~\cite{Martinez:2017ibh}  provides the most accurate macroscopic description when comparing its predictions with the ones obtained from an exact solution of the RTA Boltzmann equation. Thus, we shall consider in this work this matching procedure.

We start by introducing first the anisotropic integrals
%\
%
\bea
\label{eq:Inlq}
I_{nlq}&=&\bigl\langle(-\hat{u}\cdot\pp)^{n-l-2q} (\hat{l}\cdot\pp)^l (\hat{\Xi}_{\mu\nu}\pp^\mu\pp^\nu)^q\bigr\rangle \nonumber\\
&\equiv& \Ih_{nlq}(\hat{\Lambda},\xi)+\It_{nlq} \,.
\eea
The first term on the RHS of the previous equation comes from the leading order contribution associated to the anisotropic distribution function $f_a$~\eqref{eq:anisexp} while the second term corresponds to the subleading contribution from $\delta\tilde{f}$ in Eq.~\eqref{eq:anisexp}. In  Appendix \ref{app:anisint} we show how to perform the integrals of the leading order anisotropic distribution function for the massless case. The functional form of $f_a$ is taken to be the RS ansatz which in the LRF looks like~\cite{Romatschke:2003ms} 
\be
\label{eq:RSansatz}
f_a(\xh,\pp;\hat\Lambda,\xi)=f_\mathrm{eq}\big(E_\mathrm{RS}(\xi)/\hat{\Lambda}\big)
\ee
where $f_\mathrm{eq}(z)=e^{-z}$ is a Maxwellian distribution function evaluated for the momentum-anisotropic argument
\be
\label{eq:ERS}
E_\mathrm{RS}(\xi)
\equiv \sqrt{(\hat{u}\cdot\pp)^2+\xi\,(\hat{l}\cdot\pp)^2} 
= \sqrt{\po/(\cosh^2\rho)+(1{+}\xi)\pp_\varsigma^2}\,.
\ee
The leading order anisotropic variables contributing to the anisotropic energy-momentum tensor~\eqref{eq:MNRemt} are
\be
\label{eq:RStmn}
\tmn_\mathrm{RS}=\ene_\mathrm{RS}\,\um\,\un\,+\,\Pl^\mathrm{RS}\,\lu\lh^{\nu}\,+\,\Pt^\mathrm{RS}\,\xu,
\ee
where 
\bs
\label{eq:RSmacquant}
\begin{align}
\label{eq:RSenergydens}
\ene_\mathrm{RS}&=\left\langle(-\,\hat{u}\cdot\pp)^2\right\rangle_a=\Ih_{200}\big(\hat{\Lambda},\xi\big)\,,\\
\Pl^\mathrm{RS}&=\bigl\langle(\hat{l}\cdot\pp)^2\bigr\rangle_a=\Ih_{220}\big(\hat{\Lambda},\xi\big)\,,\\
\Pt^\mathrm{RS}&=\frac{1}{2}\bigl\langle\hat{\Xi}_{\mu\nu}\pp^\mu\pp^\nu\bigr\rangle_a
=\frac{1}{2}\Ih_{201}\big(\hat{\Lambda},\xi\big).
\end{align}
\es
%\
The traceless condition of the conformal anisotropic energy-momentum tensor implies that $\ene_\mathrm{RS}= 2\Pt^\mathrm{RS}+\Pl^\mathrm{RS}$. The conservation law gives the evolution equation for the energy density
\be
\label{eq:anisconslaw}
\partial_\rho\ene_{RS}+\frac{8}{3}\ene_{RS}\,\tanh\rho=\frac{2}{3}\left(\Pl^{RS}-\Pt^{RS}\right)\,\tanh\rho\,.
\ee

The matching condition for the energy density $\ene_\mathrm{RS}(\hat\Lambda,\xi) = \ene_\mathrm{eq}(T)$ leads to the following relation between the temperature $\tem$ and the momentum scale $\hat{\Lambda}$ 
\be
\label{eq:LMcond}
\hat{\Lambda} = \frac{\tem}{\big(\R_{200}(\xi)\big)^{1/4}}\,.
\ee
The function $\R_{200}(\xi)$ is given in Eq.~\eqref{eq:anisint2}. This relation ensures that the energy density does not receive any contribution from the residual deviation $\delta\tilde{f}$ in Eq.~\eqref{eq:anisexp}. Now, for the Gubser flow the conservation law~\eqref{eq:anisconslaw} indicates that the pressure anisotropy is the force that drives the system far from equilibrium. The microscopic origin of this force is the momentum anisotropy created by the expansion of the system which in the LRF is measured by the anisotropy parameter $\xi$. Hence, one simply adjusts the value of $\xi$ such that the leading order anisotropic distribution function $f_a$ fully captures the information of the full pressure anisotropy. For the Gubser flow this means that the longitudinal pressure $\Pl$~\eqref{eq:planis} does not receive contributions from the residual deviation $\delta\tilde{f}$ of the distribution function~\eqref{eq:anisexp}, i.e. $\Pl=\Pl^{RS}$. Now the effective shear viscous tensor is related with the total pressure anisotropy via the identity~\eqref{eq:shearreduced} which in this case gives~\cite{Martinez:2017ibh}
\be
\label{eq:RSpi}
\begin{split}
\she_\mathrm{RS}&=\frac{2}{3}\left(\Pl^{RS}-\Pt^{RS}\right)\,,\\
&=\bigl\langle(\hat{l}\cdot\pp)^2-\textstyle{\frac{1}{3}}(-\hat{u}\cdot\pp)^2\big\rangle_a\,,\\
&=\Ih_{220}\big(\hat{\Lambda},\xi\big)-\textstyle{\frac{1}{3}}\Ih_{200}\big(\hat{\Lambda},\xi\big).
\end{split}
\ee
Thus, the $\Pl$ matching prescription can be recast into the following condition for the effective shear viscous component
\be
\label{eq:PLmatch}
\she=\she_\mathrm{RS}.
\ee
Furthermore, Eq.~\eqref{eq:RSpi} allows us to rewrite the conservation law~\eqref{eq:anisconslaw} as
\be
\partial_\rho\ene_{RS}+\frac{8}{3}\ene_{RS}\,\tanh\rho=\hat\pi_{RS}\,\tanh\rho\,.
\ee
For the $\Pl$ matching, the equation for $\hat{\pi}_\mathrm{RS}$ is found from Eq. \eqref{eq:RSpi}. After some algebra one finds the following evolution equation for the normalized effective shear $\bar{\pi}$~\footnote{Technical details of this derivation are discussed in Sec.~IIIC of Ref.~\cite{Martinez:2017ibh}}
\be
\label{eq:ahydroPl}
\partial_\rho\bar\pi + \frac{\bar\pi}{\hat{\tau}_r} =
\frac{4}{3}\tanh\rho\left(\frac{5}{16} + \bar\pi - \bar\pi^2 - \frac{9}{16}\mathcal{F}(\bar\pi)\right) ,
\ee
where
\be
\label{eq:F}
\mathcal{F}({\bar\pi}) \equiv \frac{\R_{240}\bigl(\xi(\bar\pi)\bigr)}{\R_{200}\bigl(\xi(\bar\pi)\bigr)}.
\ee
The functions $\R_{240}$ and $\R_{200}$ are listed in App.~\ref{app:anisint}. Furthermore, $\xi({\bar\pi})$ in Eq.~\eqref{eq:F} is the inverse of the function
\bea
\label{eq:pibarxi}
\bar\pi(\xi) &=& 
\frac{\she}{\ene+\hat{\mathcal{P}}}=\frac{3\,\Ih_{220}{-}\Ih_{200}}{4\,\Ih_{200}}
%\nonumber\\
= \frac{1}{4}\left(\frac{3\,\R_{220}(\xi)}{\R_{200}(\xi)}-1\right).
\eea

An advantage of using aHydro is that by construction the transverse and longitudinal pressures remain positive during the entire evolution of the system as it is expected in kinetic theory~\footnote{The positive definite condition of the longitudinal and transverse pressures is not a requirement for holographic models within the AdS/CFT correspondence since the quasiparticle picture does not exist in these theories. In weakly coupled theories where the quasiparticle picture exists, the positivity of the pressure is violated in the presence of external fields.}. These constraints imply $-1/4<\bar{\pi}<1/2$ which is in agreement with the anisotropy parameter $\xi\in(-1,\infty)$. On the other hand, the LHS of Eq.~\eqref{eq:pibarxi} is identified as a term proportional to the inverse Reynolds number $Re^{-1}=\sqrt{\pi_{\mu\nu}\pi^{\mu\nu}}/\mathcal{P}_0$  (with $\mathcal{P}_0=\ene/3$ for the conformal case)~\cite{Denicol:2012cn}. Eq.~\eqref{eq:pibarxi} therefore indicates that the anisotropy parameter resums non-perturbatively not only large gradients (i.e., Knudsen number) but also large  $Re^{-1}$ numbers. An analogous relation between the effective shear and the anisotropy parameter was found in the Bjorken case~\cite{Martinez:2010sc}. The dissipative corrections $\mathcal{O}(Re^{-2})$ arise in general from the most nonlinear sector of the collisional kernel~\cite{Denicol:2012cn} and thus, the calculation of these terms is cumbersome even for the simplest kernels~\cite{Molnar:2013lta}. Moreover, some of the nonlinear terms $\mathcal{O}(Re^{-2})$ calculated in the DNMR theory lead to violations of causality~\cite{Tsumura:2015fxa}.

%%%%%%%%%%%%%%%%%%%%%%%%%%%%%%%%%%%%
\subsection{Exact solution to the Boltzmann equation in the RTA approximation}
\label{subsec:exact}
%%%%%%%%%%%%%%%%%%%%%%%%%%%%%%%%%%%%
The RTA Boltzmann equation \eqref{eq:boltzgubeq} admits the following exact solution~\cite{Denicol:2014xca,Denicol:2014tha} 
\be
\label{eq:exsol}
f_{ex}(\rho;\po,\pv) =D(\rho,\rho_0) f_0(\rho_0;\po,\pv)\,+\,\frac{1}{c}\int_{\rho_0}^\rho d\rho'\,D(\rho,\rho')\,\hat{T}(\rho')\, f_{eq}\big(\hat{E}_\pp(\rho')/\tem(\rho')\big)\,,
\ee
where $D(\rho,\rho_0){\,=\,}\exp\!\left[-\frac{1}{c}\int_{\rho_0}^\rho d\rho' \,\hat{T}(\rho')\right]$ is the damping function. For the initial condition of the distribution function $f_0$ at $\rho_0$ we shall consider the RS ansatz~\cite{Romatschke:2003ms}
\be
\label{eq:RS}
\!\!\!\!\!
f_{0}(\rho_0;\po,\pv)= 
\exp\left(-\frac{1}{\hat{\Lambda}_0}\sqrt{\frac{\po}{{\cosh^2}\rho_0}+(1{+}\xi_0)\pp_\varsigma^2}\,\right)\!.
\ee
where $\hat{\Lambda}_0$ is the initial temperature and $\xi_0$ is the initial momentum anisotropy along the $\varsigma$ direction. From the exact solution for $f$ one gets the energy density  and the only independent component of the shear stress~\cite{Denicol:2014xca,Denicol:2014tha}: 
\begin{subequations}
\label{eq:exactenepi}
\beal
\label{eq:exactene}
\ene(\rho) 
&=\Ih_{200}\,,\nonumber\\
&=D(\rho,\rho_0)\Big(\frac{\cosh\rho_0}{\cosh\rho}\Big)^{\!4}\,
\ene_\mathrm{RS}\big(\lam_0, \xi_\mathrm{FS}(\rho;\rho_0,\xi_0)\big)\nonumber\\
&+ \frac{1}{c}\int_{\rho_0}^\rho d\rho'\,D(\rho,\rho')\,\hat{T}(\rho')\Big(\frac{\cosh\rho^\prime}{\cosh\rho}
\Big)^{\!4}\,
\ene_\mathrm{RS}\big(\tem(\rho^\prime), \xi_\mathrm{FS}(\rho;\rho^\prime,0)\big), \\
\label{eq:exactpi}
\pit(\rho)&=\Ih_{220}{-}\frac{1}{3}\Ih_{200}\,,\nonumber\\
&=D(\rho,\rho_0)\Big(\frac{\cosh\rho_0}{\cosh\rho}\Big)^{\!4}\,
\pit_\mathrm{RS} \big(\lam_0, \xi_\mathrm{FS}(\rho;\rho_0,\xi_0)\big)\nonumber\\
&+ \frac{1}{c}\int_{\rho_0}^\rho d\rho'\,D(\rho,\rho')\,\hat{T}(\rho')\Big(\frac{\cosh\rho^\prime}{\cosh\rho}
\Big)^{\!4}\,\pit_\mathrm{RS}\big(\tem(\rho^\prime), \xi_\mathrm{FS}(\rho;\rho^\prime,0)\big).
\end{align}
\end{subequations}
We note that $\xi_\mathrm{FS}(\rho;\rho_\alpha,\xi_\alpha) = -1 + (1+\xi_\alpha)\big(\frac{\cosh\rho_\alpha}{\cosh\rho}\big)^2$. These integral equations are solved numerically by
 means of the method described in Refs.~\cite{Banerjee:1989by,Florkowski:2013lza,Florkowski:2013lya,Denicol:2014tha}. The temperature of the system is obtained from the energy density through the Landau matching condition, i.e., $\ene_\mathrm{RS}(\lam,\xi)=(3/\pi^2)\,\lam^4\R_{200}(\xi)=3\,\tem^4/\pi^2$.

%%%%%%%%%%%%%%%%%%%%%%%%%%%%%%%%%%%%
\section{Flow diagrams of the IS theory}
\label{sec:flow}
%%%%%%%%%%%%%%%%%%%%%%%%%%%%%%%%%%%%
In $\ds$ the expansion rate of the Gubser flow $\hat{\theta}=2\tanh\rho$ becomes a constant when $\rho\to\pm\infty$. As we shall see below this geometrical property of the velocity profile poses a challenge to find the attractors of different hydrodynamical theories. In this section we explain a method for finding the attractors based on the mathematical theory of non-autonomous systems~\cite{kloeden,tournier}. We discuss extensively the case of IS theory. 
The same method can be used in different models.  

The gist of what we are about to do in this section is to study the flow diagrams of the Gubser flow from the perspective of
nonlinear dynamical systems. The flow lines at early times {\it do} dictate the far-from-equilibrium behavior of any system governed by a set of differential equations and at late times they show what happens to the matter distribution until it evolves to a {\it steady state} (thermally non-equilibrium) at $\rho\rightarrow\infty$ \footnote{Throughout this paper the word `equilibrium' is always meant to be thermal equilibrium and a steady state expresses a state of
the system which requires energy or continual work to remain stationary and it thus does not imply thermal equilibrium.}. 

As a minor digression, we recall that for the Bjorken flow, the second order hydrodynamical equations can be entirely reduced to one single explicitly time-dependent ODE (i.e. $1d$ non-autonomous parametrized by a new time $w$) due to scaling symmetry~\cite{Heller:2014wfa,Heller:2015dha}. This suggests that its attractor is a planar 1-manifold characterized by $w$ and the basin of attraction is indeed $2d$. But what is the proper definition of attractor? {\it It is an invariant set of points and every flow line starting from a point within the field of attraction of the attracting fixed point, will always limit to this set of points at late times}. Given a set of initial values for Bjorken time, the flow lines always lie on a plane and as long as they belong to a special set of numbers, they tend to the attractor at late times. This set defines the basin of attraction: Given an initial time $\tau_0$, there is an associated $\tem(\tau_0)$ and $\bar{\pi}(\tau_0)$, and the basin of attraction is elaborated as the set of all pairs of $(\tem(\tau_0),\bar{\pi}(\tau_0))$ such that the Bjorken flow lines are doomed to limit to the equilibrium. We remind that $(\tem,\bar{\pi})$ is the actual phase space\footnote{In this section, `phase space' is
referred to as the space of independent macroscopic variables labeling a flow diagram. The latter represents the dynamics of the underlying hydrodynamical system via the connection between the velocity fields and the macroscopic variables that form the flow lines.} of the Bjorken flow which fixes the dimensionality of the basin of attraction as well.

In the case of Gubser flow, we simply identify the fixed points and discuss whether or not they are stable by analyzing the eigenvalues of the Jacobian matrix of the linearized flow equations near each fixed point. We find that there are two unstable saddle points and one stable fixed point which determines the steady state of the system in any hydro theory. This task in carried out by analyzing the nulllines of the $2d$ non-autonomous system \eqref{eq:ISGub1}-\eqref{eq:ISGub2}. The intersection of these lines gives the fixed points. By analyzing the asymptotics of flow lines close to the steady state, it is determined the fixed point is indeed exponentially asymptotically stable. One observation that we make is that the flow equations cannot be reduced to one single non-autonomous ODE by any kind of reparametrization unless at very late times such that $\rho$ is large enough that we can take $\tanh^2\rho\sim1$. But this is fine from the viewpoint of attractors since they are asymptotic or in other words limit sets. This does not affect the dimensionality of the attractor and hence Gubser flow lines still evolve into a $1$-manifold but in reality it is no longer a planar curve since the time direction is not fully decoupled from the ODE in terms of the new time $w$, which will be clarified in subsection~\ref{subsubsec:LinAtt}. What about the basin of attraction? It is indeed what we go at great length into its details in subsection \ref{subsubsec:basin} where the impossibility of this dimensional reduction to a $1d$ system is a symptom of the peculiarity of the $q$-deformed conformal symmetry $SO(3)_q$ which is not a simple scaling of variables and time as in the Bjorken case. The variable $\tau=\tanh\rho$ is promoted to be independent which lifts the phase space dimensionality by $1$, now being labeled by $(\tem,\bar{\pi},\tau)$. Therefore, we lift the dynamical system describing the flow equations to a $3d$ autonomous system where $\rho$-dependency is now implicit. So the basin of attraction is three dimensional, being identical to the dimensions of the autonomous system (or phase space).

We finally emphasize on the importance of the basin of attraction and the way to define it locally and globally. As was said before, the basin of attraction defines an effective attractive potential field for the flow lines, thus forcing them to evolve into the stable steady state at late times. This is an important problem in the far-from-equilibrium hydrodynamics to map out all the divergent flows and only probe the relevant stable ones. Similar to the path integral where a steepest descent path is one that counts as contributing to the path integral manifold, in hydrodynamics the space of all flow lines taking initial values in the basin of attraction would determine this manifold. Effectively, it is just enough to consider all the paths starting at $\rho\rightarrow -\infty$ somewhere on the boundary of basin where the attractive force field of the fixed point is the weakest. This field is determined by an effective potential function(al) known as Lyapunov function $\mathcal{V}$ that satisfies two key properties: $\mathcal{V}>0$, and $d\mathcal{V}/d\rho\le0$. The existence of $\mathcal{V}$ hints at the stability of steady state and that the linearized flow equations are asymptotically stable. The local function $\mathcal{V}_{\rm loc}$ can always be computed in the near-equilibrium region by solving Lyapunov equation given in \ref{subsubsec:basin} and we   hence solve it to estimate the shape of the local basin of attraction. The global function is hard to be precisely built but there are numerical and analytical optimization techniques that are mentioned in the same subsection. The global Lyapunov function mimics the behavior of all stable fluctuations both at early and late times which in this regard might be a very useful tool to construct an effective partition function for hydrodynamics. We will close this section by commenting on this issue.

%%%%%%%%%%%%%%%%%%%%%%%%%%%%%%%%%%%%
\subsection{From the perspective of $2d$ non-autonomous dynamical system}
\label{subsec:2d}
\begin{figure}
	\centering
	\includegraphics[width=.7\linewidth]{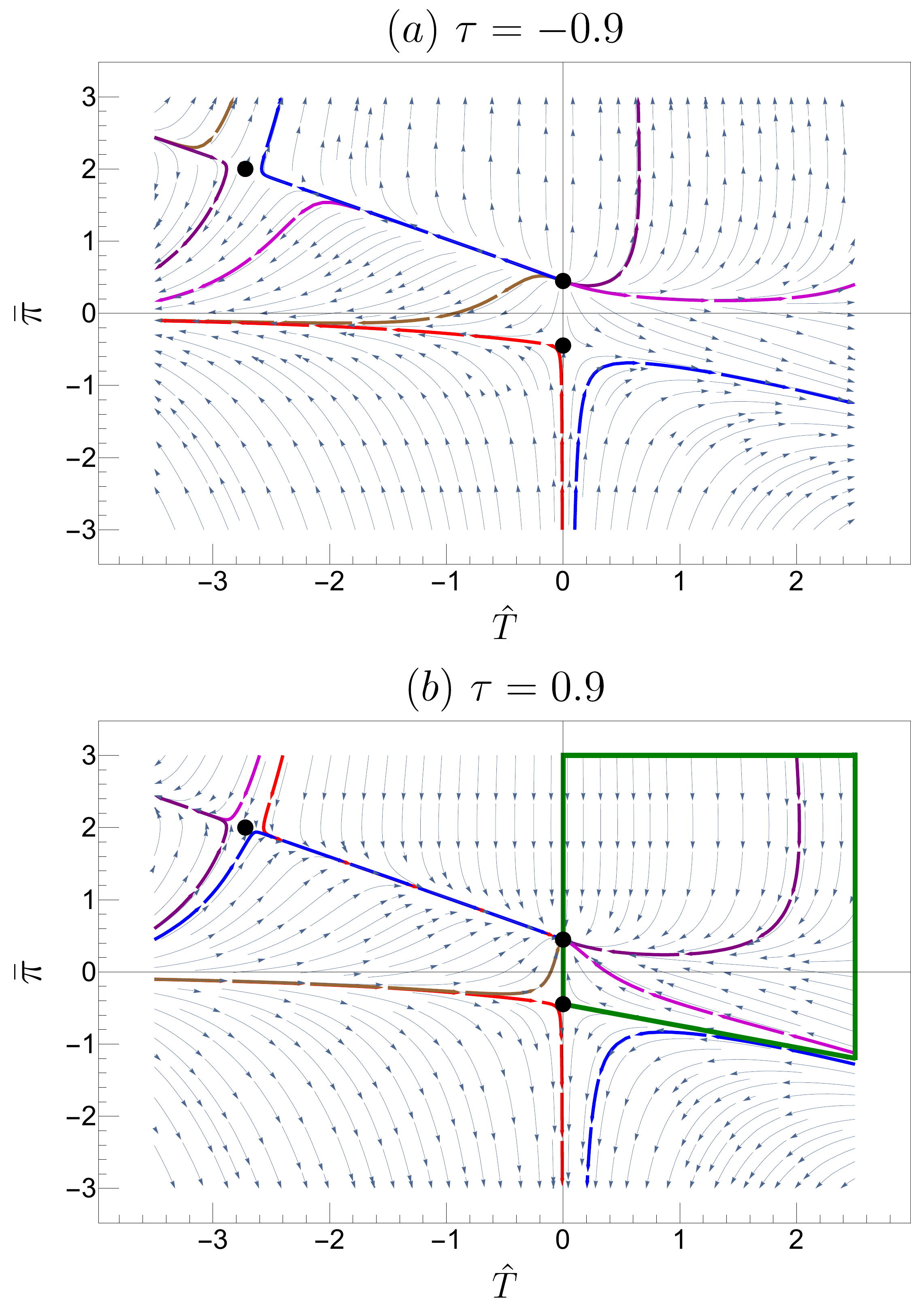}
	\caption{\scriptsize (Color online) IS flow lines of the Gubser model for (a) early-time (top panel) and (b) late-time (bottom panel). The black dots indicate the phase-space fixed points. In the panel (a) the repelling fixed point $(0,-1/\sqrt{5})$ at early times (here $\tau= -0.9$) directs the flows in the basin towards the attractor at late times. Notice that $3\mathcal{A}_s+2 = -1/\sqrt{5}$ where $\mathcal{A}_s$ defines the initial condition in Eq. \eqref{eq:initialconditions}. Physically, this is the situation where the distribution of matter is most anisotropic, thus farthest from the stable steady state. In the panel (b) the flow diagram shows that the stable fixed point \eqref{eq:fp} is approached by the flow lines at late times (here $\tau=0.9$) Note that, in terms of the new variable $\tau$ the time flows from $\tau=-1$ to $\tau=1$ and therefore an observer having a light of sight along the time direction would see the flow lines coming out of the point $(0,1/\sqrt{5})$ at early times and going in at late times, which is suggestive of
    the basin of attraction being three dimensional. 
	At late-times the flow diagram for Gubser flow in the IS theory at $\sim\tau=0.9$ shows that there are one attracting fixed point e.g. a sink, and two repelling fixed points e.g. saddle points. The area trapped between the green lines on the top right represents the physical portion of the basin of attraction at a given time slice. }
	\label{fig:flowdiagram}
\end{figure}

\begin{figure}
	\centering
	\includegraphics[width=0.7\linewidth]{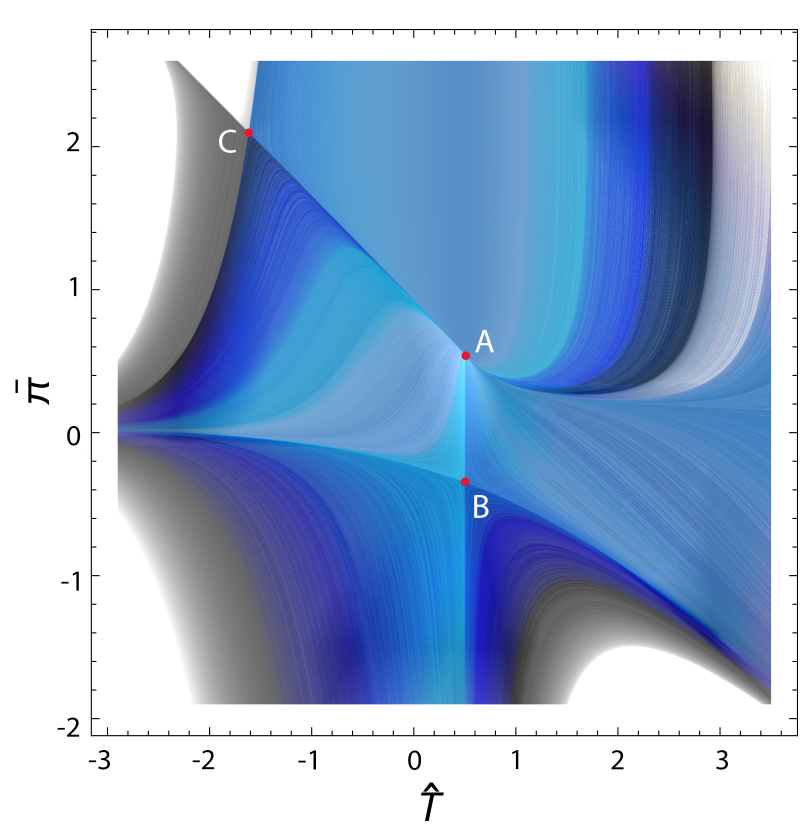}
	\caption{\scriptsize (Color online) The late-time behavior of flow lines in the IS theory of Gubser flow. The wedge near the fixed point $(0,1/\sqrt{5})$ asymptotically evolves into a $1d$ manifold that represents the behavior of Gubser attractor at late times. The other two fixed points are clearly saddle points of which the lower one in the plot feeds
		the attractor at early times. The flow line lying on the segment $BA$ has the fastest convergence among all other flows due to the repelling nature of the saddle point. The coloring of the flow lines is implemented in such a way 
		that one could perceive the depth of the $3d$ basin of attraction by looking into the flow diagram perpendicularly.}
	\label{fig:Fluiddiagram}
\end{figure}

%%%%%%%%%%%%%%%%%%%%%%%%%%%%%%%%%%%%
Using a secondary time parameter $\tau=\tanh\rho\in[-1,1]$, Eqs.~\eqref{eq:conlaw} and~\eqref{eq:ISpi}
are put into the form
\begin{subequations}
\label{eq:ISGub}
\be
\frac{d\tem}{d\tau}=\frac{\tau\tem}{3(1-\tau^2)}\left(\bar{\pi}(\tau)-2\right),\label{eq:ISGub1}
\ee
\be
\frac{d\bar\pi}{d\tau}=-\frac{1}{1-\tau^2}\left(\frac{4}{3}\,\bar{\pi}^2(\tau)\,\tau\,+\frac{1}{c}\,\bar\pi(\tau)\,{\tem}(\tau)-\frac{4}{15}\,\tau\right).
\label{eq:ISGub2}
\ee
\end{subequations}
These two equations can be combined in such a way that we are left with the following second-order nonlinear non-autonomous differential equation for the temperature $\tem(\tau)$ in the IS theory
\be
\begin{split}
135 c \tau   \left(\tau ^2-1\right)^2\tem'^2(\tau )+\tem^2(\tau ) \left[76 c\, \tau ^3-45 \tau  \left(\tau ^2-1\right) \tem'(\tau )\right]
+30\tau ^2 \tem^3(\tau )\\
-\, 15c \tem(\tau ) \left(\tau ^2-1\right) \left[\left(13 \tau ^2-3\right) \tem'(\tau )
- \frac{3 \tau  }{ \tau^2-1} \tem''(\tau )\right]=0. 
\label{eq:nulllines}
\end{split}
\ee
To find the fixed points of the system, we solve~\eqref{eq:ISGub} for the  $\bar{\pi}_N$ and $\tem_N$ along the respective nulllines (An $A$-nullline is a trajectory in the phase space along which $d A/ds=0$ where $s$ is the flow time variable.). We first consider the case where the ${\tem}$-nullline is given by the solution $\bar{\pi}_N(\tau)=2$ to Eq.~\eqref{eq:ISGub1} with vanishing temperature derivative,  and $\tem(\tau)$ is a nonzero constant.
The Gubser $\bar{\pi}$-nullline equation,
\be
\frac{4}{3}\,\bar{\pi}_N^2(\tau)\,\tau\,+\frac{1}{c}\,\bar\pi(\tau)\,{\tem}_N(\tau)-\frac{4}{15}\,\tau\,=0
\ee
then fixes the value of temperature at $\tem_N(\tau)=-38c\tau/15,$ which at large time $\tau\sim 1$ is simply $\tem_c=-38c/15$. The intersection of these nulllines are not generally reached by all flow lines since near $\tau\sim 1$ there are flow lines, for instance, that either diverge from or converge to this point. Furthermore, knowing that the system converges to a fixed point sometime at late times $\tau\gg0$, it is certainly {\it not} physical to consider the point to be reached has a negative temperature $(-38c/15)$ and thus we are led to a situation where we look for the stable steady state in the range $\bar{\pi}=\bar{\pi}_c <2$. Fig. \ref{fig:flowdiagram} indicates explicitly why this point
cannot define a stable steady state for the system.
Taking this to be the case, one may solve for temperature from \eqref{eq:ISGub1}, to only obtain
\be
\tem(\tau)=\tem_0\left({1-\tau ^2}\right)^{\frac{2-\bar{\pi}_c}{6}}, \label{eq:Tclose}
\ee
for a constant $\tem_0>0$. Note that this represents the evolution of temperature near $\tau=1$ and it is always positive for
all times. We can check that
\be
\lim_{\tau\rightarrow 1^-}(1-\tau^2) \frac{d\tem}{d\tau} =- \lim_{\tau\rightarrow 1^-} \tfrac{\tau }{6}\,(2-\bar{\pi}_c)\,\tem(\tau) =0.
\ee
Inserting this temperature into \eqref{eq:ISGub2} gives the shear along the $\bar{\pi}$-nullline toward the stable steady state,
\be
\bar{\pi}^{\pm}_N(\tau)=\frac{3}{8 c \tau } \left(\pm\sqrt{\frac{64 c^2 \tau ^2}{45}+\left({1-\tau ^2}\right)^{\frac{2-\bar{\pi}_c}{3}}}-\left({1-\tau ^2}\right)^{\frac{2-\bar{\pi}_c}{6}}\right).  \label{eq:piclose}
\ee
We also notice that $\lim_{\tau\rightarrow 1^-}(1-\tau^2) \frac{d\bar{\pi}^\pm_N}{d\tau}$ vanishes as expected.
At the limit $\tau\rightarrow 1^-$, this yields the value of $\bar{\pi}^\pm_c=\pm1/\sqrt{5}$. Similarly the ray
$\tem=0$ is a $\tem$-nullline, along which Eq. \eqref{eq:ISGub1} can be solved subject to conditions $\bar{\pi}^\pm(\infty)=\pm 1/\sqrt{5}$ to give
\be
\bar{\pi}^\pm(\tau)=\mp\frac{1}{\sqrt{5}}\pm\frac{2}{\sqrt{5} \left(e^{\frac{8}{3 \sqrt{5}}} (1-\tau^2)^{\frac{4}{3 \sqrt{5}}}+1\right)},
\ee
which demonstrates the evolution of the shear component of the trajectories near the fixed points
$(0,\pi^{\pm}_c)$. The fixed points in the IS theory of the Gubser flow are therefore determined to be
\be
{\rm Fixed \,\,points:}\quad \bar{\pi}^\pm_c = \pm\frac{1}{\sqrt{5}},\quad \tem_c=0,\quad {\rm and\,}  \quad\bar{\pi}_c = 2,\quad \tem_c=-\frac{38c}{15}. \label{eq:fp}
\ee
which is consistent with the flow diagrams plotted in Fig.~\ref{fig:Fluiddiagram}. We can also read the {\it Lyapunov exponent} \footnote{The Gubser flow as a dynamical system is deterministic in the sense that the stability of its fixed points is quantified by the eigenvalues of  the linearization matrix (so-called Jacobian matrix, cf. Eq.~\eqref{eq:Jacobianmatrixeq}) at these points. Therefore, the Lyapunov exponents are the real parts of the eigenvalues of this matrix computed along a flow trajectory that satisfies the ordinary differential equations \eqref{eq:3dproblem}. The more interesting case of deterministic {\it chaos} occurs when the maximal Lyapunov  exponent is	positive. In the current system under study, or in any known hydrodynamical system, the maximal Lyapunov
	exponent is always negative near a stable fixed point which does not give rise to chaos.} of $\tem$ from the asymptotic solution \eqref{eq:Tclose}, for both fixed points with $\bar{\pi}_c^\pm=\pm 1/\sqrt{5}$, about which temperature goes like
$\tem^\pm(\rho)\sim \tem_0\,e^{\lambda^{\pm}_{\tem}\,\rho}$ where $\lambda^\pm_{\tem} = -\tfrac{1}{3}(2\mp1/\sqrt{5})$,
thus hinting at the exponentially fast running up of the flow lines at $\rho\rightarrow\infty$ to the stable steady state and even faster convergence to the repelling point only along the attracting direction.
	 Along the shear component of the flow trajectories close to $(0,\pm1/\sqrt{5})$ the Lyapunov exponent is given by $\lambda_{\bar{\pi}}=- {8}/{(3\sqrt{5})}$, being a sign of seemingly drastic convergence  $(\sim e^{\lambda^+_{\bar{\pi}}\rho})$ to the steady state or divergence from the repeller. This provides a de facto evidence of the asymptotically exponential stability of the steady state in Gubser model which can be rigorously derived by building Lyapunov functions of the linearized system. 

We will be mainly interested in the late-time behavior of the Gubser flow near the attracting fixed point. We refer to this as ``attractor''
that points to the existence of some bounded set, $\mathcal{B}^A_e$, namely an absorbing set, in
the phase space $\mathcal{X}$ of all independent state variables $(\tem,\bar{\pi})$ and possibly time $\tau$. 
$\mathcal{B}^A_e$ is indeed invariant under the forward evolution of state equations in time such that
every solution of the system of ordinary differential equations (ODEs) has to ultimately enter $\mathcal{B}^A_e$ provided that the chosen initial conditions
for the flow trajectories put them in a bounded set $\mathcal{B}_e$ known as ``basin of attraction'' for the attracting fixed point. Notice
that by way of definition, $\mathcal{B}^A_e\subset\mathcal{B}_e$. Let $\phi^{\tau}$ be the flow defined by the Gubser dynamical system, Eqs. \eqref{eq:ISGub1}-\eqref{eq:ISGub2}, then we can define an attractor for this system by
\be
\mathcal{A} = \cap_{\tau \ge -1} \phi^\tau(\mathcal{B}^A_e),
\ee
which will be compact and invariant, and subject to the condition 
\be
\phi^{-1}\notin \mathcal{X}\setminus\mathcal{B}_e
\ee
every flow trajectory will approach this set as $\tau \rightarrow 1^-$. Since there are two other fixed points in the Gubser flow geometry, 
this condition guarantees that the flow trajectories will not start in the other basins $\subset\mathcal{X}\setminus\mathcal{B}_e$ otherwise
$\phi^{\tau>{-1}}$ would always lie in $\mathcal{X}\setminus\mathcal{B}_e$. Hence any flow line starting in the basin of attraction $\mathcal{B}_e$ independently of their whereabouts will {\it always} converge asymptotically to the attractor $\mathcal{A}$ at late times.

In general an attracting fixed point defining a basin of attraction for any (non-chaotic) model is referred to as the dynamical equilibrium state of the system and for the Gubser system this is denoted by $(\tem_e,\bar{\pi}_e)$. From a differentiable geometry point of view, in the phase space of a $2d$ autonomous dynamical system, one would expect the attractor $\mathcal{A}$ to be a manifold of co-dimension $1$. For a $2d$ non-autonomous dynamical system such as the one at hand, 
this may not be the case unless it could be reduced to a single non-autonomous ODE, by a reparametrization of state variables and $\tau$. In the case such a reduction is possible, the attractor $\mathcal{A}$ can be formally defined using, instead of the flows $\phi^\tau$, the map
$\mathcal{A}(w):\mathcal{U}\rightarrow \mathbb{R}$ for some $\mathcal{U}=[w_{\min},w_{\max}]\subseteq\mathbb{R}$, where for instance, 
$w:=w(\tem,\tau)$, and $\mathcal{A}(w)$ is an algebraic function of $\bar{\pi}$ subject to the boundary conditions
\be
\mathcal{A}(w\rightarrow w_{\max})= \mathcal{A}_e\,\,\, {\rm and}\,\,\,\mathcal{A}(w\rightarrow w_{\min})= \mathcal{A}_s \label{eq:initialconditions}
\ee
such that $(\tem_e,\mathcal{A}_e^{-1},\tau_{\max} )\in \mathcal{B}^A_e$ and $(\tem_s,\mathcal{A}_s^{-1},\tau_{\min} ) \in \partial\mathcal{B}_e$ \footnote{Two remarks are due here. First, we point out that in the Gubser flow geometry, the sink appears to be also located on the boundary of $\mathcal{B}_e$ which is related to the way the flow time is introduced
	in the state equations. We refer the reader to a different coordinate system used in Eqs.~(\ref{eq:twotime}) in which this sink is located inside the basin. Second, the saddle point is a source of propulsion and this fact plays a role in determining the rate at which the convergence of flow lines occurs.}. We keep in mind that $(\tem_s,\mathcal{A}_s^{-1},\tau_{\min} )$ is basically a saddle point in second order hydrodynamical theories in which the underlying dynamical system 
entails a term proportional to $\bar{\pi}^2$. Finally, the one dimensional manifold of $\mathcal{A}$ can be represented by
\be
\mathcal{A}\underset{w\gg w_{\min}}{ {\sim}} \mathcal{A}(w).
\ee

Let us introduce the parametrization $w=\tanh\rho/\tem$ of Gubser time and the
function $\mathcal{A}(w)$ defined as
\be
\label{eq:ffunction}
\mathcal{A}(w)=\frac{1}{\tanh\rho}\,\frac{\partial_\rho \tem}{\tem}=\frac{d\log(\tem)}{d\log(\cosh\rho)}.
\ee
Using these definitions, the evolution equations of the IS theory (Eqs.~\eqref{eq:conlaw} and~\eqref{eq:ISpi}) boil down to the following ODE
\be
3w\left(\coth^2\rho-1-\mathcal{A}(w)\right)\,\frac{d\mathcal{A}(w)}{dw} +\frac{4}{3}(3\mathcal{A}(w)+2)^2+\frac{3\mathcal{A}(w)+2}{cw}-\frac{4}{15} =0, \label{eq:ISwstate}
\ee
where $\bar{\pi}=3\mathcal{A}(w)+2$ from the conservation law~\eqref{eq:conlaw}. This reduces the dimensionality of the problem only {\it asymptotically} (e.g. $\rho\rightarrow \pm \infty$) to one as opposed to the Bjorken model where a truly one dimensional ODE was achieved via a similar trick in~\cite{Heller:2015dha}. Therefore it is expected that the attractors for the Gubser flow in all the hydrodynamical schemes are $1d$ {\it non-planar} manifolds and correspondingly the basins of attraction are three dimensional. We analyze this below briefly in the context of dynamical systems  for the IS theory but bear in mind that a similar line of thought can be applied to any other theory as well.
%
%%%%%%%%%%%%%%%%%%%%%%%%%%%%%%%%%%%%
\subsection{From the perspective of $3d$ autonomous dynamical system}
\label{subsec:3d}

In this section we study the linearization problem by enhancing the $2d$ system to a $3d$ autonomous one, determine the stability conditions, and show that the attractor is a non-planar manifold of co-dimension 2. We will sketch the proof of the exponential asymptotic stability of the steady state in Gubser flow and create a local Lyapunov function to estimate the basin of attraction that is conjectured to be crucial in the study of an effective field theory of stable hydrodynamical and non-hydrodynamical modes. 
%%%%%%%%%%%%%%%%%%%%%%%%%%%%%%%%%%%%
\subsubsection{Linearization and exponential asymptotic stability}
\label{subsubsec:LinAtt}
The non-autonomous system \eqref{eq:ISGub1}-\eqref{eq:ISGub2} can easily be extended to an autonomous system
by considering the time $\tau$ as an additional variable. Therefore the IS theory of Gubser flow is a truly $3d$ autonomous system of ODEs given by
\be
\frac{d\tem}{d\rho}  =  \frac{1}{3}\,\tem(\bar{\pi}-2)\tau,\quad\quad \frac{d\bar{\pi}}{d\rho}  =  \frac{4}{3}\left(\frac{1}{5}-\bar{\pi}^{2}\right)\tau-\frac{1}{c}\pi\tem,\quad\quad
\frac{d\tau}{d\rho}  =  1-\tau^{2}.\label{eq:3dproblem}
\ee
This is a polynomial vector field whose fixed points are given by
\be
(\tem_c,\bar{\pi}_c,\tau_c)\,\,=\,\,
(-38c/15, 2, \pm1),\quad (0 , -1/\sqrt{5}, \pm1),\quad (0, 1/{\sqrt{5}}, \pm1).
\ee
Because of the symmetry $(\tem,\bar{\pi},\tau)\rightarrow(-\tem,\bar{\pi},-\tau)$ of the $3d$ problem \eqref{eq:3dproblem}, 
we are left with only three fixed points $A=(0,1/\sqrt{5} ,1)$, $B=(0,-1/\sqrt{5} ,-1)$, and $C=(-38c/15,2, 1)$. We now 
solve the linearized system around any fixed point, namely
\be
\left(
\begin{array}{ccc}
d\tem/d\rho \\
d\bar{\pi}/d\rho\\
d\tau/d\rho \\
\end{array}
\right)
=\left(\begin{array}{ccc}
\frac{1}{3} (\bar{\pi} -2) \tau  & \frac{\tem \tau }{3} & \frac{1}{3} \tem (\bar{\pi} -2) \\
-\frac{\bar{\pi} }{c} &-\frac{\tem}{c}-\frac{8 \bar{\pi} \tau}{3} & \frac{4}{15}-\frac{4\bar{\pi}^2}{3} \\
0 & 0 & -2 \tau \\
\end{array}
\right)_{(\tem_c,\bar{\pi}_c,\tau_c)}
\left(\begin{array}{ccc}
\tem-\tem_c  \\
\bar{\pi}-\bar{\pi}_c\\
\tau-\tau_c \\
\end{array}
\right),
\label{eq:Jacobianmatrixeq}
\ee
and find the eigenvalues of the Jacobian matrix at every fixed point to be
\be
A:\left\{
-2 ,
-\tfrac{8}{3 \sqrt{5}},
-\tfrac{2}{3 } +\tfrac{1}{3\sqrt{5}}
\right\},\,\, B:\left\{
-2 ,
:\tfrac{8}{3 \sqrt{5}},
-\tfrac{2}{3 } - \tfrac{1}{3\sqrt{5}}
\right\},\,\,
C:\left\{
-2,
\tfrac{7}{5} - \tfrac{\sqrt{821}}{15},
-\tfrac{7}{5 } +  \tfrac{\sqrt{821}}{15}
\right\}.\nonumber
\ee
One can then immediately see that $A$ is a sink (all eigenvalues are negative), and $B,C$ have two positive eigenvalues, thus making them
saddle points, as expected and since the eigenvalues all are nonzero, the fixed points are hyperbolic. Hence,
we can apply the Hartman-Grobman (HG) theorem~\cite{hartman,arrowsmith}~\footnote{This mathematical theorem states the behavior of a nonlinear system of differential equations in a domain near its hyperbolic equilibrium points is qualitatively the same as the behavior of the linearized version of these differential equations near this fixed point. Hyperbolicity means that no eigenvalue of the Jacobian matrix associated with the linearized dynamical system has real part equal to zero. Therefore, one can use the linearization of the original dynamical system to analyze its behavior around equilibria.}, 
that allows the local flow structure (phase space portrait) near a
hyperbolic fixed point to be topologically equivalent to the flow diagram
of its linearized system. For the Gubser flow in the IS theory, the flow diagram shown in Fig.~\ref{fig:Fluiddiagram} confirms the HG theorem in the vicinity of all the fixed points. This figure also portrays the state of flow lines on the time slice $\tau=0.7$ of the three dimensional phase space.
Using the HG and the fact that $A$ is an exponentially asymptotically stable fixed point of Gubser flow, it is then easy to write down the Lyapunov exponents of phase space variables along all the trajectories converging to the attractor by the eigenvalues given above, that are 
\be
\lambda_{\tem}= -\tfrac{2}{3 } +\tfrac{1}{3\sqrt{5}},\quad \lambda_{\bar{\pi}}= -\tfrac{8}{3 \sqrt{5}},
\quad \lambda_{\tau}=-2.
\ee
which matches the result obtained from the $2d$ analysis previously.
The attractor may therefore be parametrized vectorially in the phase space spanned by $(\mathbf{u_1},\mathbf{u_2},\mathbf{u_3})$ as 
\be
e^{\int_{\rho_0}^\rho d\rho'\mathcal{A}(w(\rho'))\tau(\rho')}\mathbf{u_1}+ (3\mathcal{A}(w(\rho))+2)\mathbf{u_2} +\tau\rho\mathbf{u_3}, \label{eq:attapprox}
\ee
for some constant $\rho_0$. Evidently {\it only} for large $\rho$, $\tau(\rho)$ tends to one exponentially faster than do the other two converge to the fixed point i.e., $\lambda_{\tau}<\lambda_{\bar{\pi}}<\lambda_{\tem}$ such that $\mathbf{u}_3$ is just a point in the phase space and Eq.~\eqref{eq:attapprox} at $\rho\rightarrow\infty$
lies on the plane spanned by $\{\mathbf{u_1},\mathbf{u_2}\}$, and is well-approximated by
\be
 \tem_0\,e^{\lambda_{\tem}\rho}\mathbf{u_1}+(\tfrac{1}{\sqrt{5}}-\bar{\pi}_0\,e^{\lambda_{\bar{\pi}}\rho})\mathbf{u_2}+\mathbf{u_3},
\ee
given the initial conditions recovered from the asymptotics.

Since the non-autonomous system \eqref{eq:ISGub1}-\eqref{eq:ISGub2} is exponentially asymptotically autonomous for the attracting fixed point, with the limiting functions $g_i(\tem,\bar{\pi})$ being the RHS of the two equations resulted from  $\tanh^2\rho\sim1$, Theorem 2.2 of \cite{numbasin} applies where solving for the attractor of either $2d$ or $3d$ systems would yield the same result. The main assumption is to suppose a 1-to-2 transformation of 
the form $t=\pm e^{-2{\rho}}$ where for each value of $\rho$ there is a pair of values for $t$, such that
the steady state is not on the boundary of the basin of attraction, $\mathcal{B}_e$. Then ${\rho} = -\tfrac{1}{2}\log|t|$ in both cases. Also, let $F_3(\tem,\bar{\pi},t) := -2t$ where the index $3$ denotes the 
RHS of the 3rd equation in the $3d$ problem. We define for $i=1,2$ (the RHS of $i$th state equation)
\be
F_i(\tem,\bar{\pi},t) =\bigg\{ \begin{array}{cc}
f_{i}(\tem,\bar{{\pi}}+\tfrac{1}{\sqrt{5}},-\tfrac{1}{2}\log|t|) & ;\quad t\ne0\\
g_{i}(\tem,\bar{{\pi}}+\tfrac{1}{\sqrt{5}}) & ;\quad t=0. \label{eq:twotime}
\end{array}
\ee
where we have shifted the $\bar{\pi}$ for later purposes. With this new time parametrization, Eqs.
\eqref{eq:3dproblem} may be cast into the form
\be
\label{eq:3dproblemtildetau2}
\begin{split}
\frac{d\tem}{d\rho}  &=  \frac{1}{3}\,\tem\left(\bar{\pi}-2+\tfrac{1}{\sqrt{5}}\right)\frac{1-|t|}{1+|t|},\\ \frac{d\bar{\pi}}{d\rho}  &= -\frac{4\bar{\pi}}{3}\left(\tfrac{2}{\sqrt{5}}+\bar{\pi}\right) \frac{1-|t|}{1+|t|}-\frac{\tem}{c}\left(\bar{\pi}+\tfrac{1}{\sqrt{5}}\right),\\
\frac{d t}{d\rho}  &=  -2t .
\end{split}
\ee
We note that the Jacobian matrix of this $3d$ system, ${\rm Jac}(F_i)$ has a block form at $(\tem,\bar{\pi},0)$ and the origin is a fixed point, i.e. $F_i(0,0,0)=0$, and ${\rm Jac}(g_i)_{(0,0)}$ has negative eigenvalues. Since the original system is exponentially asymptotically autonomous, and $g_i(0,0) = 0$ the exponential asymptotic stability immediately follows. 

Lastly, a few remarks are due. There is a sense in which one must consider the aforementioned coordinate transformation inspired by the asymptotics of the series solutions to the flow equations \eqref{eq:ISGub1}-\eqref{eq:ISGub2} mainly the study of resurgence properties and transseries \cite{Basar:2015ava,Aniceto:2015mto}. Focusing on just a simple series solution is problematic on its own because of divergence issues but above all else lies the fact that there is a challenge to pick a good expansion variable for the Gubser flow due to the peculiarities of de Sitter time. The natural choice for such a series expansion should at first glance be either  $\tanh\rho$ or $\coth\rho$, but they come with a caveat; they never grow bigger than $1$ as $\rho\rightarrow\infty$ thus not useful from the perspective of series asymptotics. It turns out that the exponential asymptotic stability of the fixed point offers a suitable and rather {\it natural} choice of the expansion variable which will be briefly touched upon in Sec.~\ref{subsec:resurgence}. It suffices to state that the rate at which the converging flow trajectories approach the sink is known from the analysis above to be exponentially fast and therefore one can expand around this fixed point 
by considering a variable of the type $1/t$ or any arbitrary positive real power of it.

%%%%%%%%%%%%%%%%%%%%%%%%%%%%%%%%%%%%%%%%%%%%%%%%%%%%%%%%%%%%%%%%%%%%%%%%
\subsubsection{Estimate of basin of attraction locally and globally}
\label{subsubsec:basin}
%%%%%%%%%%%%%%%%%%%%%%%%%%%%%%%%%%%%%%%%%%%%%%%%%%%%%%%%%%%%%%%%%%%%%%%%%5

\noindent{\bf A path integral analogy:} Before going into any details, we seek to motivate the reader to think about the question of why the basin of attraction and Lyapunov functions are extremely important concepts from a physical standpoint. 

In the Feynman path integral formalism of quantum field theory and quantum mechanics, one often encounters integrals of the sort
\be
Z=\int_{M} D\phi\, e^{-S[\phi]} \label{eq:path_int}
\ee
where $S[\phi]$ is the action functional of the underlying theory that involves some field $\phi$ and the manifold $M$ defines the space of fields or {\it paths} over which the integral is performed. For keeping the generality of the problem, we complexify $\phi$ and thus $S[\phi]$ is complex-valued and $M$ is a middle-dimensional manifold in the space of complex fields. We will come to the dynamical system interpretation of this integral in a moment. But before, we have to note that the main burden is to understand what $M$ is made of and how to find it in a general field theory is extremely difficult. 
Qualitatively, $M$ is considered to be built out of the union of all the paths each connected to a critical point (stable fixed point of $S[\phi]$ that satisfies some equation of motion subject to stationary action principle) which are the physically relevant paths (steepest descent paths) in the phase space or configuration space over which the field $\phi$ takes values. So the most favored configuration is the one along which ${\rm Re}(S[\phi])$ increases as one approaches the critical point attached to it by some dynamical flow involving the action functional and an appropriate flow time $t_f$ defined over the real line, for instance. This is described by an ODE of the form \footnote{We note that even if the $\phi$ is complexified,
	the degrees of the freedom of the complex theory is same as the original one on the manifold $M$,
	namely it is a middle-dimensional manifold in the complex space as dictated by the form of flow equation \eqref{eq:floweq}.}
\be
\frac{d\phi}{d t_f} = -\overline{\frac{\delta S[\phi]}{\delta \phi}},\label{eq:floweq}
\ee 
where the RHS is nothing but the variation of action functional with respect to the dynamical field $\phi$ which now depends on $t_f$ and 
the bar represents complex conjugation. The set of fixed points of this `flow equation' now describes the saddle solutions to some equation of motion governed by the action principle $\delta S[\phi]/{\delta \phi}=0$.
This simple-looking flow equation resembles the more complicated dynamical system given by \eqref{eq:ISGub1}-\eqref{eq:ISGub2} but the idea is that 
all the flow lines along which ${\rm Re}(S[\phi])\rightarrow \infty$ contribute to the construction of $M$ \footnote{Morse theory and its complex generalization, also called Picard-Lefschetz theory are recent attempts toward understating the means of building this $M$ at least in calculable cases where it is indeed of finite dimensions such as quantum mechanics and certain quantum field theories with {\it nice} properties \cite{Witten:2010cx,Behtash:2015loa,Behtash:2017rqj,Dunne:2015eaa}.} where the stability condition is guaranteed in \eqref{eq:floweq} as $t_f\rightarrow \infty$ since then a fixed point is reached
   where $\phi$ remains constant for all times thereafter. So $M$ is a stable manifold of integration shaped by the solutions to 
  \eqref{eq:floweq} that by construction is the manifold of steepest descent paths. We note that the initial conditions to solve this equation is picked by the convergence properties of the \eqref{eq:path_int}
  at $t_f\rightarrow -\infty$. Namely, $\phi=\phi_0$ if ${\rm Re}(S[\phi_0])\rightarrow \infty$ at this limit, otherwise the path integral would be divergent. The collective space of such initial values is called a ``good space" denoted by $G_c$. Formally speaking, $G_c=\cup_i G_i$ where each good subspace $G_i$ the real part of action functional remains always positive. Hence the flow must begin from the points in $G_c$ where the paths can converge to the critical points along any direction in $M$. In a hydrodynamical system, we can then regard $M$ as a ``Lefschetz thimble'' made out of all the flow lines attached to the attracting fixed point for the field $\phi$ in the path integral \eqref{eq:path_int} starting at time $\rho\rightarrow-\infty$ and the collective space $G_c$ as the boundary of the basin of attraction. In other words,
  \be
   \label{eq:manifolddef}
  \begin{split}
  G_c &:= \{\cup_i\phi_{i,0}:\phi_{i,0}=\phi_i(\rho\rightarrow -\infty)\in \partial \mathcal{B}_e\},\\
   M&:= \{{\cup}_i\phi_i(\rho):\phi_i(\rho\rightarrow -\infty)\in G_c,\,\phi_i(\rho\rightarrow \infty) ={\rm fixed\,\, point}\in \mathcal{B}_e\}.
  \end{split}
  \ee 
There is in principle no universally well-established picture where we have an effective field theory approach to hydrodynamics, we rather choose to go to the phase space of state variables and flow time to make our analogies with the picture given above. But what about the action functional in the phase space? It is kind of obvious that the Lyapunov function(al) $\mathcal{V}$ that depends on the phase space parameters of the underlying hydrodynamical system acts like an effective action for all the stable thus {\it relevant} hydro modes. Once defined, $\mathcal{V}$ is always positive definite, and more importantly its derivative with respect to the flow time has to be always {\it non-positive} i.e. \be
\frac{d\mathcal{V}_{\rm}}{d\rho}\le 0,
\ee
 similarly to the negative of real part of action functional decreasing on $M$ due to stability conditions set by the properties of \eqref{eq:floweq} \footnote{\label{foot12} One can think of the effective action as a {\it Morse-Bott} functional, which satisfies
 	$d{\rm Re}(-S)/d\rho \le 0$.}. Given a global Lyapunov function, we therefore can write an effective action functional for the Gubser flow in any theory  \footnote{
	For a similar discussion on the relation between the effective potentials and Lyapunov potential functions based on gradient flow equations, check \cite{tirapegui2000instabilities}.}
\be
S_{\rm eff}(\mathbf{x},c):=\int d\rho \left[\left(\frac{d\mathbf{x}}{d\rho}\right)^2 -\mathcal{V}(\mathbf{x},c)\right],
\ee
where $\mathbf{x}(\rho) = (\tem,\bar{\pi},t)$, and $c$ is a `coupling constant'. Note that this effectively describes the underlying hydrodynamical system as a theory of phase space variables $\bar{\pi},\tem,t$ taking values in $\mathcal{B}_e$ - the basin of attraction for the stable fixed point. Finally the partition function can be formulated as
\be
Z_{\rm eff}(c)=\int_{M} D\tem D\bar{\pi}Dt \exp\left[-\int d\rho \,\left(\left(\frac{d\mathbf{x}}{d\rho}\right)^2 -\mathcal{V}(\mathbf{x},c)\right)\right].
\ee
where $M$ was defined in \eqref{eq:manifolddef}. This can be generalized to any theory of hydrodynamics which probes far-equilibrium aspects 
as well. In what follows, we create a local Lyapunov function and review a few techniques of obtaining a global one that practically speaking is the ultimate goal of taking this approach. In an upcoming work, we explore this effective partition function and to what extent it captures the right
properties of the original (relativistic) hydrodynamics.

\noindent{\bf Local Lyapunov function:} 
We showed in the previous subsection that the dynamical ODEs in \eqref{eq:3dproblemtildetau2} limit to a $2d$ autonomous system close to $t=0$,
\be
\frac{d\tem}{d\rho}  =  \frac{1}{3}\,\tem\left(\bar{\pi}-2+\tfrac{1}{\sqrt{5}}\right),\quad\quad \frac{d\bar{\pi}}{d\rho}  = -\frac{4\bar{\pi}}{3}\left(\tfrac{2}{\sqrt{5}}+\bar{\pi}\right)-\frac{\tem}{c}\left(\bar{\pi}+\tfrac{1}{\sqrt{5}}\right),\label{eq:2dproblem}
\ee
where $(0,0)$ is an exponentially asymptotically stable fixed point similarly to $(0,0,0)$ being the same for
the $3d$ autonomous system.

A local Lyapunov function $\mathcal{V}_{\rm loc}(\tem,\bar{\pi},t)$ can estimate the basin of attraction near the 
fixed point $(0,0,0)$ of \eqref{eq:3dproblemtildetau2} that contains the origin inside its basin. Like the global Lyapunov function,
$\mathcal{V}_{\rm loc}(\tem,\bar{\pi},t)$ is a continuous positive-definite scalar function $\mathcal{V}_{\rm loc}(\tem,\bar{\pi},t)$ defined on the set $D=\{|t|>0, \tem \in \mathbb{R}, \bar{\pi} \in \mathbb{R}\}$. $\mathcal{V}_{\rm loc}$ has continuous first-order partial derivatives at every point of $D$. Here we first focus on a local construction of the basin of attraction, where a local Lyapunov function $\mathcal{V}_{\rm loc}=\mathbf{x}^T P \mathbf{x}$ should exist such that $\mathbf{x}^T=(\tem,\bar{\pi},t)$ and $P$ satisfies the Lyapunov equation \cite{giesl}
\be
\begin{split}
&{\rm Jac}(F_i)_{(0,0,0)}^T\, P + P\,{\rm Jac}(F_i)_{(0,0,0)} = -\text{diag.}(1,1,1),\\
 &{\rm Jac}(F_i)_{(0,0,0)} = \left(
\begin{array}{ccc}
	\frac{1}{3} \left(1/\sqrt{5}-2\right) & 0 & 0 \\
	-1/(\sqrt{5} c) & -8/(3 \sqrt{5}) & 0 \\
	0 & 0 & -2 \\
\end{array}
\right).
\end{split}
\ee
Solving this equation for $P$ and inserting the resulting matrix in the formula for $\mathcal{V}_{\rm loc}$, we obtain 
\be\label{eq:Lyapunov}
\begin{split}
\mathcal{V}_{\rm loc}(\tem,\bar{\pi},t)&=\frac{1}{8816 c^2}\left[29 c^2 \left(57 \sqrt{5} \bar{\pi}^2+76 t^2+24 \left(10+\sqrt{5}\right)\tem^2\right)\right.\\
&\left.+\,342 \left(10-7 \sqrt{5}\right) c \bar{\pi}\tem +27 \left(60-13 \sqrt{5}\right) \tem^2\right]. 
\end{split}
\ee
As discussed before, along every flow line in the basin of attraction, $\mathcal{V}_{\rm loc}$ should effectively decrease with time, therefore 
\be
\begin{split}
\frac{d\mathcal{V}_{\rm loc}(\tem,\bar{\pi},t)}{d\rho}=\frac{1}{2204 c^2 (| t| +1)}\bigg[-58 c^2 \big(38 t^2 (| t| +1)+(| t| -1) \big(2 ((10+\sqrt{5}) \bar{\pi} -19) \tem^2 -\\-~
19 \bar{\pi}^2 (\sqrt{5} \bar{\pi} +2)\big)\big)
-57 c \bar{\pi}  \tem ((5 (5 \sqrt{5}-3) \bar{\pi} +29) | t| +(15+4 \sqrt{5}) \bar{\pi} )
+\\+~9 \tem^2 (((73 \sqrt{5}-125) \bar{\pi} -38 \sqrt{5}+133) | t| +(60 \sqrt{5}-65) \bar{\pi} )
\bigg]<0. \nonumber
\end{split}
\ee
This provides the main restriction on the shape of the basin.

\begin{figure}
\centering
\includegraphics[width=0.8\linewidth]{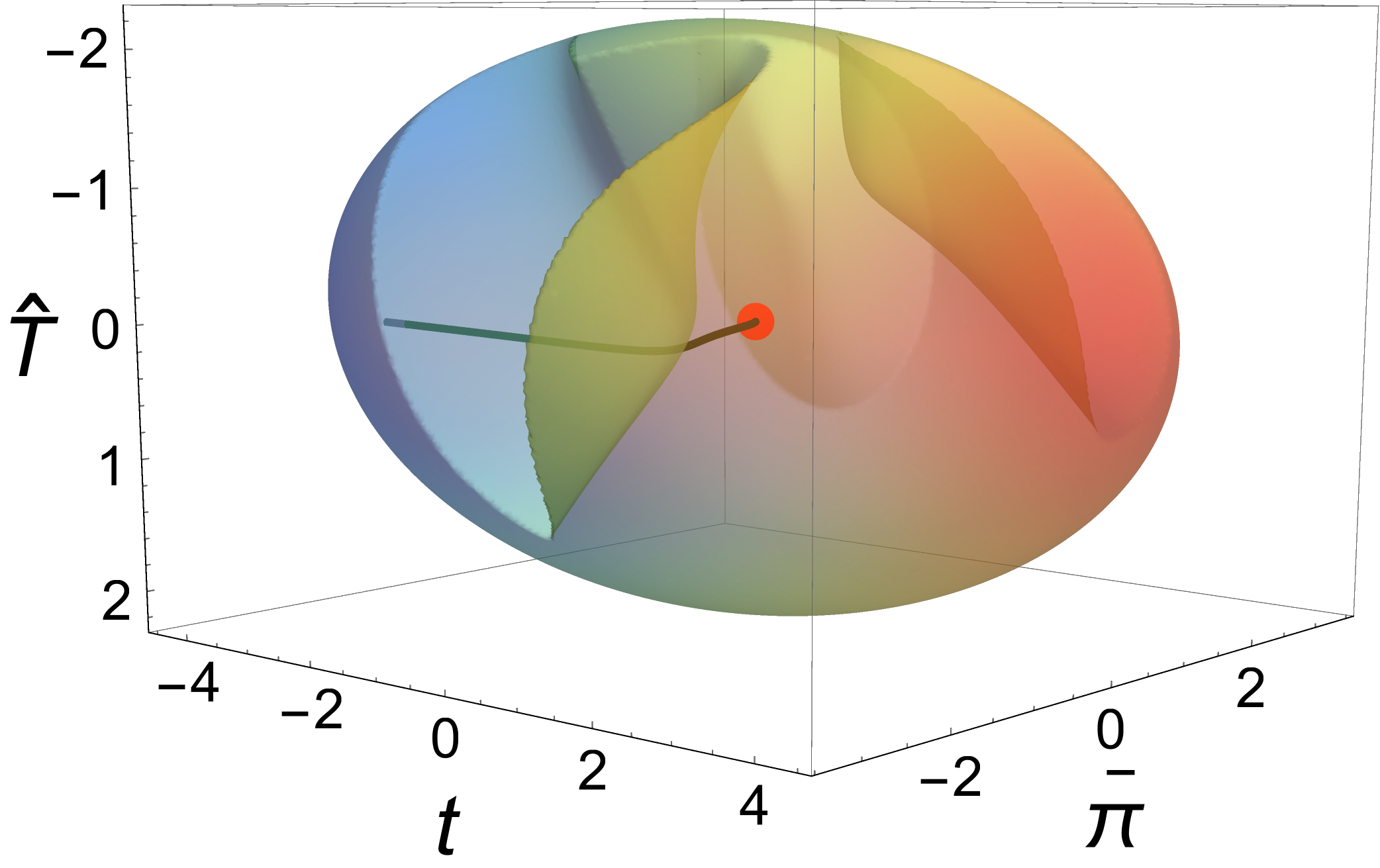}
\caption{\scriptsize (Color online) The local basin of attraction for the $3d$ system \eqref{eq:3dproblemtildetau2} for the Gubser flow in the IS theory using the Lyapunov function \eqref{eq:Lyapunov}. The boundary of the basin is topologically isomorphic to a deformed two-sphere which is locally a $\rho={\rm const}$ hypersurface. For better visibility, the basin shape and size are set by the condition $-100<{d\mathcal{V}_{\rm loc}}/{d\rho}\le 0$ and $\mathcal{V}_{\rm loc}<5$ respectively,
	which can be set, for instance, based on a near-equilibrium energy scale of the hydrodynamical modes. In the new coordinate system the origin (red dot) is the stable steady state  that corresponds to the point $(0,1/\sqrt{5},1)$ in the original coordinate system and, locally speaking, all the initial conditions chosen from the inside of this basin will yield converging trajectories towards the origin at late times. We remark that the heart-shaped gap mimics the asymptotics of boundary flow lines converging to the stable fixed point from below on the boundary of the basin 
	of attraction in Fig.~\ref{fig:flowdiagram}. The black line represents the IS asymptotic attractor solved numerically.}
\label{fig:basin}
\end{figure}

We note that the function $\mathcal{V}_{\rm loc}$ defines a symmetric basin of attraction under $t\rightarrow-t$ which is consistent with Proposition 2.9 of \cite{numbasin}. The two-time coordinate patch $(\tem,\bar{\pi},\pm e^{-\rho})$ does in fact tell us that from the eyes of an observer sitting at the origin, there is a mirror symmetric copy of the attractor, say $A_R$, with respect to the $t=0$ plane to which all the flow lines starting at $t\gg 0$ will be seen to converge as well. This is nothing new other than a symmetry of the new coordinate system. The left copy $A_L$ is depicted in Fig. \ref{fig:basin} along with the basin of attraction corresponding to the Lyapunov function in \eqref{eq:Lyapunov}.
Finding a Lyapunov function that captures the global basin of attraction for non-autonomous systems of higher dimensions ($>1$) is in general a hard problem and in most cases an analytic result cannot be obtained unless the dynamical system entails some nice properties due to hidden symmetries that could give rise to further simplifications.

Sum of squares (SOS) polynomials is another recent method developed based on global optimization for many applications including Lyapunov stability analysis and control theory \cite{chesi2009homogeneous}. This method attempts at seeking a sum of the form
\be
f(\mathbf{x}=x_1,\dots,x_n)=\sum_{i=0}^N g^2_i(\mathbf{x})\ge0  \label{eq:SOS}
\ee
solving an $n$ dimensional optimization problem for some positive integer $N$ where $g_i(\mathbf{x})$
are all polynomials. Here all $x_i$ depend on a time parameter $\rho$. If there is a $\lambda\in \mathbb{R}$ that allows a quadratic expansion of $f(\mathbf{x}) =\mathbf{x}^T P(\lambda) \mathbf{x}$ in terms of monomials for some $n\times n$ matrix $P(\lambda)$ such that $P$ factorizes as $P(\lambda)=QQ(\lambda)^T$ for a rank-$N$ matrix $Q$, then $f$ is said to have an SOS decomposition of the form \eqref{eq:SOS}. For $f$ to be a Lyapunov function $\mathcal{V}$, one more condition comes from the fact that $-d\mathcal{V}/d\rho$ has to be SOS as well. 

\begin{figure}[t]
	\centering   
	\includegraphics[scale=0.4]{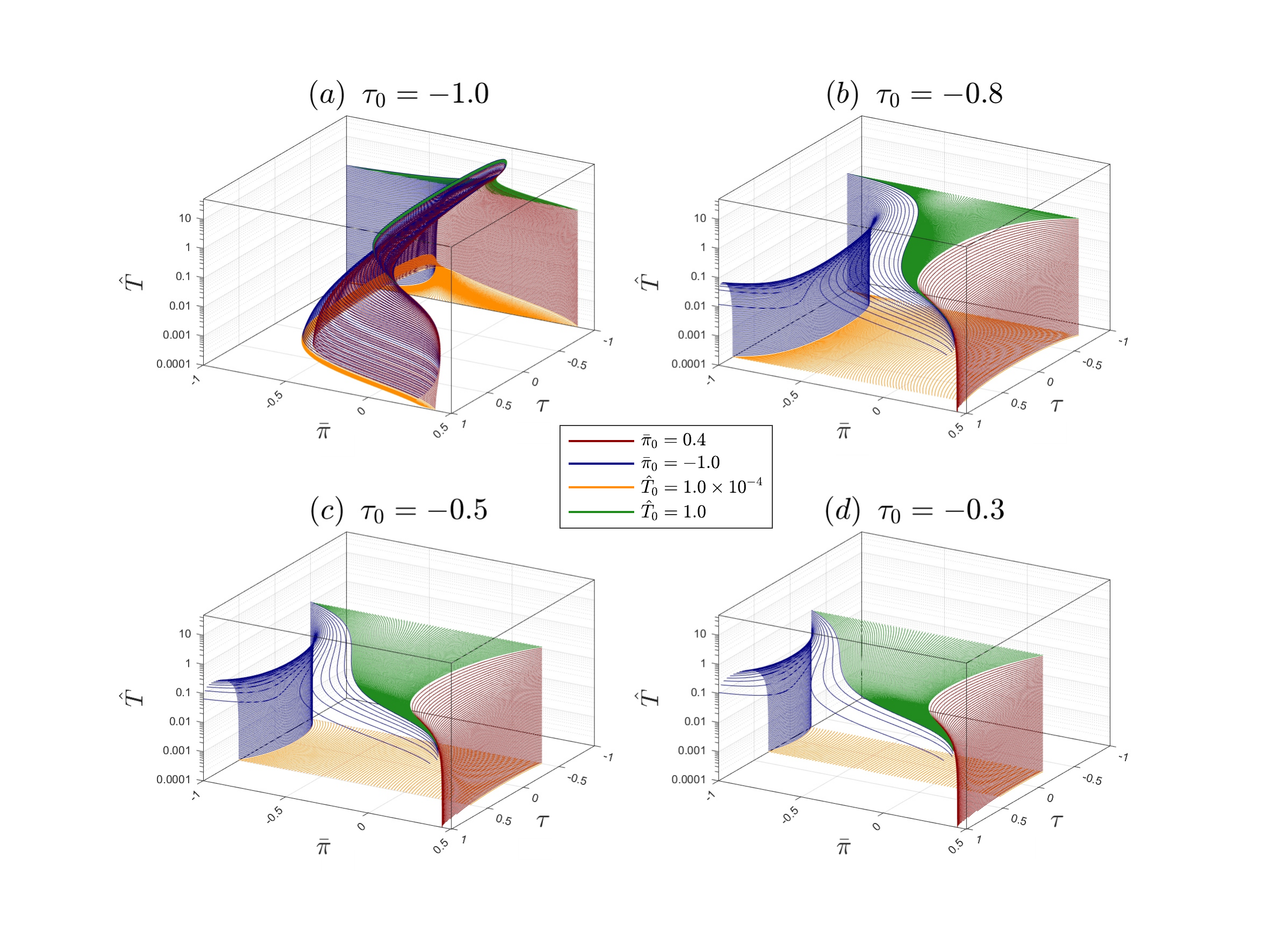}  
	\caption{\scriptsize (Color online) Basin of attraction for different initial times $\tau$. We used as initial conditions a rectangular boundary in variables $(\hat{T},\bar{\pi})=\left[1\times10^{-4},\,1\right]\times\left[-1.0,\,0.4\right]$ then chose an initial time, and let the $3d$ system evolve as \eqref{eq:3dproblem}. (a) In this plot the flow begins on the boundary of the basin of attraction at $\tau=-1$, the surfaces made out of all the flow lines merge onto the attractor before finishing up at the stable fixed point. In the plots (b), (c) and (d), there are two different behaviors observed for the initial conditions; one for when the flows start inside the basin of attraction, in which case the system evolves toward the stable fixed point asymptotically, and the other when the initial condition trigger the flows outside the basin of attraction, in which case the flow lines diverge. The separating blue flow lines determine a portion of the boundary of the basin of attraction. }
	\label{fig:basinnumerical}
\end{figure}

Another method that is similar in spirit to SOS programming calls for a solution of the PDE \cite{numbasin,numbasin2}
\be
\mathcal{V}'(t, \mathbf{x}) = -t^2-||\mathbf{x}||^2, \label{eq:globalLya}
\ee
where $||..||$ shows the usual distance from the origin assuming that it defines a stable fixed point for the dynamical system. The differentiation is implemented with respect to a flow time $\rho$. It has been comprehensively studied in literature that in various (non)linear dynamical systems, global Lyapunov functions can be well approximated by solving Eq. \eqref{eq:globalLya}
using radial basis functions, see for instance \cite{giesl}. We leave this to a future work and rather show a numerical plot of a $2d$ surfaces in the basin of attraction for different initial 
values of $\tau$ in Fig.~\ref{fig:basinnumerical}. 
Let us summarize the key take-home lessons learned in this section:
\begin{enumerate}
	\item For any value of flow time, as long as the initial values initiate 
	the flows in the basin of attraction, the flow lines will always 
	go toward the attractor;
\item The attractor is an invariant set of numbers toward which the flow lines
 evolve inside the basin of attraction at late times. If there are no repelling fixed points
around, with an underlying regular geometry, the rate of convergence to the attractor for all the flows should be uniform;
\item The flow line that begins at the saddle point located on the boundary of the basin,
goes to the attractor at the fastest possible rate among all other flow lines because it gets propelled by the repulsive force of this point. For a flow starting from $(\tem(\rho_0),\bar{\pi}(\rho_0),\tau(\rho_0))$, this rate can be quantified by
\be
r:=\int_{\rho_0}^{\infty}\sqrt{(\partial_\rho\tem)^2+\left(\partial_\rho\bar{\pi}\right)^2+\left(\partial_\rho\tau\right)^2}\,d\rho,
\ee 
which is literally the length of the curve of velocity vector field or flow line.
\item Out of all the flows starting at the same time-slice, say $\tau=-1$, the ones closer to the saddle points around, will have more repelling force and thus a faster convergence;
\item Less often it is mentioned that the role of basin of attraction is important in determining
the (multi)stability and strength of the attractors as well as the usefulness of the dynamical systems in consideration even in the regimes far from (thermal) equilibrium. So knowing the basin of attraction, its topology and size, is {\it necessary} for the study of hydrodynamical flows and the search for attractors per se in a dynamical system with a stable fixed point and some unstable fixed points would not be illuminating. Things evolve toward stability, fast or slow as long as they are initiated with values in the basin of attraction;
\item The attractor line has the fastest rate of convergence due to the propulsion of saddle point it starts from and therefore {\it naturally} there is always a fast convergence of modes nearby even in the far-from-equilibrium regime of hydrodynamics;
\item Any effective theory of hydrodynamics can be written as a simple kinetic term with a
Lyapunov potential functional that describes the relevant physical modes that only belong to the basin of attraction. In the effective field theory language, anything outside of the basin of attraction is completely integrated out.
\end{enumerate}

%%%%%%%%%%%%%%%%%%%%%%%%%%%%%%%%%%%%
\section{Universal asymptotic attractors for different dynamical models}
\label{sec:attractors}
%%%%%%%%%%%%%%%%%%%%%%%%%%%%%%%%%%%%
The relaxation equations of the different components of the energy-momentum tensor derived from a particular microscopic theory do not lead necessarily to the same attractor. In this section we present the numerical results of  calculating the attractors for different hydrodynamical models and exact solution of the RTA Boltzmann equation. 

In order to determine the attractors we follow closely the procedure outlined in the work of Heller and Spalinski~\cite{Heller:2015dha} which is based on the slow roll-down approximation~\cite{Liddle:1994dx}. The reader should bear in mind that  the attractors calculated in this section are asymptotical since the basin of attraction is three dimensional as we argue in the previous section. 

The equations of IS, DNMR and aHydro can be combined into an unique ODE for the function $\mathcal{A}(w)$~\eqref{eq:ffunction} which reads as
\be
\label{eq:genericfw}
3w\left(\coth^2\rho-1-\mathcal{A}(w)\right)\,\frac{d\mathcal{A}(w)}{dw} +H(\mathcal{A}(w),w) =0\,,
\ee
where the functional form of $H(\mathcal{A}(w),w)$ depends on the hydrodynamical model under consideration. For the hydrodynamical schemes studied in this work $H(\mathcal{A}(w),w)$ takes three functional forms given by respectively
\bs
\label{eq:hfunchydro}
\beal
\label{eq:hfuncIS}
\text{IS:}&\hspace{1cm}H(\mathcal{A}(w),w)=\tfrac{4}{3}\left(3\mathcal{A}(w)+2\right)^2+\tfrac{3\mathcal{A}(w)+2}{cw}-\tfrac{4}{15}\,,\\
\label{eq:hfuncDNMR}
\text{DNMR:}&\hspace{1cm}H(\mathcal{A}(w),w)=\tfrac{4}{3}\left(3\mathcal{A}(w)+2\right)^2+\left(3\mathcal{A}(w)+2\right)\left[\tfrac{1}{c\,w}-\tfrac{10}{7}\right]-\tfrac{4}{15}\,,\\
\label{eq:hfuncahydro}
\text{aHydro:}&\hspace{1cm}H(\mathcal{A}(w),w)=\tfrac{4}{3}\left(3\mathcal{A}(w)+2\right)^2+\left(3\mathcal{A}(w)+2\right)\left[\tfrac{1}{c\,w}-\tfrac{4}{3}\right]\nonumber\\
&\hspace{3.14cm}-\tfrac{5}{12}+\tfrac{3}{4}\mathcal{F}\left(3\mathcal{A}(w)+2\right)\,.
\end{align}
\es
These expressions were determined by using the conservation law~\eqref{eq:conlaw} together with Eqs~\eqref{eq:ISpi} and~\eqref{eq:DNMRpi} for IS and DNMR respectively while for anisotropic hydro we consider Eqs.~\eqref{eq:anisconslaw} and~\eqref{eq:RSpi}. We point out that for aHydro it is necessary to rewrite the function $\mathcal{F}(\bar{\pi})$ ~\eqref{eq:F} in terms of $\mathcal{A}(w)$. 

When applying the slow roll down approximation $d\mathcal{A}/dw=0$ in Eq.~\eqref{eq:genericfw} one needs to find the roots of $H(\mathcal{A}(w),w)\equiv 0$. We take the late time or asymptotic limit $\tanh^2\rho\sim1$ which was explained before and it poses no problem in the universality of the tor due to exponentially fast convergence of flow lines to the attractor at late times.  In the case of IS and DNMR this constraint gives two different solutions so one chooses only the stable one $\mathcal{A}_+(w)$~\cite{Heller:2015dha,Basar:2015ava}. For these theories we get
\bs
\label{eq:fminus}
\beal
\text{IS:}&\hspace{.7cm}\mathcal{A}_+(w)=-\tfrac{1}{24}\left[\left(16+\tfrac{3}{c\,w}\right)+\sqrt{\left(16+\tfrac{3}{c\,w}\right)^{2}-48\left(\tfrac{76}{15}+\tfrac{2}{c\,w}\right)}\right],\\
\text{DNMR:}&\hspace{.7cm}\mathcal{A}_+(w)=-\tfrac{1}{24}\left[\left(\tfrac{82}{7}+\tfrac{3}{c\,w}\right)+\sqrt{\left(\tfrac{82}{7}+\tfrac{3}{c\,w}\right)^{2}-48\left(\tfrac{232}{105}+\tfrac{2}{c\,w}\right)}\right]\,.
\end{align}
\es
The initial condition for solving the differential equation~\eqref{eq:genericfw} for IS and DNMR is obtained by evaluating $\lim_{w\to-\infty}\mathcal{A}_+(w)\equiv \mathcal{A}_{i}$~\footnote{Alternatively, one can also predict the value of the initial condition for $\mathcal{A}_+({w\to-\infty})$ from the method discussed in Sec.~\ref{sec:flow}. In this case one equates the value of $\bar{\pi}({\rho\to-\infty})$ at the stable fixed point into the conservation law~\eqref{eq:conlaw}. For instance, for $c=15/(4\pi)$ the values of $\bar{\pi}({\rho\to-\infty})$ at the stable fixed points for IS and DNMR are $1/\sqrt{5}$ and $15/28-1/28 \sqrt(1909/5)$, respectively, so  $\mathcal{A}_+({\rho\to-\infty})$ in turn takes the approximate values $-0.8157$ and $-0.7207$ for the former and the latter hydro model. These numbers coincide exactly with the $w\to-\infty$ limit of Eqs.~\eqref{eq:fminus}.}. Afterwards, as mentioned before, one simply solves Eq.~\eqref{eq:genericfw} while taking $\rho\to-\infty$ (and thus $\coth^2\rho-1\to0$). For the case of aHydro the initial condition is determined by finding numerically the roots of $H(\mathcal{A}(w),w)$~\eqref{eq:hfuncahydro}  which gives $\mathcal{A}_i\approx-0.75$. In the next subsection we discuss the numerical solutions of Eq.~\eqref{eq:genericfw} for each approximation scheme.

%%%%%%%%%%%%%%%%%%%%%%%%%%%%%%%%%%%%%%%%%%%%%%%%%%%%%%%%%
\subsection{Numerical results}
\label{subsec:num}
%%%%%%%%%%%%%%%%%%%%%%%%%%%%%%%%%%%%%%%%%%%%%%%%%%%%%%%%%%%%
%
\begin{figure}
	\centering
\hspace{-.7cm}	\centerline{\includegraphics[scale=.74]{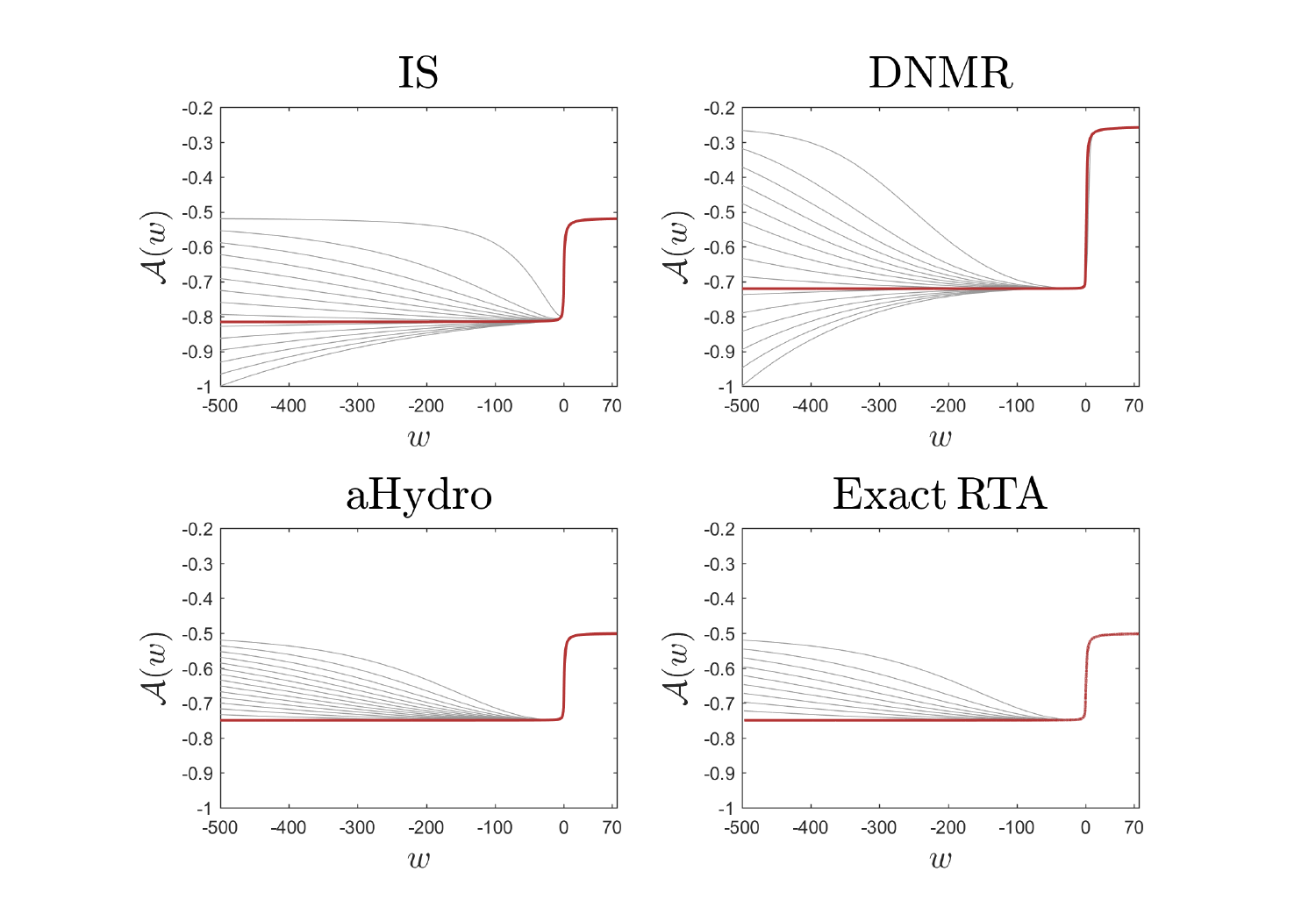}}
	\caption{\scriptsize (Color online) Attractors for different theories (solid red line) and numerical solutions for a large set of initial conditions (solid gray lines). The (steady state) non-equilibrium attractors shown in this figure correspond to IS (top left panel), DNMR (top right panel), aHydro (bottom left panel) and the exact solution of the Gubser flow (bottom right panel). In each case we use $c=15/(4\pi)$.}
	\label{fig:attractor1}
\end{figure}
We are now ready to discuss the numerical solutions of the attractors (late-time asymptotic behavior) for each truncation scheme together with the exact attractor of the Gubser solution~\eqref{eq:exsol}. The exact location of the attractor was found by following the technique explained in Ref.~\cite{Romatschke:2017vte}. For completeness we briefly explain it in App.~\ref{app:attrexact}. When solving numerically we used $\rho_0=-10$, which is good enough for our purposes because this value avoids unphysical behaviour e.g. negative temperatures~\cite{Denicol:2014tha,Heinz:2015cda}. 

In Fig.~\ref{fig:attractor1} the asymptotic attractors (late-time tail of solid red line) of the IS (top left panel), DNMR (top right panel), anisotropic hydro (bottom left panel) and the exact solution of the Gubser flow (bottom right panel) for a variety of initial conditions (gray lines) are shown. In each plot we chose $c=5\eta/s$ with $\eta/s=3/(4\pi)$. The set of initial conditions were chosen by allowing the initial condition of the effective shear to be either prolate ($\bar{\pi}<0$ and thus $\Pl<\Pt$), or oblate ($\bar{\pi}>0$ and thus $\Pl>\Pt$). The former configuration corresponds to $\mathcal{A}_i<-2/3$ while the latter indicates $\mathcal{A}_i>-2/3$. The initial values $\mathcal{A}_i$'s of the different asymptotic attractors (late-time tail of solid red line) in Fig.~\ref{fig:attractor1} are always below  $\mathcal{A}_i<-2/3$ so that the steady state attractor of fluids undergoing Gubser flow corresponds to a prolate configuration. This result is valid independently of the hydrodynamical model and in agreement with the methods discussed in Sec.~\ref{sec:flow} since the saddle fix points are located precisely when $\bar{\pi}\bigl|_{\rho\to-\infty}<0$. We also notice that independently of the theory, prolate configurations merge faster to their corresponding attractors than the oblate ones. Furthermore, the positive definite condition of the transverse and longitudinal pressures implies that $-3/4<\mathcal{A}(w)<-1/2$. The positivity of the pressures is satisfied only by aHydro and the exact Gubser solution and thus one cannot initialize $\mathcal{A}(w)$ below these values. This statement was tested numerically and explains why there are no gray lines below the early time line of attractor in the bottom panels of Fig.~\ref{fig:attractor1}. However, for the IS and DNMR evolution equations this condition can be slightly broken by initializing $\mathcal{A}(w)$ below the attractor as it is shown in the top panels of Fig.~\ref{fig:attractor1}. This indicates that the basin of attraction of IS and DNMR theories is larger than aHydro and the exact solution while being physically invalid. If one restricts IS and DNMR to satisfy the positivity condition of the pressures, it would physically smallen the phase space of initial conditions and thus, the basin of attraction~\cite{Martinez:2009mf}.
\begin{figure}
	\centering
	\includegraphics[width=.9\linewidth]{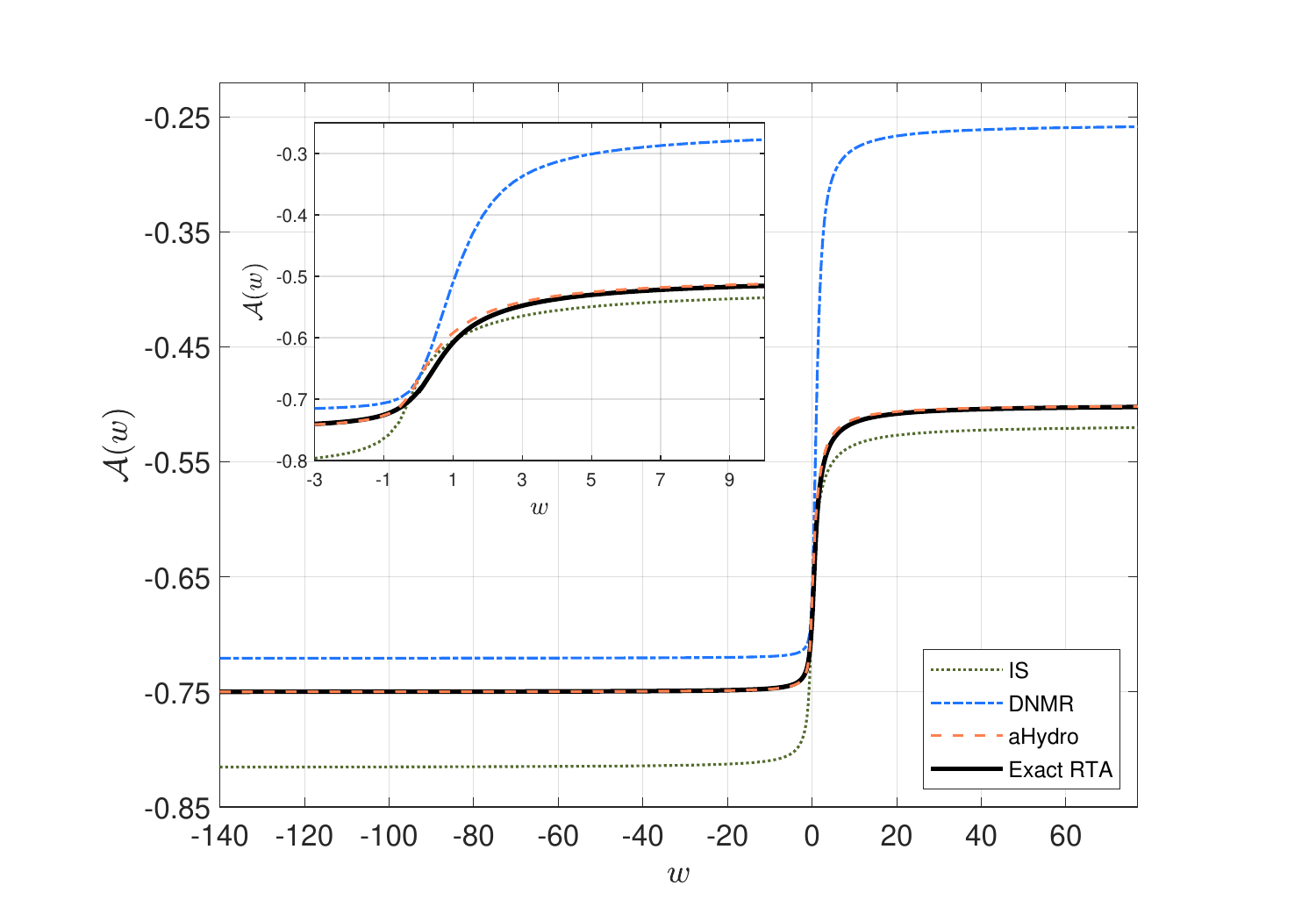}
	\caption{\scriptsize (Color online) Asymptotic attractors of the IS (dotted green lines), DNMR (dash-dotted blue lines), aHydro (short-dashed orange lines) and exact Gubser solution (solid black line), DNMR (top right panel). For all cases we use $c=15/(4\pi)$.}
	\label{fig:attractor2}
\end{figure}

A natural question which arises in our context is: {\it What is the best model that matches the underlying exact microscopic kinetic theory?} We answer this question by plotting in Fig.~\ref{fig:attractor2} the attractors of different hydrodynamical theories together with the one obtained for the exact Gubser solution~\eqref{eq:exsol}. First, we observe that none of the truncated approximated schemes - DNMR and IS - are able to be in good agreement with the exact attractor over the entire $w$ regime studied here. One might be at first surprised that none of the hydrodynamical truncation schemes do work even at large $w$ when the system supposedly reaches its thermal state. However, the Gubser flow does not reach this state asymptotically since the expansion rate $\hat{\theta}=2\tanh\rho$ saturates at large $\pm\rho$ without vanishing exactly. Among these two hydrodynamical truncation schemes, we find the IS to be closer to the numerical values of the exact attractor than DNMR albeit still unable to match it to high numerical accuracy. Now, the best theory to describe the exact attractor shown Fig.~\ref{fig:attractor2} is aHydro. A closer look shows that all the hydrodynamical models are not able to match the exact result in the small $w$ region ($-0.8\lesssim w\lesssim 2$) as it is shown in the inset of Fig.~\ref{fig:attractor2}. However, the  numerical difference of IS and aHydro with respect to the exact result is no larger than $4\%$ in this $w$ interval while DNMR deviates entirely in this regime of $w$. In the large or intermediate regime, on the other hand, we verify numerically that the largest numerical deviation between the aHydro attractor and the exact one does not exceed $0.06\%$. The numerical results presented here provide a conclusive prove that aHydro resums effectively the Knudsen and inverse Reynolds numbers to all orders independent of the initial conditions. 

We point out again that the notion of ``attractor solution'' is ill-defined and one should not care about what occurs in mid-range or early time regime of $w$ because the attractor is actually an statement about late-time asymptotics of the flow lines. We can only say that the attractor solution is just a solution to some system of ODEs with a {\it given} initial value that is located exactly at the saddle point on the boundary of the basin of attraction. But because there are two stable and one unstable directions at this point (one of the eigenvalues of Jacobian matrix being positive means unstable direction), then being a metastable, the saddle point will turn out to initiate a flow inside the basin of attraction and finally the attractor will absorb it. We point out that {\it this flow line is the fastest to converge to the stable fixed point and thus the attractor}. In general this is an invariant set of numbers that flow lines converge to at very large $w$ and as can be seen in Fig.~\ref{fig:attractor2}, even for $w>0$ there is no separation between flow lines due to the exponentially fast convergence. This is way better than what we could have asked for from the approximation $\tanh^2\rho\sim 1$ that was applied to get $\mathcal{A}(w)$ in \eqref{eq:genericfw}.
%%%%%%%%%%%%%%%%%%%%%%%%%%%%%%%%%%%%%%%%%%%%%%%%%%%%%%
\subsection{Asymptotic perturbative series expansion}
\label{subsec:divergence}
%%%%%%%%%%%%%%%%%%%%%%%%%%%%%%%%%%%%%%%%%%%%%%%%%%%%%%%
The numerical comparisons between different models discussed previously demonstrated that the IS and DNMR theories cannot describe the exact steady state attractor. This failure is somehow expected since both theories are derived from an asymptotic series expansion of the distribution function. Here we show strong evidence of the divergence of this series and we briefly comment on how to fix it using resurgence techniques. 
\subsubsection{Divergence of IS  and DNMR theories}
\label{subsec:divergence1}
\begin{figure}[t]
	\centering
	\includegraphics[width=0.7\linewidth]{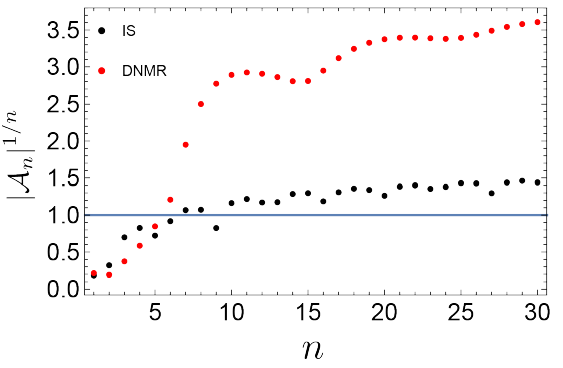}
	\caption{\scriptsize (Color online) $|\mathcal{A}_n|^{1/n}$ as a function of $n$ for both IS (black) and DNMR (red) theories.  We use $c=15/(4\pi)$.}
	\label{fig:roottest}
\end{figure}

Since $w\in(-\infty,\infty)$, one can propose an asymptotic series ansatz of the form
\be
\label{eq:serans}
\mathcal{A}(w) = \sum_{n=0}^{\infty} \mathcal{A}_n w^{-n},
\ee
that is claimed to solve Eq.~\eqref{eq:genericfw} in the asymptotic limit $\rho\to\pm\infty$ with $H(\mathcal{A}(w),w)$ given by Eqs.~\eqref{eq:hfuncIS} and~\eqref{eq:hfuncDNMR}. After equating this ansatz to Eq.~\eqref{eq:genericfw} one obtains the following recursive relation for the $k$th coefficient ($k\ge2$)
\bs
\label{eq:recrel}
\beal
\label{eq:ISrecrel}
\text{IS:}&\hspace{1cm}\sum _{n=0}^k (n+4) \mathcal{A}_n \mathcal{A}_{k-n}+\frac{\mathcal{A}_{k-1}}{c}+\frac{16 \mathcal{A}_k}{3}=0\,,\\
\label{eq:DNMRrecrel}
\text{DNMR:}&\hspace{1cm} \sum _{n=0}^k (n+4) \mathcal{A}_n \mathcal{A}_{k-n}+\frac{\mathcal{A}_{k-1}}{c}+\frac{82 \mathcal{A}_k}{21}=0\,.
\end{align}
\es
For the IS theory $\mathcal{A}_0 = \frac{1}{15} \left(\sqrt{5}-10\right)$ and $\mathcal{A}_1 = -\frac{9+2 \sqrt{5}}{61 c}$ while for the DNMR theory $\mathcal{A}_0=-\frac{1}{84\sqrt{5}} \left(\sqrt{1909}+41\sqrt{5}\right)$ and $\mathcal{A}_1=-\frac{1266+11 \sqrt{9545}}{9139 c}$. The truncation of this power series at $n=30$ is sufficient for the purpose of convergence tests in general. In Fig.~\ref{fig:roottest} we present the numerical solution of the recursive equations~\eqref{eq:recrel} for both IS and DNMR theories when $c=15/(4\pi)$. From this figure we observe that for both of these hydrodynamical models 
\be
1<\lim_{n\rightarrow 30}\sup \left(\mathcal{A}_n\right)^{1/n}<\lim_{n\rightarrow\infty}\sup \left(\mathcal{A}_n\right)^{1/n},
\ee 
so the root test~\cite{apostol} fails for the ansatz of $\mathcal{A}(w)$~\eqref{eq:serans} and thus, the asymptotic series expansions of the IS and DNMR theories are divergent. Fig.~\ref{fig:roottest} also indicates that in this asymptotic regime the divergence in the DNMR case is more severe. This is due to the new term in Eq.~\eqref{eq:hfuncDNMR} which survives in the large $w$ limit instead of converging to zero and thus any divergence in $\mathcal{A}(w)$ in the IS theory will be magnified in the DNMR theory. This result also confirms earlier numerical studies of the Gubser flow (see for instance Fig. 3 in Ref.~\cite{Denicol:2014tha}) where it was found that in the large $\rho^+$ limit the DNMR solution does a poorer job than IS when comparing the predictions of both theories with the RTA exact solution~\eqref{eq:exsol}. 
\subsubsection{Resurgence to the rescue}
\label{subsec:resurgence}
%%%%%%%%%%%%%%%%%%%%%%%%%%%
The true nature of the attractor from the perspective of solving for a series solution
would be unveiled by studying the resurgent aysmptotic expansions around this solution which mimics 
the instanton corrections on top of perturbation theory in quantum field theories. This can be packaged into a formal exponential series solution known as transseries accounting for all sorts of corrections such as exponential, logarithmic and so on \cite{Costin2001}. For the system of equations \eqref{eq:ISGub1}-\eqref{eq:ISGub1} with the substitution $\tau=(1-e^{2\rho})/(1+e^{2\rho})$ an analytic solution of this sort can be written as \cite{costin1998}
\bea
\tem(e^{\rho})
&=& \tem^{(0,0)}(e^{\rho}) + \sigma_1^{1} \sigma_2^{0}\, e^{m_1\,\rho} \,e^{-|\lambda_{\tem}| e^{\rho}} \,\tem^{(1,0)}(e^{\rho}) + \sigma_1^{0} \sigma_2^{1}\, e^{m_2\,\rho} \,e^{-|\lambda_{\bar{\pi}}| e^{\rho}} \,\tem^{(0,1)}(e^{\rho})+\dots \nonumber
\label{eq:trans-ansatz}\\
&=&  \sum_{n_1,n_2=0}^\infty \sigma_1^{n_1} \sigma_2^{n_2}\, e^{\vec{m}\cdot{\vec{n}}\,\rho} \,e^{-\vec{n}\cdot{\vec{\lambda} e^{\rho}}} \,\tem^{(n_1,n_2)}(e^{\rho}),
\label{eq:trans2}
\eea
where $e^{\rho}$ is a good variable for the transseries since at large $\rho$, $w\sim 1/\tem(\rho)\sim \cosh(\rho)^{-\lambda_{\tem}}\sim (e^{\rho})^{-\lambda_{\tem}}$ from the asymptotic solution 
\eqref{eq:Tclose}, with $\vec{\lambda}=(|\lambda_{\tem}|,|\lambda_{\bar{\pi}}|)=( \tfrac{1}{3}(2-\tfrac{1}{\sqrt{5}}),8/(3\sqrt{5}))$ being the absolute value of the Lyapunov exponents of trajectories converging to the attracting fixed point. Here, 
\be
\tem^{(n_1,n_2)}= e^{-\vec{n}\cdot(\vec{m}+\vec{\beta})}\sum_{l=0}^\infty \tem^{(n_1,n_2)}_l\,e^{-l\rho},
\ee
 is the formal power series and 
$\vec{m}=(m_1,m_2)=(1,1)-{\rm Int}[\vec{\beta}]$ with ${\rm Int}[.]$ meaning to be the `integer part of' where $\vec{\beta}$ is a constant real vector field whose components are the coefficients of the term proportional to $({e^{-\rho}}\tem,{e^{-\rho}}\bar{\pi})$ after linearization. $\sigma_{1,2}\in\mathbb{C}$ are
the transseries expansion parameters and the real part of $\sigma_1$ imposes the initial condition for the temperature. Similarly, for $\bar{\pi}$ we propose a transseries of the form
\bea
\bar{\pi}(e^{\rho})
&=& \bar{\pi}^{(0,0)}(e^{\rho}) + \sigma_1^{1} \sigma_2^{0}\, e^{m_1\,\rho} \,e^{-|\lambda_{\tem}| e^{\rho}} \,\bar{\pi}^{(1,0)}(e^{\rho}) + \sigma_1^{0} \sigma_2^{1}\, e^{m_2\,\rho} \,e^{-|\lambda_{\bar{\pi}}| e^{\rho}} \,\bar{\pi}^{(0,1)}(e^{\rho})+\dots \nonumber
\label{eq:trans-ansatzv1}\\
&=&  \sum_{n_1,n_2=0}^\infty \sigma_1^{n_1} \sigma_2^{n_2}\, e^{\vec{m}\cdot{\vec{n}}\,\rho} \,e^{-\vec{n}\cdot{\vec{\lambda} e^{\rho}}} \,\bar{\pi}^{(n_1,n_2)}(e^{\rho}).
\label{eq:transv2}
\eea
  The real part of $\sigma_2\in \mathbb{C}$ encodes the initial data for $\bar{\pi}$ in solving for the full set of transseries solutions.

But what about the imaginary parts? In reality, the leading large-order terms in both factorially divergent series $\bar{\pi}^{(0,0)}$ and $\tem^{(0,0)}$,
\be
\tem^{(0,0)}_{l}\propto \frac{(l-1+m_1)!}{|\lambda_{\tem}|^{l+m_1}},\quad
\bar{\pi}^{(0,0)}_{l}\propto \frac{(l-1+m_2)!}{|\lambda_{\bar{\pi}}|^{l+m_2}}
\ee
where $l\rightarrow \infty$, can be transformed into functions with singularities in the complex plane that need to be dodged via careful contour deformations. We introduce the {\it Borel transform},
\be
\mathcal{B}[\tem^{(0,0)}](s) = \sum_{l=0}^\infty \frac{\tem^{(0,0)}_{l} s^l}{l!},\quad \mathcal{B}[\bar{\pi}^{(0,0)}](s) = \sum_{l=0}^\infty \frac{\bar{\pi}^{(0,0)}_{l} s^l}{l!} \label{eq:Borel}
\ee
that has a finite radius of convergence by construction. To capture the divergence of the original series as
a tangible object, one performs a Borel summation of the series in \eqref{eq:Borel}, of the preferred form 
\be
\mathcal{T}_\epsilon[\tem^{(0,0)}](e^{\rho}) =\int^{\infty+i\epsilon}_0 e^{-e^{\rho}s} \mathcal{B}[\tem^{(0,0)}](s) \,ds,\quad \,\,
\mathcal{T}_\epsilon[\bar{\pi}^{(0,0)}](e^{\rho}) =\int^{\infty+i\epsilon}_0 e^{-e^{\rho}s} \mathcal{B}[\bar{\pi}^{(0,0)}](s) \,ds, 
\ee
which are nothing but the inverse Laplace transform of the Borel functions.
Since there are in general singularities on the real line in the Borel plane ($s$-plane) that capture the divergence of the original series,
there is a slight shift of the integration contour $(\epsilon=+0)$ or $(\epsilon=-0)$ and depending on how this is qualitatively done,
we would end up with a discontinuity along the contour of integration that by Cauchy residue theorem would 
generate a pure imaginary part once we jump from $+i0$ to $-i0$ contour or vice versa. This is related to the fact that 
the real line is simply a ``Stokes line'', and the jump associated with the discontinuity across it, is called the ``Stokes jump'' performed by a Stokes constant $ iS_1$ with $S_1$ being real. These jumps are calculated via the formula 
\be
\vec{\sigma}(\epsilon)_I = 
\Bigg\{ \begin{array}{cc}
	\vec{\sigma}^{-}_I=\vec{\sigma}_I(-0) & {\rm for}\quad\epsilon<0\\
	\vec{\sigma}^{0}_I=\vec{\sigma}_I(-0)+\tfrac{1}{2}S_{1}\hat{{e}}_{1} & {\rm for}\quad\epsilon=0\\
	\vec{\sigma}^{+}_I=\vec{\sigma}_I(-0)+S_{1}\hat{{e}}_{1} & {\rm for}\quad \epsilon>0
\end{array}
\ee
where $\hat{e}_1=(1,0)$ and $\vec{\sigma}_I=({\rm Im}\,\sigma_1,{\rm Im}\,\sigma_2)$, which was obtained originally by a ``balanced average" summation of the formal divergent series \cite{costin1998} \footnote{We thank O. Costin for kindly explaining this summation technique.}. Therefore, the reality condition on the transseries exactly on the positive real axis would imply that
\be
\vec{\sigma}_I^0=0\Rightarrow \vec{\sigma}_I^\pm = (\pm\tfrac{1}{2}S_{1},0).
\ee

%%%%%%%%%%%%%%%%%%%%%%%%%%%%%%%%%%%%
\section{Conclusions}
\label{sec:conclusion}%%%%%%%%%%%%%%%%%%%%%%%%%%%%%%%%%%%%

The properties of the attractors of different hydrodynammical systems undergoing Gubser flow were studied within relativistic kinetic theory. Our work extended previous studies of attractors by incorporating techniques and tools of nonlinear dynamical systems. We extensively discuss the application of these methods to the IS evolution equations to investigate the stability properties of the fixed points, the flow lines around those and the Lyapunov exponents. These techniques were proven to be strikingly useful in the understanding of the reason behind the impossibility to reduce the $2d$ system of ODEs of each hydrodynamical scheme into a single one universally as it was done in the Bjorken case~\cite{Heller:2015dha}. It was shown that this non-reduction is intrinsically related with the dimensionality of the basin of attraction and that the $\tanh\rho$ also acts as an independent variable of $\rho$ that is responsible for the extra dimension of the phase space compared to the Bjorken case. Inspired by the exponential asymptotic stability and resurgent transseries type arguments, we defined a natural time variable, $t=\pm e^{-2\rho}$, which was expounded in subsection~\ref{subsubsec:basin}. In this coordinate system, the observer sees two copies of the same flow started at $t\rightarrow \pm \infty$ that approach to the point $(0,0,0)$ which is now the stable fixed point of the system, hence making the basin symmetric under $t\rightarrow-t$. This method is a very well-known trick used in usually exponentially asymptotically stable systems to connect the basin of attraction in non-autonomous systems to that of autonomous ones as well as compactifying the basin in case the time-dependency was in the form of an unbounded function. It also brings the fixed point to the origin that is crucial for the  construction of Lyapunov functions. This helped us build such a function for the IS theory and discuss the shape of the basin of attraction at least locally near the steady state equilibrium. This notion of the Lyapunov function from dynamical systems was shown to be related to an effective-action description of hydrodynamics motivated by Picard-Lefschetz theory.

The attractors of the IS, DNMR and aHydro models were obtained via the slow roll-down approximation while the attractor associated to the Gubser RTA solution was determined through Romatschke's method~\cite{Romatschke:2017vte}. From the numerical comparisons between the attractors of different theories with the exact solution we conclude: (a) Hydrodynamical models based on an asymptotic series expansion of the distribution function, aka IS and DNMR, are unable to provide a quantitative description of the attractor of the exact solution and, (b) the best agreement with the exact attractor is achieved by aHydro up to high numerical accuracy. For the IS and DNMR approximation schemes we showed that their corresponding asymptotic solutions to the respective differential equations are divergent. The origin of the divergences and how to resum them was briefly discussed by applying resurgence techniques. 

The success of aHydro to describe the asymptotic attractor of the exact Gubser solution demonstrates that this theory is able to take into account both, large inhomogeneities in the fluid due to collisions -quantified by the Knudsen number- as well as big spacetime inhomogeneities of the macroscopic fluid variables -quantified by the inverse Reynolds number-. As a matter of fact, the match between aHydro attractor and the exact one demonstrates that aHydro resums the $Kn$ and $Re^{-1}$ to all orders. This resummation is carried out in a non-perturbative way by including the largest momentum-space anisotropies present in the plasma into the leading order anisotropic distribution function. 

The very common techniques in the context of nonlinear dynamical systems presented in this work and applied to hydrodynamics, can be extended to study a long list of properties of other relevant more complex physical systems than the one studied here. This list may include an investigation of hydrodynamization processes and dynamics of attractors for more general nonlinear collisional kernels~\cite{Bazow:2015dha,Kurkela:2015qoa,Bazow:2016oky}, holographic models~\cite{Critelli:2017euk,Chesler:2009cy,Heller:2011ju,Wu:2011yd,vanderSchee:2012qj,Casalderrey-Solana:2013aba}, spatially non-homogeneous expanding fluids~\cite{Romatschke:2017acs}. On the other hand, it is also interesting to investigate the rich structure and topology of the basin of attractors in turbulent flows and other chaotic systems of interest. On a more theoretical subject, the possibility of formulating effective actions for hydrodynamics by exploring the analogy between the steepest descent directions in the path integrals and the flow lines starting at the boundary of the basin of attraction for the attractors opens a new perspective to re-formulate this old problem. Moreover, the issues addressed in this work shed more light on new questions that remain to be answered within the aHydro framework. For instance, in the resurgence program, the attractor is understood as the leading-order asymptotic transseries~\cite{tournier,costin}. This mathematical statement together with our results suggest a highly non-trivial relation between the nonperturbative resummation of large Knudsen and inverse Reynolds numbers carried out by aHydro and a certain class of solutions of the nonlinear Boltzmann equation that can be expressed as a transseries. We leave these matters to future works.

%%%%%%%%%%%%%%%%%%%%%%%%%%%%%%%%%%%%%%%%%%%%%%%%%
\section*{Acknowledgements} 
We would like to thank Gokce Basar, Gerald Dunne, Behnam Kia,  Thomas Sch\"afer and Mithat Unsal for useful discussions. We would also like to specially thank David Sauzin and Ovidiu Costin for very useful discussions and helpful comments. C.~N.~C.~C. would like to thank E. C. Pinilla for her continuous support and useful discussions. We thank Michal Heller for reading the draft and his comments/suggestions. M. M. was supported in part by the US Department of Energy grant DE-FG02-03ER41260 and by the BEST (Beam Energy Scan Theory) DOE Topical Collaboration. A. B. was supported by the DOE grant DE-SC0013036. A. B.'s portion of the research was supported in part by the National Science Foundation under Grant No. NSF PHY-1125915. A.~B. thanks the local organizers of ``Resurgent Asymptotics in Physics and Mathematics'' workshop at KITP for the warm hospitality.

%%%%%%%%%%%%%%%%%%%%%%%%%%%%%%%%%%%%%%%%%%%%%%%%

\appendix
%%%%%%%%%%%%%%%%%%%%%%%%%%%%%%%%%%%%%%%%%%%%%%%%%%%%%%%%
\section{Anisotropic integrals}
\label{app:anisint}
%%%%%%%%%%%%%%%%%%%%%%%%%%%%%%%%%%%%%%%%%%%%%%%%%%%%%%%%

Here we calculate the anisotropic integrals $\Ih_{nlq}$ that appear in this paper. First we define
\bea
\label{eq:anisint}
\Ih_{nlq}\big(\hat{\Lambda},\xi\big)
&=&\langle\,(-\hat{u}\cdot\pp)^{n-l-2q}\,(\hat{l}\cdot\pp)^l\,(\hat{\Xi}_{\mu\nu}\pp^\mu\pp^\nu)^q\,\rangle_a
\nonumber\\
&=& \int_\pp \, (\pp^\rho)^{n-l-2q}\,\pp_\eta^l\, \left(\frac{\po}{\cosh^2\rho}\right)^q\,f_a.
\eea
By considering the following change of variables
\bs
\label{eq:changecoor}
\beal
&\frac{\pth}{\cosh\rho}=\lambda\,\sin\alpha\,\cos\beta\,,\\
&\frac{\pph}{\cosh\rho\,\sin\theta}=\lambda\,\sin\alpha\,\sin\beta\,,\\
&\pp_\eta=\lambda\,(1+\xi)^{-1/2}\cos\alpha
\end{align}
\es
one is able to factorize the integral~\eqref{eq:anisint} as 
\be\label{eq:anisint2}
\Ih_{nlq}\big(\hat{\Lambda},\xi \big)
=\,\J_{n}\big(\hat{\Lambda}\big)\,\R_{nlq}\left(\xi\right)\,,
\ee
where
\bea
\label{eq:Jfunc}
\J_{n}\big(\hat{\Lambda}\big)
&=& \int_0^\infty \frac{d\lambda}{2\pi^2}\,\lambda^{n+1}\, e^{-\lambda/\hat{\Lambda}} 
= \frac{(n{+}1)!}{2\pi^2}\,\hat{\Lambda}^{n+2},
\\
\label{eq:Rfunc}
\R_{nlq}\left(\xi\right)
&=&\frac{1}{2(1{+}\xi)^{\left(n-2q\right)/2}} \int_{-1}^1 dx\,\left(1-x^2\right)^{q}\,x^l\,
\,\left[(1+\xi)\,(1-x^2)\,+\,x^2\right]^{(n-l-2q-1)/2}\!\!\!\!.\ \ 
\eea
The explicit forms of the functions $\R_{nlq}$ used in this paper are given by
\bs
\label{eq:Rfunctions}
\beal
\label{eq:R}
\R_{200}(\xi)&=\frac{1}{2}\left(
\frac{1}{1+\xi}+\frac{{\arctan}\sqrt{\xi}}
{\sqrt{\xi}}
\right) \\
\R_{220}(\xi)&=\frac{1}{2\xi}\left(
-\frac{1}{1+\xi}+\frac{{\arctan}\sqrt{\xi}}
{\sqrt{\xi}} 
\right) \\
\R_{240}(\xi)&=\frac{1}{2\xi^2}\left(
\frac{3+2\xi}{1+\xi}-3\frac{{\arctan}\sqrt{\xi}}
{\sqrt{\xi}} 
\right) 
\end{align}
\es
The moments $\Ih_{nlq}^\mathrm{eq}$ associated with the equilibrium distribution function are obtained by considering the $\xi\to 0$ limit of Eq.~\eqref{eq:anisint2}:
\be
\label{eq:equilmom}
\Ih_{nlq}^\mathrm{eq}(\tem)\equiv \lim_{\xi\to 0} \Ih_{nlq}\big(\hat{\Lambda},\xi \big)= \Ih_{nlq}\big(\tem,0)\,.
\ee
%

%%%%%%%%%%%%%%%%%%%%%%%%%%%%
\section{Attractor of the RTA Boltzmann equation for the Gubser flow}
\label{app:attrexact}
%%%%%%%%%%%%%%%%%%%%%%%%%%%%
The attractor for the exact solution~\eqref{eq:exsol} was found by considering Romatschke's technique~\cite{Romatschke:2017vte}. For the Gubser case the slow roll condition, cf. Refs.~\cite{Heller:2015dha,Liddle:1994dx,Romatschke:2017vte},~ $d\mathcal{A}(w)/dw=0$ gives the following condition
\be
\label{eq:attrexactsol}
\frac{d \mathcal{A}(w)}{dw}=0\Rightarrow\,\frac{1}{4}\left[\frac{1}{\ene^2\,\tanh\rho}\left(\ene\,\partial_\rho^2\ene-\left(\partial_\rho\ene\right)^2\right)-\frac{1}{\sinh^2\rho}\frac{\partial_\rho\ene}{\ene}\right]\Biggl|_{\rho=\rho_0,\xi=\xi_0}\,=\,0.
\ee
As we pointed out in Sec.~\ref{sec:flow} and~\ref{sec:attractors} one eliminates the $\rho$ dependence by considering the asymptotic behavior at $\rho\to-\infty$. For numerical purposes it is enough to take $\rho_0=-10$ and thus $\lim_{\rho\to-10}\tanh\rho\approx-1$. By doing this each term entering in the constraint~\eqref{eq:attrexactsol} is given by
\be
\begin{split}
&\ene\bigl|_{\rho=-10,\xi=\xi_0}= \Ih_{200}(\hat{\Lambda}_0,\xi_0) \,,\hspace{1cm}\partial_\rho\ene\bigl|_{\rho=-10,\xi=\xi_0}=\,3\,\Ih_{200}(\hat{\Lambda}_0,\xi_0)-\Ih_{220}(\hat{\Lambda}_0,\xi_0)\,,\\
&\partial^2_\rho\ene|_{\rho=-10,\xi=\xi_0}=9\Ih_{200}(\hat{\Lambda}_0,\xi_0)-4\Ih_{220}(\hat{\Lambda}_0,\xi_0)+\Ih_{240}(\hat{\Lambda}_0,\xi_0)\\\
&\hspace{2cm}+\,\frac{\left[\R_{200}(\xi_0)\right]^{1/4}\hat{\Lambda}_0}{c}\left[\Ih_{220}(\hat{\Lambda}_0,\xi_0)-\Ih^\mathrm{eq}_{220}\left(\left[\R_{200}(\xi_0)\right]^{1/4}\hat{\Lambda}_0\right)\right]\,.
\end{split}
\ee
%

%%%%%%%%%%%%%%%%%%%%%%%%%%%%
\bibliography{nonequilattractor}

\begin{thebibliography}{91}
\expandafter\ifx\csname natexlab\endcsname\relax\def\natexlab#1{#1}\fi
\expandafter\ifx\csname bibnamefont\endcsname\relax
  \def\bibnamefont#1{#1}\fi
\expandafter\ifx\csname bibfnamefont\endcsname\relax
  \def\bibfnamefont#1{#1}\fi
\expandafter\ifx\csname citenamefont\endcsname\relax
  \def\citenamefont#1{#1}\fi
\expandafter\ifx\csname url\endcsname\relax
  \def\url#1{\texttt{#1}}\fi
\expandafter\ifx\csname urlprefix\endcsname\relax\def\urlprefix{URL }\fi
\providecommand{\bibinfo}[2]{#2}
\providecommand{\eprint}[2][]{\url{#2}}

\bibitem[{\citenamefont{Landau and Lifshitz}(1987)}]{landaufluid}
\bibinfo{author}{\bibfnamefont{L.}~\bibnamefont{Landau}} \bibnamefont{and}
  \bibinfo{author}{\bibfnamefont{E.}~\bibnamefont{Lifshitz}},
  \emph{\bibinfo{title}{Fluid Mechanics}}, \bibinfo{number}{Course of
  Theoretical Physics, Vol. 6} (\bibinfo{publisher}{Pergamon Press, Oxford},
  \bibinfo{year}{1987}).

\bibitem[{\citenamefont{Aad et~al.}(2016)}]{Aad:2015gqa}
\bibinfo{author}{\bibfnamefont{G.}~\bibnamefont{Aad}} \bibnamefont{et~al.}
  (\bibinfo{collaboration}{ATLAS}), \bibinfo{journal}{Phys. Rev. Lett.}
  \textbf{\bibinfo{volume}{116}}, \bibinfo{pages}{172301}
  (\bibinfo{year}{2016}), \eprint{1509.04776}.

\bibitem[{\citenamefont{Khachatryan et~al.}(2016)}]{Khachatryan:2015lva}
\bibinfo{author}{\bibfnamefont{V.}~\bibnamefont{Khachatryan}}
  \bibnamefont{et~al.} (\bibinfo{collaboration}{CMS}), \bibinfo{journal}{Phys.
  Rev. Lett.} \textbf{\bibinfo{volume}{116}}, \bibinfo{pages}{172302}
  (\bibinfo{year}{2016}), \eprint{1510.03068}.

\bibitem[{\citenamefont{Weller and Romatschke}(2017)}]{Weller:2017tsr}
\bibinfo{author}{\bibfnamefont{R.~D.} \bibnamefont{Weller}} \bibnamefont{and}
  \bibinfo{author}{\bibfnamefont{P.}~\bibnamefont{Romatschke}},
  \bibinfo{journal}{Phys. Lett.} \textbf{\bibinfo{volume}{B774}},
  \bibinfo{pages}{351} (\bibinfo{year}{2017}), \eprint{1701.07145}.

\bibitem[{\citenamefont{Bozek}(2011)}]{Bozek:2010pb}
\bibinfo{author}{\bibfnamefont{P.}~\bibnamefont{Bozek}}, \bibinfo{journal}{Eur.
  Phys. J.} \textbf{\bibinfo{volume}{C71}}, \bibinfo{pages}{1530}
  (\bibinfo{year}{2011}), \eprint{1010.0405}.

\bibitem[{\citenamefont{Werner et~al.}(2011)\citenamefont{Werner, Karpenko, and
  Pierog}}]{Werner:2010ss}
\bibinfo{author}{\bibfnamefont{K.}~\bibnamefont{Werner}},
  \bibinfo{author}{\bibfnamefont{I.}~\bibnamefont{Karpenko}}, \bibnamefont{and}
  \bibinfo{author}{\bibfnamefont{T.}~\bibnamefont{Pierog}},
  \bibinfo{journal}{Phys. Rev. Lett.} \textbf{\bibinfo{volume}{106}},
  \bibinfo{pages}{122004} (\bibinfo{year}{2011}), \eprint{1011.0375}.

\bibitem[{\citenamefont{Kurkela and Zhu}(2015)}]{Kurkela:2015qoa}
\bibinfo{author}{\bibfnamefont{A.}~\bibnamefont{Kurkela}} \bibnamefont{and}
  \bibinfo{author}{\bibfnamefont{Y.}~\bibnamefont{Zhu}},
  \bibinfo{journal}{Phys. Rev. Lett.} \textbf{\bibinfo{volume}{115}},
  \bibinfo{pages}{182301} (\bibinfo{year}{2015}), \eprint{1506.06647}.

\bibitem[{\citenamefont{Critelli et~al.}(2017)\citenamefont{Critelli,
  Rougemont, and Noronha}}]{Critelli:2017euk}
\bibinfo{author}{\bibfnamefont{R.}~\bibnamefont{Critelli}},
  \bibinfo{author}{\bibfnamefont{R.}~\bibnamefont{Rougemont}},
  \bibnamefont{and} \bibinfo{author}{\bibfnamefont{J.}~\bibnamefont{Noronha}}
  (\bibinfo{year}{2017}), \eprint{1709.03131}.

\bibitem[{\citenamefont{Denicol
  et~al.}(2014{\natexlab{a}})\citenamefont{Denicol, Heinz, Martinez, Noronha,
  and Strickland}}]{Denicol:2014xca}
\bibinfo{author}{\bibfnamefont{G.~S.} \bibnamefont{Denicol}},
  \bibinfo{author}{\bibfnamefont{U.~W.} \bibnamefont{Heinz}},
  \bibinfo{author}{\bibfnamefont{M.}~\bibnamefont{Martinez}},
  \bibinfo{author}{\bibfnamefont{J.}~\bibnamefont{Noronha}}, \bibnamefont{and}
  \bibinfo{author}{\bibfnamefont{M.}~\bibnamefont{Strickland}},
  \bibinfo{journal}{Phys.Rev.Lett.} \textbf{\bibinfo{volume}{113}},
  \bibinfo{pages}{202301} (\bibinfo{year}{2014}{\natexlab{a}}),
  \eprint{1408.5646}.

\bibitem[{\citenamefont{Florkowski et~al.}(2014)\citenamefont{Florkowski,
  Maksymiuk, Ryblewski, and Strickland}}]{Florkowski:2014sfa}
\bibinfo{author}{\bibfnamefont{W.}~\bibnamefont{Florkowski}},
  \bibinfo{author}{\bibfnamefont{E.}~\bibnamefont{Maksymiuk}},
  \bibinfo{author}{\bibfnamefont{R.}~\bibnamefont{Ryblewski}},
  \bibnamefont{and}
  \bibinfo{author}{\bibfnamefont{M.}~\bibnamefont{Strickland}}
  (\bibinfo{year}{2014}), \eprint{1402.7348}.

\bibitem[{\citenamefont{Florkowski
  et~al.}(2013{\natexlab{a}})\citenamefont{Florkowski, Ryblewski, and
  Strickland}}]{Florkowski:2013lza}
\bibinfo{author}{\bibfnamefont{W.}~\bibnamefont{Florkowski}},
  \bibinfo{author}{\bibfnamefont{R.}~\bibnamefont{Ryblewski}},
  \bibnamefont{and}
  \bibinfo{author}{\bibfnamefont{M.}~\bibnamefont{Strickland}},
  \bibinfo{journal}{Nucl.Phys.} \textbf{\bibinfo{volume}{A916}},
  \bibinfo{pages}{249} (\bibinfo{year}{2013}{\natexlab{a}}),
  \eprint{1304.0665}.

\bibitem[{\citenamefont{Florkowski
  et~al.}(2013{\natexlab{b}})\citenamefont{Florkowski, Ryblewski, and
  Strickland}}]{Florkowski:2013lya}
\bibinfo{author}{\bibfnamefont{W.}~\bibnamefont{Florkowski}},
  \bibinfo{author}{\bibfnamefont{R.}~\bibnamefont{Ryblewski}},
  \bibnamefont{and}
  \bibinfo{author}{\bibfnamefont{M.}~\bibnamefont{Strickland}},
  \bibinfo{journal}{Phys. Rev.} \textbf{\bibinfo{volume}{C88}},
  \bibinfo{pages}{024903} (\bibinfo{year}{2013}{\natexlab{b}}),
  \eprint{1305.7234}.

\bibitem[{\citenamefont{Denicol
  et~al.}(2014{\natexlab{b}})\citenamefont{Denicol, Heinz, Martinez, Noronha,
  and Strickland}}]{Denicol:2014tha}
\bibinfo{author}{\bibfnamefont{G.~S.} \bibnamefont{Denicol}},
  \bibinfo{author}{\bibfnamefont{U.~W.} \bibnamefont{Heinz}},
  \bibinfo{author}{\bibfnamefont{M.}~\bibnamefont{Martinez}},
  \bibinfo{author}{\bibfnamefont{J.}~\bibnamefont{Noronha}}, \bibnamefont{and}
  \bibinfo{author}{\bibfnamefont{M.}~\bibnamefont{Strickland}},
  \bibinfo{journal}{Phys. Rev.} \textbf{\bibinfo{volume}{D90}},
  \bibinfo{pages}{125026} (\bibinfo{year}{2014}{\natexlab{b}}),
  \eprint{1408.7048}.

\bibitem[{\citenamefont{Chesler and Yaffe}(2010)}]{Chesler:2009cy}
\bibinfo{author}{\bibfnamefont{P.~M.} \bibnamefont{Chesler}} \bibnamefont{and}
  \bibinfo{author}{\bibfnamefont{L.~G.} \bibnamefont{Yaffe}},
  \bibinfo{journal}{Phys. Rev.} \textbf{\bibinfo{volume}{D82}},
  \bibinfo{pages}{026006} (\bibinfo{year}{2010}), \eprint{0906.4426}.

\bibitem[{\citenamefont{Heller et~al.}(2012)\citenamefont{Heller, Janik, and
  Witaszczyk}}]{Heller:2011ju}
\bibinfo{author}{\bibfnamefont{M.~P.} \bibnamefont{Heller}},
  \bibinfo{author}{\bibfnamefont{R.~A.} \bibnamefont{Janik}}, \bibnamefont{and}
  \bibinfo{author}{\bibfnamefont{P.}~\bibnamefont{Witaszczyk}},
  \bibinfo{journal}{Phys. Rev. Lett.} \textbf{\bibinfo{volume}{108}},
  \bibinfo{pages}{201602} (\bibinfo{year}{2012}), \eprint{1103.3452}.

\bibitem[{\citenamefont{Wu and Romatschke}(2011)}]{Wu:2011yd}
\bibinfo{author}{\bibfnamefont{B.}~\bibnamefont{Wu}} \bibnamefont{and}
  \bibinfo{author}{\bibfnamefont{P.}~\bibnamefont{Romatschke}},
  \bibinfo{journal}{Int. J. Mod. Phys.} \textbf{\bibinfo{volume}{C22}},
  \bibinfo{pages}{1317} (\bibinfo{year}{2011}), \eprint{1108.3715}.

\bibitem[{\citenamefont{van~der Schee}(2013)}]{vanderSchee:2012qj}
\bibinfo{author}{\bibfnamefont{W.}~\bibnamefont{van~der Schee}},
  \bibinfo{journal}{Phys. Rev.} \textbf{\bibinfo{volume}{D87}},
  \bibinfo{pages}{061901} (\bibinfo{year}{2013}), \eprint{1211.2218}.

\bibitem[{\citenamefont{Casalderrey-Solana
  et~al.}(2013)\citenamefont{Casalderrey-Solana, Heller, Mateos, and van~der
  Schee}}]{Casalderrey-Solana:2013aba}
\bibinfo{author}{\bibfnamefont{J.}~\bibnamefont{Casalderrey-Solana}},
  \bibinfo{author}{\bibfnamefont{M.~P.} \bibnamefont{Heller}},
  \bibinfo{author}{\bibfnamefont{D.}~\bibnamefont{Mateos}}, \bibnamefont{and}
  \bibinfo{author}{\bibfnamefont{W.}~\bibnamefont{van~der Schee}},
  \bibinfo{journal}{Phys. Rev. Lett.} \textbf{\bibinfo{volume}{111}},
  \bibinfo{pages}{181601} (\bibinfo{year}{2013}), \eprint{1305.4919}.

\bibitem[{\citenamefont{Heller et~al.}(2014)\citenamefont{Heller, Janik,
  Spaliński, and Witaszczyk}}]{Heller:2014wfa}
\bibinfo{author}{\bibfnamefont{M.~P.} \bibnamefont{Heller}},
  \bibinfo{author}{\bibfnamefont{R.~A.} \bibnamefont{Janik}},
  \bibinfo{author}{\bibfnamefont{M.}~\bibnamefont{Spaliński}},
  \bibnamefont{and}
  \bibinfo{author}{\bibfnamefont{P.}~\bibnamefont{Witaszczyk}},
  \bibinfo{journal}{Phys. Rev. Lett.} \textbf{\bibinfo{volume}{113}},
  \bibinfo{pages}{261601} (\bibinfo{year}{2014}), \eprint{1409.5087}.

\bibitem[{\citenamefont{Heller et~al.}(2016)\citenamefont{Heller, Kurkela, and
  Spalinski}}]{Heller:2016rtz}
\bibinfo{author}{\bibfnamefont{M.~P.} \bibnamefont{Heller}},
  \bibinfo{author}{\bibfnamefont{A.}~\bibnamefont{Kurkela}}, \bibnamefont{and}
  \bibinfo{author}{\bibfnamefont{M.}~\bibnamefont{Spalinski}}
  (\bibinfo{year}{2016}), \eprint{1609.04803}.

\bibitem[{\citenamefont{Denicol and Noronha}(2016)}]{Denicol:2016bjh}
\bibinfo{author}{\bibfnamefont{G.~S.} \bibnamefont{Denicol}} \bibnamefont{and}
  \bibinfo{author}{\bibfnamefont{J.}~\bibnamefont{Noronha}}
  (\bibinfo{year}{2016}), \eprint{1608.07869}.

\bibitem[{\citenamefont{Bazow et~al.}(2016{\natexlab{a}})\citenamefont{Bazow,
  Denicol, Heinz, Martinez, and Noronha}}]{Bazow:2016oky}
\bibinfo{author}{\bibfnamefont{D.}~\bibnamefont{Bazow}},
  \bibinfo{author}{\bibfnamefont{G.~S.} \bibnamefont{Denicol}},
  \bibinfo{author}{\bibfnamefont{U.}~\bibnamefont{Heinz}},
  \bibinfo{author}{\bibfnamefont{M.}~\bibnamefont{Martinez}}, \bibnamefont{and}
  \bibinfo{author}{\bibfnamefont{J.}~\bibnamefont{Noronha}},
  \bibinfo{journal}{Phys. Rev.} \textbf{\bibinfo{volume}{D94}},
  \bibinfo{pages}{125006} (\bibinfo{year}{2016}{\natexlab{a}}),
  \eprint{1607.05245}.

\bibitem[{\citenamefont{Santos et~al.}(1986)\citenamefont{Santos, Brey, and
  Dufty}}]{santosdufty}
\bibinfo{author}{\bibfnamefont{A.}~\bibnamefont{Santos}},
  \bibinfo{author}{\bibfnamefont{J.~J.} \bibnamefont{Brey}}, \bibnamefont{and}
  \bibinfo{author}{\bibfnamefont{J.~W.} \bibnamefont{Dufty}},
  \bibinfo{journal}{Phys. Rev. Lett.} \textbf{\bibinfo{volume}{56}},
  \bibinfo{pages}{1571} (\bibinfo{year}{1986}).

\bibitem[{\citenamefont{Santos}(2008)}]{sant}
\bibinfo{author}{\bibfnamefont{A.}~\bibnamefont{Santos}},
  \bibinfo{journal}{Phys. Rev. Lett.} \textbf{\bibinfo{volume}{100}},
  \bibinfo{pages}{078003} (\bibinfo{year}{2008}).

\bibitem[{\citenamefont{Heller and Spalinski}(2015)}]{Heller:2015dha}
\bibinfo{author}{\bibfnamefont{M.~P.} \bibnamefont{Heller}} \bibnamefont{and}
  \bibinfo{author}{\bibfnamefont{M.}~\bibnamefont{Spalinski}},
  \bibinfo{journal}{Phys. Rev. Lett.} \textbf{\bibinfo{volume}{115}},
  \bibinfo{pages}{072501} (\bibinfo{year}{2015}), \eprint{1503.07514}.

\bibitem[{\citenamefont{Basar and Dunne}(2015)}]{Basar:2015ava}
\bibinfo{author}{\bibfnamefont{G.}~\bibnamefont{Basar}} \bibnamefont{and}
  \bibinfo{author}{\bibfnamefont{G.~V.} \bibnamefont{Dunne}},
  \bibinfo{journal}{Phys. Rev.} \textbf{\bibinfo{volume}{D92}},
  \bibinfo{pages}{125011} (\bibinfo{year}{2015}), \eprint{1509.05046}.

\bibitem[{\citenamefont{Aniceto and Spaliński}(2016)}]{Aniceto:2015mto}
\bibinfo{author}{\bibfnamefont{I.}~\bibnamefont{Aniceto}} \bibnamefont{and}
  \bibinfo{author}{\bibfnamefont{M.}~\bibnamefont{Spaliński}},
  \bibinfo{journal}{Phys. Rev.} \textbf{\bibinfo{volume}{D93}},
  \bibinfo{pages}{085008} (\bibinfo{year}{2016}), \eprint{1511.06358}.

\bibitem[{\citenamefont{Berges et~al.}(2014{\natexlab{a}})\citenamefont{Berges,
  Boguslavski, Schlichting, and Venugopalan}}]{Berges:2013eia}
\bibinfo{author}{\bibfnamefont{J.}~\bibnamefont{Berges}},
  \bibinfo{author}{\bibfnamefont{K.}~\bibnamefont{Boguslavski}},
  \bibinfo{author}{\bibfnamefont{S.}~\bibnamefont{Schlichting}},
  \bibnamefont{and}
  \bibinfo{author}{\bibfnamefont{R.}~\bibnamefont{Venugopalan}},
  \bibinfo{journal}{Phys. Rev.} \textbf{\bibinfo{volume}{D89}},
  \bibinfo{pages}{074011} (\bibinfo{year}{2014}{\natexlab{a}}),
  \eprint{1303.5650}.

\bibitem[{\citenamefont{Micha and Tkachev}(2004)}]{Micha:2004bv}
\bibinfo{author}{\bibfnamefont{R.}~\bibnamefont{Micha}} \bibnamefont{and}
  \bibinfo{author}{\bibfnamefont{I.~I.} \bibnamefont{Tkachev}},
  \bibinfo{journal}{Phys. Rev.} \textbf{\bibinfo{volume}{D70}},
  \bibinfo{pages}{043538} (\bibinfo{year}{2004}), \eprint{hep-ph/0403101}.

\bibitem[{\citenamefont{Baier et~al.}(2001)\citenamefont{Baier, Mueller,
  Schiff, and Son}}]{Baier:2000sb}
\bibinfo{author}{\bibfnamefont{R.}~\bibnamefont{Baier}},
  \bibinfo{author}{\bibfnamefont{A.~H.} \bibnamefont{Mueller}},
  \bibinfo{author}{\bibfnamefont{D.}~\bibnamefont{Schiff}}, \bibnamefont{and}
  \bibinfo{author}{\bibfnamefont{D.~T.} \bibnamefont{Son}},
  \bibinfo{journal}{Phys. Lett.} \textbf{\bibinfo{volume}{B502}},
  \bibinfo{pages}{51} (\bibinfo{year}{2001}), \eprint{hep-ph/0009237}.

\bibitem[{\citenamefont{Mehtar-Tani}(2017)}]{Mehtar-Tani:2016bay}
\bibinfo{author}{\bibfnamefont{Y.}~\bibnamefont{Mehtar-Tani}},
  \bibinfo{journal}{Nucl. Phys.} \textbf{\bibinfo{volume}{A966}},
  \bibinfo{pages}{241} (\bibinfo{year}{2017}), \eprint{1611.01527}.

\bibitem[{\citenamefont{Berges et~al.}(2014{\natexlab{b}})\citenamefont{Berges,
  Boguslavski, Schlichting, and Venugopalan}}]{Berges:2013fga}
\bibinfo{author}{\bibfnamefont{J.}~\bibnamefont{Berges}},
  \bibinfo{author}{\bibfnamefont{K.}~\bibnamefont{Boguslavski}},
  \bibinfo{author}{\bibfnamefont{S.}~\bibnamefont{Schlichting}},
  \bibnamefont{and}
  \bibinfo{author}{\bibfnamefont{R.}~\bibnamefont{Venugopalan}},
  \bibinfo{journal}{Phys. Rev.} \textbf{\bibinfo{volume}{D89}},
  \bibinfo{pages}{114007} (\bibinfo{year}{2014}{\natexlab{b}}),
  \eprint{1311.3005}.

\bibitem[{\citenamefont{Spaliński}(2017)}]{Spalinski:2017mel}
\bibinfo{author}{\bibfnamefont{M.}~\bibnamefont{Spaliński}}
  (\bibinfo{year}{2017}), \eprint{1708.01921}.

\bibitem[{\citenamefont{Buchel et~al.}(2016)\citenamefont{Buchel, Heller, and
  Noronha}}]{Buchel:2016cbj}
\bibinfo{author}{\bibfnamefont{A.}~\bibnamefont{Buchel}},
  \bibinfo{author}{\bibfnamefont{M.~P.} \bibnamefont{Heller}},
  \bibnamefont{and} \bibinfo{author}{\bibfnamefont{J.}~\bibnamefont{Noronha}},
  \bibinfo{journal}{Phys. Rev.} \textbf{\bibinfo{volume}{D94}},
  \bibinfo{pages}{106011} (\bibinfo{year}{2016}), \eprint{1603.05344}.

\bibitem[{\citenamefont{Romatschke}(2017{\natexlab{a}})}]{Romatschke:2017vte}
\bibinfo{author}{\bibfnamefont{P.}~\bibnamefont{Romatschke}}
  (\bibinfo{year}{2017}{\natexlab{a}}), \eprint{1704.08699}.

\bibitem[{\citenamefont{Strickland et~al.}(2017)\citenamefont{Strickland,
  Noronha, and Denicol}}]{Strickland:2017kux}
\bibinfo{author}{\bibfnamefont{M.}~\bibnamefont{Strickland}},
  \bibinfo{author}{\bibfnamefont{J.}~\bibnamefont{Noronha}}, \bibnamefont{and}
  \bibinfo{author}{\bibfnamefont{G.}~\bibnamefont{Denicol}}
  (\bibinfo{year}{2017}), \eprint{1709.06644}.

\bibitem[{\citenamefont{Florkowski et~al.}(2017)\citenamefont{Florkowski,
  Maksymiuk, and Ryblewski}}]{Florkowski:2017jnz}
\bibinfo{author}{\bibfnamefont{W.}~\bibnamefont{Florkowski}},
  \bibinfo{author}{\bibfnamefont{E.}~\bibnamefont{Maksymiuk}},
  \bibnamefont{and} \bibinfo{author}{\bibfnamefont{R.}~\bibnamefont{Ryblewski}}
  (\bibinfo{year}{2017}), \eprint{1710.07095}.

\bibitem[{\citenamefont{Romatschke}(2017{\natexlab{b}})}]{Romatschke:2016hle}
\bibinfo{author}{\bibfnamefont{P.}~\bibnamefont{Romatschke}},
  \bibinfo{journal}{Eur. Phys. J.} \textbf{\bibinfo{volume}{C77}},
  \bibinfo{pages}{21} (\bibinfo{year}{2017}{\natexlab{b}}),
  \eprint{1609.02820}.

\bibitem[{\citenamefont{Romatschke}(2017{\natexlab{c}})}]{Romatschke:2017acs}
\bibinfo{author}{\bibfnamefont{P.}~\bibnamefont{Romatschke}}
  (\bibinfo{year}{2017}{\natexlab{c}}), \eprint{1710.03234}.

\bibitem[{\citenamefont{Gubser}(2010)}]{Gubser:2010ze}
\bibinfo{author}{\bibfnamefont{S.~S.} \bibnamefont{Gubser}},
  \bibinfo{journal}{Phys.Rev.} \textbf{\bibinfo{volume}{D82}},
  \bibinfo{pages}{085027} (\bibinfo{year}{2010}), \eprint{1006.0006}.

\bibitem[{\citenamefont{Gubser and Yarom}(2011)}]{Gubser:2010ui}
\bibinfo{author}{\bibfnamefont{S.~S.} \bibnamefont{Gubser}} \bibnamefont{and}
  \bibinfo{author}{\bibfnamefont{A.}~\bibnamefont{Yarom}},
  \bibinfo{journal}{Nucl.Phys.} \textbf{\bibinfo{volume}{B846}},
  \bibinfo{pages}{469} (\bibinfo{year}{2011}), \eprint{1012.1314}.

\bibitem[{\citenamefont{Kloeden and Rasmussen}(2011)}]{kloeden}
\bibinfo{author}{\bibfnamefont{P.}~\bibnamefont{Kloeden}} \bibnamefont{and}
  \bibinfo{author}{\bibfnamefont{M.}~\bibnamefont{Rasmussen}},
  \emph{\bibinfo{title}{Nonautonomous Dynamical Systems}}, Mathematical surveys
  and monographs (\bibinfo{publisher}{American Mathematical Soc.},
  \bibinfo{year}{2011}).

\bibitem[{\citenamefont{Tournier}(1994)}]{tournier}
\bibinfo{author}{\bibfnamefont{E.}~\bibnamefont{Tournier}},
  \emph{\bibinfo{title}{Computer Algebra and Differential Equations}}, Lecture
  note series / London Mathematical Society (\bibinfo{publisher}{Cambridge
  University Press}, \bibinfo{year}{1994}).

\bibitem[{\citenamefont{Israel and Stewart}(1979)}]{Israel:1979wp}
\bibinfo{author}{\bibfnamefont{W.}~\bibnamefont{Israel}} \bibnamefont{and}
  \bibinfo{author}{\bibfnamefont{J.~M.} \bibnamefont{Stewart}},
  \bibinfo{journal}{Ann. Phys.} \textbf{\bibinfo{volume}{118}},
  \bibinfo{pages}{341} (\bibinfo{year}{1979}).

\bibitem[{\citenamefont{Denicol et~al.}(2012)\citenamefont{Denicol, Niemi,
  Moln\'ar, and Rischke}}]{Denicol:2012cn}
\bibinfo{author}{\bibfnamefont{G.~S.} \bibnamefont{Denicol}},
  \bibinfo{author}{\bibfnamefont{H.}~\bibnamefont{Niemi}},
  \bibinfo{author}{\bibfnamefont{E.}~\bibnamefont{Moln\'ar}}, \bibnamefont{and}
  \bibinfo{author}{\bibfnamefont{D.~H.} \bibnamefont{Rischke}},
  \bibinfo{journal}{Phys. Rev. D} \textbf{\bibinfo{volume}{85}},
  \bibinfo{pages}{114047} (\bibinfo{year}{2012}).

\bibitem[{\citenamefont{Martinez and Strickland}(2010)}]{Martinez:2010sc}
\bibinfo{author}{\bibfnamefont{M.}~\bibnamefont{Martinez}} \bibnamefont{and}
  \bibinfo{author}{\bibfnamefont{M.}~\bibnamefont{Strickland}},
  \bibinfo{journal}{Nucl. Phys.} \textbf{\bibinfo{volume}{A848}},
  \bibinfo{pages}{183} (\bibinfo{year}{2010}), \eprint{1007.0889}.

\bibitem[{\citenamefont{Ryblewski and Florkowski}(2011)}]{Ryblewski:2010ch}
\bibinfo{author}{\bibfnamefont{R.}~\bibnamefont{Ryblewski}} \bibnamefont{and}
  \bibinfo{author}{\bibfnamefont{W.}~\bibnamefont{Florkowski}},
  \bibinfo{journal}{Acta Phys. Polon.} \textbf{\bibinfo{volume}{B42}},
  \bibinfo{pages}{115} (\bibinfo{year}{2011}), \eprint{1011.6213}.

\bibitem[{\citenamefont{Martinez et~al.}(2017)\citenamefont{Martinez, McNelis,
  and Heinz}}]{Martinez:2017ibh}
\bibinfo{author}{\bibfnamefont{M.}~\bibnamefont{Martinez}},
  \bibinfo{author}{\bibfnamefont{M.}~\bibnamefont{McNelis}}, \bibnamefont{and}
  \bibinfo{author}{\bibfnamefont{U.}~\bibnamefont{Heinz}},
  \bibinfo{journal}{Phys. Rev.} \textbf{\bibinfo{volume}{C95}},
  \bibinfo{pages}{054907} (\bibinfo{year}{2017}), \eprint{1703.10955}.

\bibitem[{\citenamefont{Alqahtani
  et~al.}(2017{\natexlab{a}})\citenamefont{Alqahtani, Nopoush, Ryblewski, and
  Strickland}}]{Alqahtani:2017tnq}
\bibinfo{author}{\bibfnamefont{M.}~\bibnamefont{Alqahtani}},
  \bibinfo{author}{\bibfnamefont{M.}~\bibnamefont{Nopoush}},
  \bibinfo{author}{\bibfnamefont{R.}~\bibnamefont{Ryblewski}},
  \bibnamefont{and}
  \bibinfo{author}{\bibfnamefont{M.}~\bibnamefont{Strickland}}
  (\bibinfo{year}{2017}{\natexlab{a}}), \eprint{1705.10191}.

\bibitem[{\citenamefont{Alqahtani
  et~al.}(2017{\natexlab{b}})\citenamefont{Alqahtani, Nopoush, Ryblewski, and
  Strickland}}]{Alqahtani:2017jwl}
\bibinfo{author}{\bibfnamefont{M.}~\bibnamefont{Alqahtani}},
  \bibinfo{author}{\bibfnamefont{M.}~\bibnamefont{Nopoush}},
  \bibinfo{author}{\bibfnamefont{R.}~\bibnamefont{Ryblewski}},
  \bibnamefont{and}
  \bibinfo{author}{\bibfnamefont{M.}~\bibnamefont{Strickland}},
  \bibinfo{journal}{Phys. Rev. Lett.} \textbf{\bibinfo{volume}{119}},
  \bibinfo{pages}{042301} (\bibinfo{year}{2017}{\natexlab{b}}),
  \eprint{1703.05808}.

\bibitem[{\citenamefont{Bluhm and Schäfer}(2015)}]{Bluhm:2015raa}
\bibinfo{author}{\bibfnamefont{M.}~\bibnamefont{Bluhm}} \bibnamefont{and}
  \bibinfo{author}{\bibfnamefont{T.}~\bibnamefont{Schäfer}},
  \bibinfo{journal}{Phys. Rev.} \textbf{\bibinfo{volume}{A92}},
  \bibinfo{pages}{043602} (\bibinfo{year}{2015}), \eprint{1505.00846}.

\bibitem[{\citenamefont{Bluhm and Schaefer}(2016)}]{Bluhm:2015bzi}
\bibinfo{author}{\bibfnamefont{M.}~\bibnamefont{Bluhm}} \bibnamefont{and}
  \bibinfo{author}{\bibfnamefont{T.}~\bibnamefont{Schaefer}},
  \bibinfo{journal}{Phys. Rev. Lett.} \textbf{\bibinfo{volume}{116}},
  \bibinfo{pages}{115301} (\bibinfo{year}{2016}), \eprint{1512.00862}.

\bibitem[{\citenamefont{Alqahtani
  et~al.}(2017{\natexlab{c}})\citenamefont{Alqahtani, Nopoush, and
  Strickland}}]{Alqahtani:2016rth}
\bibinfo{author}{\bibfnamefont{M.}~\bibnamefont{Alqahtani}},
  \bibinfo{author}{\bibfnamefont{M.}~\bibnamefont{Nopoush}}, \bibnamefont{and}
  \bibinfo{author}{\bibfnamefont{M.}~\bibnamefont{Strickland}},
  \bibinfo{journal}{Phys. Rev.} \textbf{\bibinfo{volume}{C95}},
  \bibinfo{pages}{034906} (\bibinfo{year}{2017}{\natexlab{c}}),
  \eprint{1605.02101}.

\bibitem[{\citenamefont{Molnar et~al.}(2016)\citenamefont{Molnar, Niemi, and
  Rischke}}]{Molnar:2016vvu}
\bibinfo{author}{\bibfnamefont{E.}~\bibnamefont{Molnar}},
  \bibinfo{author}{\bibfnamefont{H.}~\bibnamefont{Niemi}}, \bibnamefont{and}
  \bibinfo{author}{\bibfnamefont{D.~H.} \bibnamefont{Rischke}},
  \bibinfo{journal}{Phys. Rev.} \textbf{\bibinfo{volume}{D93}},
  \bibinfo{pages}{114025} (\bibinfo{year}{2016}), \eprint{1602.00573}.

\bibitem[{\citenamefont{Molnár et~al.}(2016)\citenamefont{Molnár, Niemi, and
  Rischke}}]{Molnar:2016gwq}
\bibinfo{author}{\bibfnamefont{E.}~\bibnamefont{Molnár}},
  \bibinfo{author}{\bibfnamefont{H.}~\bibnamefont{Niemi}}, \bibnamefont{and}
  \bibinfo{author}{\bibfnamefont{D.~H.} \bibnamefont{Rischke}}
  (\bibinfo{year}{2016}), \eprint{1606.09019}.

\bibitem[{\citenamefont{Bazow et~al.}(2015)\citenamefont{Bazow, Heinz, and
  Martinez}}]{Bazow:2015cha}
\bibinfo{author}{\bibfnamefont{D.}~\bibnamefont{Bazow}},
  \bibinfo{author}{\bibfnamefont{U.~W.} \bibnamefont{Heinz}}, \bibnamefont{and}
  \bibinfo{author}{\bibfnamefont{M.}~\bibnamefont{Martinez}},
  \bibinfo{journal}{Phys. Rev.} \textbf{\bibinfo{volume}{C91}},
  \bibinfo{pages}{064903} (\bibinfo{year}{2015}), \eprint{1503.07443}.

\bibitem[{\citenamefont{Nopoush et~al.}(2014)\citenamefont{Nopoush, Ryblewski,
  and Strickland}}]{Nopoush:2014pfa}
\bibinfo{author}{\bibfnamefont{M.}~\bibnamefont{Nopoush}},
  \bibinfo{author}{\bibfnamefont{R.}~\bibnamefont{Ryblewski}},
  \bibnamefont{and}
  \bibinfo{author}{\bibfnamefont{M.}~\bibnamefont{Strickland}},
  \bibinfo{journal}{Phys. Rev.} \textbf{\bibinfo{volume}{C90}},
  \bibinfo{pages}{014908} (\bibinfo{year}{2014}), \eprint{1405.1355}.

\bibitem[{\citenamefont{Tinti and Florkowski}(2014)}]{Tinti:2013vba}
\bibinfo{author}{\bibfnamefont{L.}~\bibnamefont{Tinti}} \bibnamefont{and}
  \bibinfo{author}{\bibfnamefont{W.}~\bibnamefont{Florkowski}},
  \bibinfo{journal}{Phys. Rev.} \textbf{\bibinfo{volume}{C89}},
  \bibinfo{pages}{034907} (\bibinfo{year}{2014}), \eprint{1312.6614}.

\bibitem[{\citenamefont{Bazow et~al.}(2013)\citenamefont{Bazow, Heinz, and
  Strickland}}]{Bazow:2013ifa}
\bibinfo{author}{\bibfnamefont{D.}~\bibnamefont{Bazow}},
  \bibinfo{author}{\bibfnamefont{U.~W.} \bibnamefont{Heinz}}, \bibnamefont{and}
  \bibinfo{author}{\bibfnamefont{M.}~\bibnamefont{Strickland}}
  (\bibinfo{year}{2013}), \eprint{1311.6720}.

\bibitem[{\citenamefont{Ryblewski and Florkowski}(2012)}]{Ryblewski:2012rr}
\bibinfo{author}{\bibfnamefont{R.}~\bibnamefont{Ryblewski}} \bibnamefont{and}
  \bibinfo{author}{\bibfnamefont{W.}~\bibnamefont{Florkowski}},
  \bibinfo{journal}{Phys. Rev.} \textbf{\bibinfo{volume}{C85}},
  \bibinfo{pages}{064901} (\bibinfo{year}{2012}), \eprint{1204.2624}.

\bibitem[{\citenamefont{Martinez et~al.}(2012)\citenamefont{Martinez,
  Ryblewski, and Strickland}}]{Martinez:2012tu}
\bibinfo{author}{\bibfnamefont{M.}~\bibnamefont{Martinez}},
  \bibinfo{author}{\bibfnamefont{R.}~\bibnamefont{Ryblewski}},
  \bibnamefont{and}
  \bibinfo{author}{\bibfnamefont{M.}~\bibnamefont{Strickland}},
  \bibinfo{journal}{Phys. Rev.} \textbf{\bibinfo{volume}{C85}},
  \bibinfo{pages}{064913} (\bibinfo{year}{2012}), \eprint{1204.1473}.

\bibitem[{\citenamefont{Mintzer}(1965)}]{mintzer}
\bibinfo{author}{\bibfnamefont{D.}~\bibnamefont{Mintzer}},
  \bibinfo{journal}{The Physics of Fluids} \textbf{\bibinfo{volume}{8}},
  \bibinfo{pages}{1076} (\bibinfo{year}{1965}).

\bibitem[{\citenamefont{Grad}(1949)}]{Grad}
\bibinfo{author}{\bibfnamefont{H.}~\bibnamefont{Grad}},
  \bibinfo{journal}{Commun.Pure Appl.Math.} \textbf{\bibinfo{volume}{2}},
  \bibinfo{pages}{331} (\bibinfo{year}{1949}).

\bibitem[{\citenamefont{Marrochio et~al.}(2013)\citenamefont{Marrochio,
  Noronha, Denicol, Luzum, Jeon et~al.}}]{Marrochio:2013wla}
\bibinfo{author}{\bibfnamefont{H.}~\bibnamefont{Marrochio}},
  \bibinfo{author}{\bibfnamefont{J.}~\bibnamefont{Noronha}},
  \bibinfo{author}{\bibfnamefont{G.~S.} \bibnamefont{Denicol}},
  \bibinfo{author}{\bibfnamefont{M.}~\bibnamefont{Luzum}},
  \bibinfo{author}{\bibfnamefont{S.}~\bibnamefont{Jeon}}, \bibnamefont{et~al.}
  (\bibinfo{year}{2013}), \eprint{1307.6130}.

\bibitem[{\citenamefont{Huang et~al.}(2010)\citenamefont{Huang, Huang, Rischke,
  and Sedrakian}}]{Huang:2009ue}
\bibinfo{author}{\bibfnamefont{X.-G.} \bibnamefont{Huang}},
  \bibinfo{author}{\bibfnamefont{M.}~\bibnamefont{Huang}},
  \bibinfo{author}{\bibfnamefont{D.~H.} \bibnamefont{Rischke}},
  \bibnamefont{and}
  \bibinfo{author}{\bibfnamefont{A.}~\bibnamefont{Sedrakian}},
  \bibinfo{journal}{Phys. Rev.} \textbf{\bibinfo{volume}{D81}},
  \bibinfo{pages}{045015} (\bibinfo{year}{2010}), \eprint{0910.3633}.

\bibitem[{\citenamefont{Huang et~al.}(2011)\citenamefont{Huang, Sedrakian, and
  Rischke}}]{Huang:2011dc}
\bibinfo{author}{\bibfnamefont{X.-G.} \bibnamefont{Huang}},
  \bibinfo{author}{\bibfnamefont{A.}~\bibnamefont{Sedrakian}},
  \bibnamefont{and} \bibinfo{author}{\bibfnamefont{D.~H.}
  \bibnamefont{Rischke}}, \bibinfo{journal}{Annals Phys.}
  \textbf{\bibinfo{volume}{326}}, \bibinfo{pages}{3075} (\bibinfo{year}{2011}),
  \eprint{1108.0602}.

\bibitem[{\citenamefont{Gedalin}(1991)}]{Gedalin1}
\bibinfo{author}{\bibfnamefont{M.}~\bibnamefont{Gedalin}},
  \bibinfo{journal}{Physics of Fluids B} \textbf{\bibinfo{volume}{3}},
  \bibinfo{pages}{1871} (\bibinfo{year}{1991}).

\bibitem[{\citenamefont{Gedalin and Oiberman}(1995)}]{Gedalin2}
\bibinfo{author}{\bibfnamefont{M.}~\bibnamefont{Gedalin}} \bibnamefont{and}
  \bibinfo{author}{\bibfnamefont{I.}~\bibnamefont{Oiberman}},
  \bibinfo{journal}{Phys. Rev. E} \textbf{\bibinfo{volume}{51}},
  \bibinfo{pages}{4901} (\bibinfo{year}{1995}).

\bibitem[{\citenamefont{Romatschke and Strickland}(2003)}]{Romatschke:2003ms}
\bibinfo{author}{\bibfnamefont{P.}~\bibnamefont{Romatschke}} \bibnamefont{and}
  \bibinfo{author}{\bibfnamefont{M.}~\bibnamefont{Strickland}},
  \bibinfo{journal}{Phys. Rev.} \textbf{\bibinfo{volume}{D68}},
  \bibinfo{pages}{036004} (\bibinfo{year}{2003}), \eprint{hep-ph/0304092}.

\bibitem[{\citenamefont{Molnár et~al.}(2014)\citenamefont{Molnár, Niemi,
  Denicol, and Rischke}}]{Molnar:2013lta}
\bibinfo{author}{\bibfnamefont{E.}~\bibnamefont{Molnár}},
  \bibinfo{author}{\bibfnamefont{H.}~\bibnamefont{Niemi}},
  \bibinfo{author}{\bibfnamefont{G.~S.} \bibnamefont{Denicol}},
  \bibnamefont{and} \bibinfo{author}{\bibfnamefont{D.~H.}
  \bibnamefont{Rischke}}, \bibinfo{journal}{Phys. Rev.}
  \textbf{\bibinfo{volume}{D89}}, \bibinfo{pages}{074010}
  (\bibinfo{year}{2014}), \eprint{1308.0785}.

\bibitem[{\citenamefont{Tsumura et~al.}(2015)\citenamefont{Tsumura, Kikuchi,
  and Kunihiro}}]{Tsumura:2015fxa}
\bibinfo{author}{\bibfnamefont{K.}~\bibnamefont{Tsumura}},
  \bibinfo{author}{\bibfnamefont{Y.}~\bibnamefont{Kikuchi}}, \bibnamefont{and}
  \bibinfo{author}{\bibfnamefont{T.}~\bibnamefont{Kunihiro}},
  \bibinfo{journal}{Phys. Rev.} \textbf{\bibinfo{volume}{D92}},
  \bibinfo{pages}{085048} (\bibinfo{year}{2015}), \eprint{1506.00846}.

\bibitem[{\citenamefont{Banerjee et~al.}(1989)\citenamefont{Banerjee, Bhalerao,
  and Ravishankar}}]{Banerjee:1989by}
\bibinfo{author}{\bibfnamefont{B.}~\bibnamefont{Banerjee}},
  \bibinfo{author}{\bibfnamefont{R.~S.} \bibnamefont{Bhalerao}},
  \bibnamefont{and}
  \bibinfo{author}{\bibfnamefont{V.}~\bibnamefont{Ravishankar}},
  \bibinfo{journal}{Phys. Lett.} \textbf{\bibinfo{volume}{B224}},
  \bibinfo{pages}{16} (\bibinfo{year}{1989}).

\bibitem[{\citenamefont{Hartman}(1960)}]{hartman}
\bibinfo{author}{\bibfnamefont{P.}~\bibnamefont{Hartman}},
  \bibinfo{journal}{Proceedings of the American Mathematical Society}
  \textbf{\bibinfo{volume}{11}}, \bibinfo{pages}{610} (\bibinfo{year}{1960}).

\bibitem[{\citenamefont{Arrowsmith and Place}(1992)}]{arrowsmith}
\bibinfo{author}{\bibfnamefont{D.}~\bibnamefont{Arrowsmith}} \bibnamefont{and}
  \bibinfo{author}{\bibfnamefont{C.}~\bibnamefont{Place}},
  \emph{\bibinfo{title}{Dynamical Systems: Differential Equations, Maps, and
  Chaotic Behaviour}}, Chapman Hall/CRC Mathematics Series
  (\bibinfo{publisher}{Taylor \& Francis}, \bibinfo{year}{1992}).

\bibitem[{\citenamefont{Giesl and Wendland}(2011)}]{numbasin}
\bibinfo{author}{\bibfnamefont{P.}~\bibnamefont{Giesl}} \bibnamefont{and}
  \bibinfo{author}{\bibfnamefont{H.}~\bibnamefont{Wendland}},
  \bibinfo{journal}{Nonlinear Analysis: Theory, Methods $\&$ Applications}
  \textbf{\bibinfo{volume}{74}}, \bibinfo{pages}{3191 } (\bibinfo{year}{2011}).

\bibitem[{\citenamefont{Witten}(2011)}]{Witten:2010cx}
\bibinfo{author}{\bibfnamefont{E.}~\bibnamefont{Witten}},
  \bibinfo{journal}{AMS/IP Stud. Adv. Math.} \textbf{\bibinfo{volume}{50}},
  \bibinfo{pages}{347} (\bibinfo{year}{2011}), \eprint{1001.2933}.

\bibitem[{\citenamefont{Behtash et~al.}(2015)\citenamefont{Behtash, Dunne,
  Schaefer, Sulejmanpasic, and Unsal}}]{Behtash:2015loa}
\bibinfo{author}{\bibfnamefont{A.}~\bibnamefont{Behtash}},
  \bibinfo{author}{\bibfnamefont{G.~V.} \bibnamefont{Dunne}},
  \bibinfo{author}{\bibfnamefont{T.}~\bibnamefont{Schaefer}},
  \bibinfo{author}{\bibfnamefont{T.}~\bibnamefont{Sulejmanpasic}},
  \bibnamefont{and} \bibinfo{author}{\bibfnamefont{M.}~\bibnamefont{Unsal}}
  (\bibinfo{year}{2015}), \eprint{1510.03435}.

\bibitem[{\citenamefont{Behtash}(2017)}]{Behtash:2017rqj}
\bibinfo{author}{\bibfnamefont{A.}~\bibnamefont{Behtash}}
  (\bibinfo{year}{2017}), \eprint{1703.00511}.

\bibitem[{\citenamefont{Dunne and Ünsal}(2016)}]{Dunne:2015eaa}
\bibinfo{author}{\bibfnamefont{G.~V.} \bibnamefont{Dunne}} \bibnamefont{and}
  \bibinfo{author}{\bibfnamefont{M.}~\bibnamefont{Ünsal}},
  \bibinfo{journal}{PoS} \textbf{\bibinfo{volume}{LATTICE2015}},
  \bibinfo{pages}{010} (\bibinfo{year}{2016}), \eprint{1511.05977}.

\bibitem[{\citenamefont{Tirapegui et~al.}(2000)\citenamefont{Tirapegui,
  Martinez, and Tiemann}}]{tirapegui2000instabilities}
\bibinfo{author}{\bibfnamefont{E.}~\bibnamefont{Tirapegui}},
  \bibinfo{author}{\bibfnamefont{J.}~\bibnamefont{Martinez}}, \bibnamefont{and}
  \bibinfo{author}{\bibfnamefont{R.}~\bibnamefont{Tiemann}},
  \emph{\bibinfo{title}{Instabilities and Nonequilibrium Structures VI}},
  Nonlinear Phenomena and Complex Systems (\bibinfo{publisher}{Springer
  Netherlands}, \bibinfo{year}{2000}).

\bibitem[{\citenamefont{Giesl}(2007)}]{giesl}
\bibinfo{author}{\bibfnamefont{P.}~\bibnamefont{Giesl}},
  \emph{\bibinfo{title}{Construction of Global Lyapunov Functions Using Radial
  Basis Functions}}, Lecture Notes in Mathematics (\bibinfo{publisher}{Springer
  Berlin Heidelberg}, \bibinfo{year}{2007}).

\bibitem[{\citenamefont{Chesi et~al.}(2009)\citenamefont{Chesi, Garulli, Tesi,
  and Vicino}}]{chesi2009homogeneous}
\bibinfo{author}{\bibfnamefont{G.}~\bibnamefont{Chesi}},
  \bibinfo{author}{\bibfnamefont{A.}~\bibnamefont{Garulli}},
  \bibinfo{author}{\bibfnamefont{A.}~\bibnamefont{Tesi}}, \bibnamefont{and}
  \bibinfo{author}{\bibfnamefont{A.}~\bibnamefont{Vicino}},
  \emph{\bibinfo{title}{Homogeneous Polynomial Forms for Robustness Analysis of
  Uncertain Systems}}, Lecture Notes in Control and Information Sciences
  (\bibinfo{publisher}{Springer London}, \bibinfo{year}{2009}).

\bibitem[{\citenamefont{Giesl and Wendland}(2012)}]{numbasin2}
\bibinfo{author}{\bibfnamefont{P.}~\bibnamefont{Giesl}} \bibnamefont{and}
  \bibinfo{author}{\bibfnamefont{H.}~\bibnamefont{Wendland}},
  \bibinfo{journal}{Nonlinear Analysis: Theory, Methods \& Applications}
  \textbf{\bibinfo{volume}{75}}, \bibinfo{pages}{2823 } (\bibinfo{year}{2012}).

\bibitem[{\citenamefont{Liddle et~al.}(1994)\citenamefont{Liddle, Parsons, and
  Barrow}}]{Liddle:1994dx}
\bibinfo{author}{\bibfnamefont{A.~R.} \bibnamefont{Liddle}},
  \bibinfo{author}{\bibfnamefont{P.}~\bibnamefont{Parsons}}, \bibnamefont{and}
  \bibinfo{author}{\bibfnamefont{J.~D.} \bibnamefont{Barrow}},
  \bibinfo{journal}{Phys. Rev.} \textbf{\bibinfo{volume}{D50}},
  \bibinfo{pages}{7222} (\bibinfo{year}{1994}), \eprint{astro-ph/9408015}.

\bibitem[{\citenamefont{Heinz and Martinez}(2015)}]{Heinz:2015cda}
\bibinfo{author}{\bibfnamefont{U.~W.} \bibnamefont{Heinz}} \bibnamefont{and}
  \bibinfo{author}{\bibfnamefont{M.}~\bibnamefont{Martinez}},
  \bibinfo{journal}{Nucl. Phys.} \textbf{\bibinfo{volume}{A943}},
  \bibinfo{pages}{26} (\bibinfo{year}{2015}), \eprint{1506.07500}.

\bibitem[{\citenamefont{Martinez and Strickland}(2009)}]{Martinez:2009mf}
\bibinfo{author}{\bibfnamefont{M.}~\bibnamefont{Martinez}} \bibnamefont{and}
  \bibinfo{author}{\bibfnamefont{M.}~\bibnamefont{Strickland}},
  \bibinfo{journal}{Phys. Rev.} \textbf{\bibinfo{volume}{C79}},
  \bibinfo{pages}{044903} (\bibinfo{year}{2009}), \eprint{0902.3834}.

\bibitem[{\citenamefont{Apostol}(1974)}]{apostol}
\bibinfo{author}{\bibfnamefont{T.}~\bibnamefont{Apostol}},
  \emph{\bibinfo{title}{Mathematical Analysis}}, Addison-Wesley series in
  mathematics (\bibinfo{publisher}{Addison-Wesley}, \bibinfo{year}{1974}).

\bibitem[{\citenamefont{Costin and Costin}(2001)}]{Costin2001}
\bibinfo{author}{\bibfnamefont{O.}~\bibnamefont{Costin}} \bibnamefont{and}
  \bibinfo{author}{\bibfnamefont{R.}~\bibnamefont{Costin}},
  \bibinfo{journal}{Inventiones mathematicae} \textbf{\bibinfo{volume}{145}},
  \bibinfo{pages}{425} (\bibinfo{year}{2001}).

\bibitem[{\citenamefont{Costin}(1998)}]{costin1998}
\bibinfo{author}{\bibfnamefont{O.}~\bibnamefont{Costin}},
  \bibinfo{journal}{Duke Math. J.} \textbf{\bibinfo{volume}{93}},
  \bibinfo{pages}{289} (\bibinfo{year}{1998}).

\bibitem[{\citenamefont{Bazow et~al.}(2016{\natexlab{b}})\citenamefont{Bazow,
  Denicol, Heinz, Martinez, and Noronha}}]{Bazow:2015dha}
\bibinfo{author}{\bibfnamefont{D.}~\bibnamefont{Bazow}},
  \bibinfo{author}{\bibfnamefont{G.~S.} \bibnamefont{Denicol}},
  \bibinfo{author}{\bibfnamefont{U.}~\bibnamefont{Heinz}},
  \bibinfo{author}{\bibfnamefont{M.}~\bibnamefont{Martinez}}, \bibnamefont{and}
  \bibinfo{author}{\bibfnamefont{J.}~\bibnamefont{Noronha}},
  \bibinfo{journal}{Phys. Rev. Lett.} \textbf{\bibinfo{volume}{116}},
  \bibinfo{pages}{022301} (\bibinfo{year}{2016}{\natexlab{b}}),
  \eprint{1507.07834}.

\bibitem[{\citenamefont{Costin}(2008)}]{costin}
\bibinfo{author}{\bibfnamefont{O.}~\bibnamefont{Costin}},
  \emph{\bibinfo{title}{Asymptotics and Borel Summability}}, Monographs and
  Surveys in Pure and Applied Mathematics (\bibinfo{publisher}{CRC Press},
  \bibinfo{year}{2008}).

\end{thebibliography}

\end{document}